\documentclass[onecolumn,authoryear]{els-mrw} 

\usepackage{amsmath,amssymb,amsfonts,amsthm,makeidx,graphicx}
\usepackage{txfonts}
\usepackage{helvet}
\usepackage[colorlinks=true, citecolor=blue, linkcolor=blue, urlcolor=blue]{hyperref}
\usepackage{doi}
\usepackage{orcidlink}

\newcommand{\mnras}{MNRAS}                
\newcommand{\apj}{ApJ}                    
\newcommand{\apjl}{ApJL}                  
\newcommand{\apjs}{ApJS}                  
\newcommand{\aj}{AJ}                      
\newcommand{\aap}{A\&A}                   
\newcommand{\araa}{ARA\&A}                
\newcommand{\pasp}{PASP}                  
\newcommand{\nat}{Nature}                 
\newcommand{\jcap}{JCAP}                  
\newcommand{\rmxaa}{Rev.\ Mex.\ Astron.\ Astrofis.} 
\newcommand{\nar}{New Astron.\ Rev.}      

\begin{document}
\hypersetup{colorlinks=true, citecolor=blue, linkcolor=blue, urlcolor=blue}

\chapter{Cosmological Simulations of Galaxies}\label{chap1}

\author[1]{Robert Feldmann\orcidlink{0000-0002-1109-1919}\footnote{Email: robert.feldmann@uzh.ch}}
\author[1]{Rebekka Bieri\orcidlink{0000-0002-4554-4488}}

\address[1]{\orgname{Department of Astrophysics}, \orgdiv{Universität Zürich}, \orgaddress{Winterthurerstrasse 190, CH-8057 Zurich}}


\maketitle

\begin{abstract}[Abstract]
Galaxy simulations have come a long way from the early days of simple $N$-body calculations, which considered only gravitational interactions, to the complex, multi-physics models used today. Beginning with initial conditions representative of the Universe shortly after the Big Bang, these modern simulations integrate the relevant physical processes involved in galaxy formation, such as gravity, gas dynamics, cooling, star formation, and feedback, while accounting for cosmic expansion and structure formation. This review provides an introductory overview of cosmological galaxy simulations, outlining the essential components and methods used to model the formation and evolution of galaxies on the computer. It also discusses common steps in the post-processing analysis, essential for extracting physical insights from these numerical experiments, along with basic tests to assess simulation validity and accuracy. 
Looking forward, next-generation simulations aim to push resolution boundaries, incorporate additional physical processes, and improve the robustness of the numerical models, promising to lead to a deeper understanding of how galaxies emerged and evolved over cosmic time.
\end{abstract}

\begin{BoxTypeA}[chap1:box1]{Key points}
\begin{itemize}
\item \textbf{Advances in Computing Power:} Increased computational capabilities have transformed galaxy simulations, enabling the modeling of multi-scale physical processes such star formation and feedback in a fully cosmological context.
\item \textbf{Cosmological Context:} Accurate initial conditions, based on measurements of the cosmic microwave background, along with a well-defined cosmological model, provide the framework for tracing galaxy formation and evolution from the early Universe to the present day.
\item \textbf{Solving for Dark Matter and Gas:} 
Advanced numerical solvers efficiently compute dark matter and gas dynamics, accurately modeling gravitational interactions and fluid behavior.
\item \textbf{Baryonic Physics Modeling:} Baryonic processes such as star formation, feedback, and cooling are fundamental to galaxy formation and must be accurately represented in the simulations. 
\item \textbf{Role of Feedback:} Feedback from stars and active galactic nuclei (AGN) plays a critical role in regulating star formation and shaping galactic morphology by driving gas flows and altering mass distributions.
\item \textbf{Comparison with Observations:} Robust comparisons with observations are crucial for evaluating the accuracy and reliability of simulations and their predictions.
\item \textbf{Next-Generation Simulations:} Future simulations aim to enhance the physical realism of sub-grid models and incorporate more detailed physics, providing a deeper understanding of the complexities of galaxy evolution.
\end{itemize}
\end{BoxTypeA}

\section{Introduction}\label{chap1:Motivation}

The study of galaxies has entered an unprecedented era with the advent of high-fidelity observations across multiple wavelengths with ground- and space-based facilities such as the James Webb Space Telescope (JWST), the Euclid satellite, the Atacama Large Millimeter/submillimeter Array (ALMA), and even gravitational wave observatories such as the Laser Interferometer Gravitational-Wave Observatory (LIGO). These instruments enable the study of galaxy evolution and their components across most of cosmic history, from the birth of the first galaxies at Cosmic Dawn a few hundred million years after the Big Bang to the present day, significantly advancing the understanding of galaxy formation. 

Cosmological simulations of galaxies offer a principled, theoretical framework to comprehend these observational findings and address key questions in galaxy formation \citep[see][for reviews]{Somerville+2015,
Naab+Ostriker2017,  Crain+VanDeVoort2023}. At their core, these simulations are numerical experiments that can be used to explore how galaxies form, grow, and interact over cosmic time by following the three-dimensional, dynamical evolution of gas, stars, and dark matter in an expanding Universe. The computer programs that create these simulations implement many of the relevant physical processes such as gravity, gas dynamics, star formation, and feedback from stars and active galactic nuclei (AGNs).

\begin{figure}[t]
\centering
\includegraphics[width=0.98\textwidth]{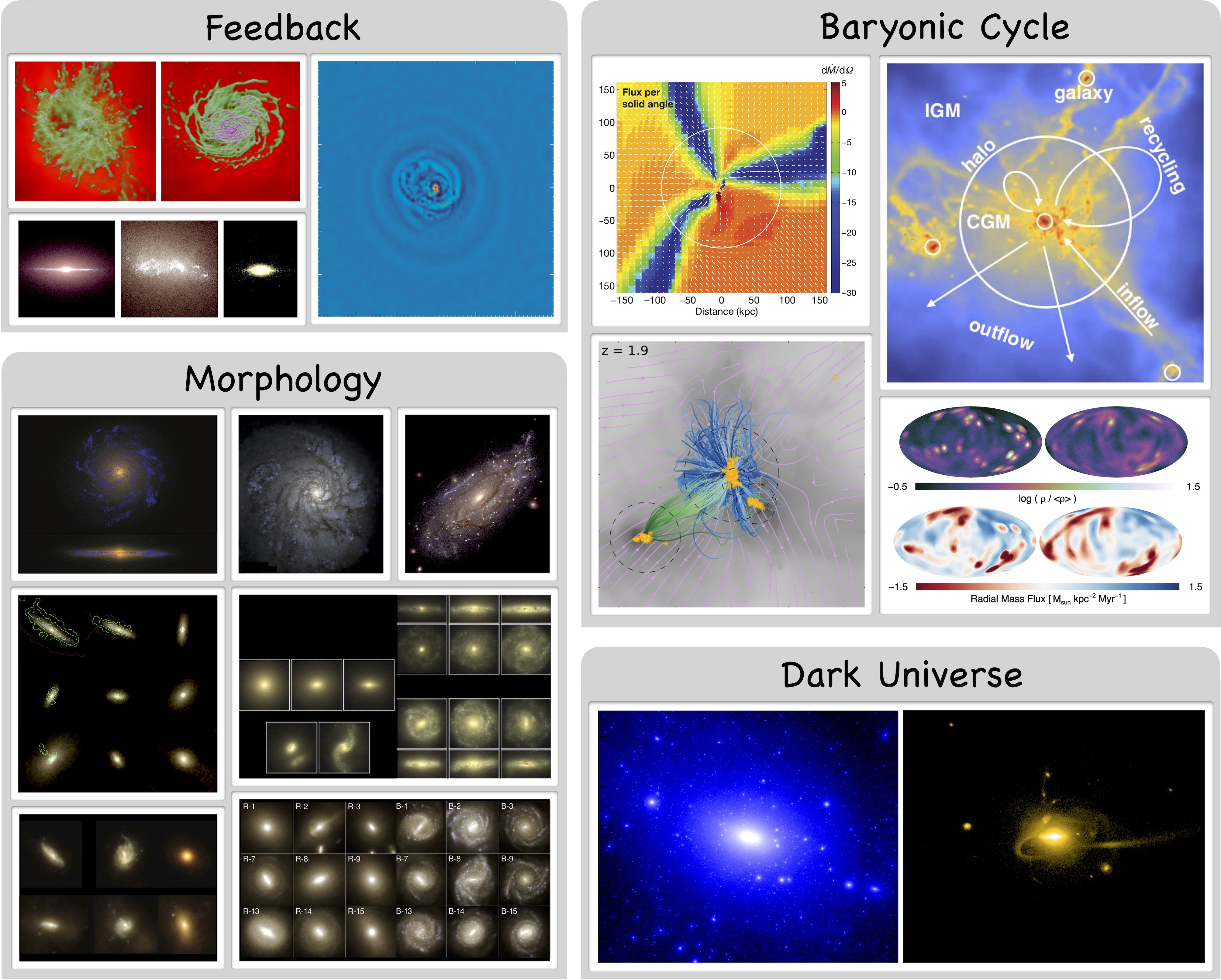}
\caption{
Cosmological galaxy simulations have led to major advancements in our understanding of how galaxies form and evolve. (\textbf{Top left quadrant}) Simulations have demonstrated that many galaxy properties are sensitive to the level of stellar and AGN feedback, highlighting their critical role in forming galaxies. Images (starting from the top left, moving counterclockwise) show the gas surface density of a simulated galaxy shortly after a feedback-driven gas ejection and the subsequent settling of gas into a disk \citep{Hopkins+2014}, an edge-on view of a disk galaxy simulated with standard, high, and low levels of stellar feedback \citep{Agertz+2016}, and the visible shells created by AGN feedback around the central galaxy in a simulated galaxy cluster \citep{Sijacki+2007}. (\textbf{Top right quadrant}) Simulations provide direct insights into how gas is accreted by galaxies. Images (starting from the top left, moving counterclockwise) show the inflow of cool gas onto a massive galaxy surrounded by a hot halo \citep{Dekel+2009}, the cycling of gas between galaxies and their neighbors \citep{Angles-Alcazar+2017}, a Mollweide projection of gas accreted onto a Milky Way-like halo \citep{Nelson+2013}, and the different gas components involved in this gas cycling process \citep{vandeVoort2017}. (\textbf{Bottom left quadrant}) Simulations have shown that the observed morphologies of galaxies, such as disk and elliptical shapes, are a natural outcome of astrophysical processes in a $\Lambda{}$CDM Universe. 
Milestones include the simulation of disk galaxies with realistic mass and light distributions (\citealt{Guesdes+2011, Hopkins+2018, Agertz+2021}, top left to right), reproducing the range of observed morphologies in a single cosmological simulation (\citealt{Feldmann+2011}, center left), and linking galaxy morphology to AGN feedback and environment in cosmological volume simulations such as EAGLE (\citealt{Schaye+2015}, center right), Horizon-AGN (\citealt{Dubois+2014}, bottom left), and Illustris (\citealt{Vogelsberger+2014}, bottom right).
 (\textbf{Bottom right quadrant}) The physics of galaxy formation significantly alters many predictions based on collisionless (dark matter-only) simulations. The image compares the total mass distribution from a dark-matter-only simulation (left) with the stellar mass distribution from a corresponding simulation with baryonic physics (right), illustrating that the inclusion of baryons results in fewer visible satellite galaxies compared to the number of substructures predicted by dark-matter-only simulations \citep{Wetzel+2016}.}
\label{fig:Highlights}
\end{figure}

Fig.~\ref{fig:Highlights} showcases four research areas where cosmological galaxy simulations have made critical contributions. For instance, such simulations have demonstrated that the diverse morphologies of observed galaxies arise naturally from astrophysical processes such as galaxy merging, environmental effects, gas cooling, and feedback \citep[e.g.,][] {Naab+2007, Mayer+2008, Guesdes+2011, Feldmann+2011, Hopkins+2014, Vogelsberger+2014, Schaye+2015, Dubois+2016, Correa+2017, El-Badry+2018, Rodriguez-Gomez+2019, Pillepich+2019, Dubois+2021, Hopkins+2023}. Additionally, they provided theoretical predictions for the flow of cosmic gas into and out of galaxies, the so-called baryonic cycle \citep{Tumlinson+2017, FaucherGiguere+Peng2023}, and studied how this flow depends on the adopted physics \citep[e.g.,][]{Keres+2005, Ocvirk+2008, Dekel+2009, vandeVoort+2011, Nelson+2013, Muratov+2015, Angles-Alcazar+2017, Roca-Fabrega+2019, Tollet+2019, Hafen+2019, Grand+2019, Mitchell+2020, Pandya+2021, Bassini+2023, Wright+2024}. Cosmological simulations of galaxies also underscored the critical role of feedback from stars \citep[e.g.,][]{Stinson+2013, Hopkins+2014, Agertz+2016, Hopkins+2018, Ocvirk+2020, Agertz+2020} and AGNs \citep{Sijacki+2007, Booth+2009, Dubois+2012, Hirschmann+2014, Weinberger+2017, Habouzit+2021, Wellons+2023} in regulating the star formation activity of galaxies. Furthermore, such simulations have been key in addressing challenges within the Lambda Cold Dark Matter ($\Lambda$CDM) framework, raised by dark matter-only simulations \citep[e.g.,][]{Bullock+2017, Sales+2022, Perivolaropoulos+2022}. The ``missing satellites'' problem, for example, arises from the discrepancy between the high number of orbiting substructures predicted by $\Lambda$CDM and the fewer visible satellites observed around galaxies like the Milky Way \citep{Kauffmann+1993, Klypin+1999, Moore+1999}. This issue has been largely resolved by incorporating stellar feedback and environmental effects \citep{Kravtsov+2004, Brooks+2013}, with recent galaxy simulations predicting satellite counts and densities that align more closely with observations \citep{Chan+2015, Dutton+2016, Sawala+2016, Wetzel+2016, Revaz2018, Garrison-Kimmel2019, Engler+2021, Jung+2024}.

The range of applicability of cosmological galaxy simulations is vast. They serve as a critical tool for testing theoretical models of galaxy formation by comparing predicted properties, such as masses, sizes, and morphologies, with observational data. Furthermore, they provide a means to study the evolution of individual galaxies across time and grant access to unobserved scales and processes. Simulations also offer a controlled environment for exploring new and speculative ideas, including alternative models of dark matter and dark energy, or the potential role of primordial black holes in shaping the Universe. Finally, simulations are a powerful tool for making predictions about what future observations might reveal. As we look toward upcoming instruments such as the Square Kilometer Array (SKA), the Vera Rubin Observatory, the Extremely Large Telescope (ELT), and the Laser Interferometer Space Antenna (LISA), cosmological simulations help guide future observational campaigns by predicting where and what we should look for--whether it is the earliest galaxies forming at high redshift or signatures of non-standard physics in low mass dwarf galaxies.

\section{Galaxy formation on the computer}\label{chap1:NumericalModeling}

\subsection{Overview}
 
The first galaxy simulations were conducted in the 1940s using an analog approach involving lightbulbs, pioneered by Erik Holmberg. Using the light intensity to model gravitational forces, these experiments provided a practical method for investigating galaxy interactions, demonstrating, for example, that tidal forces can induce spiral arms \citep{Holmberg1941}. In the 1970s and 1980s, with the advent of ever more powerful computers, digital $N$-body simulations were employed to study both the structure of galaxies \citep{Toomre+1972} and the clustering of matter in the Universe on large scales \citep{Aarseth+1979, Efstathiou+1981, Davis+1985}. By the 1990s, hydrodynamics solvers based on, e.g., smoothed particle hydrodynamics \citep[SPH,][]{Gingold+1977, Lucy1977} allowed for the modeling of cosmic gas dynamics, enabling the first cosmological galaxy simulations \citep{Katz+1991, Katz+1996}.

These early galaxy simulations suffered from excessive gas cooling and artificial angular momentum redistribution, resulting in galaxies that were too compact, contained too many baryons, and rotated too quickly \citep{Katz1992, Navarro+1994, Navarro+2000, Abadi+2003}. These issues motivated the development of more robust and efficient numerical solvers to improve integration accuracy and achieve higher resolution. Furthermore, the challenges in reproducing galaxy properties underscored the importance of carefully modeling sub-grid processes, such as gas cooling, star formation, and feedback, which are usually unresolved in cosmological simulations but critically influence galaxy evolution. In particular, the introduction of effective feedback mechanisms, which counteract excessive cooling and regulate galaxy growth, marked a breakthrough in the ability of simulations to produce galaxies with more realistic properties \citep[e.g.][]{Guesdes+2011, Hopkins+2014, Vogelsberger+2014, Schaye+2015, Agertz+2016}. 

Today, cosmological galaxy simulations are heavily used to study a wide range of open questions in galaxy formation and evolution \citep[e.g.][]{Naab+Ostriker2017, ReviewVogelsberger+2020, Crain+VanDeVoort2023}. 
The simulations differ in the numerical solvers they use, the volumes and timescales they probe, the resolution they achieve, and the physics they implement. For instance, large volume simulations such as EAGLE \citep{Schaye+2015}, Romulus \citep{Tremmel+2017}, Illustris-TNG \citep{Springel+2018, Pillepich+2018}, Simba \citep{Dave+2019}, and Millennium-TNG \citep{Pakmor+2023} and Flamingo \citep{Schaller+2024} target sizable samples of galaxies with sub-grid models adjusted for the corresponding lower numerical resolution. Other simulations, such as Latte \citep{Wetzel+2016} and MassiveFIRE \citep{Feldmann+2016} from the Feedback in Realistic Environments project \citep{Hopkins+2014, Hopkins+2018}, EDGE \citep{Agertz+2020}, VINTERGATAN \citep{Agertz+2021}, LYRA \citep{Gutcke+2021}, VELA \citep{Ceverino+2023}, and EDGE-INFERNO \citep{Andersson+2024}, zoom in on individual galaxies of different masses, allowing them to directly resolve many of the physical processes occurring in the interstellar medium (ISM) of galaxies.
High resolution simulations of moderate cosmological volumes or extended contiguous regions, such as NewHorizon \citep{Dubois+2021} and FIREbox \citep{Feldmann+2023}, bridge the gap between both approaches. Finally, simulations of galaxies in the high-redshift Universe, such as Renaissance \citep{Hao+2016}, FirstLight \citep{Ceverino+2017}, SPHINX \citep{Rosdahl+2018}, CoDa II \citep{Ocvirk+2020}, Obelisk \citep{Trebitsch+2021}, the SERRA suite \citep{Pallottini+2022}, THESAN \citep{Kannan+2022}, and FIREbox$^{\it HR}$ \citep{Feldmann+2025}, can afford higher numerical resolution, given the shorter cosmic time they cover, and they may incorporate additional physics relevant to these early epochs.

In this section, we discuss the main methods and components involved in cosmological galaxy simulations. Such simulations begin with initial conditions constrained by measurements of temperature fluctuations in the cosmic microwave background, which establish the density and velocity field of matter in the early Universe (see Section~\ref{sect:InitialConditions}). From these initial conditions, simulations use differential equations to capture the complex interplay of gravitational, hydrodynamical, and astrophysical processes driving the formation and evolution of cosmic structures. The integration of these equations requires accurate numerical solvers to handle both gravity and hydrodynamics; these are covered in Section~\ref{sect:NumericalSolvers}. Sub-grid models for unresolved astrophysical processes are discussed in Section \ref{sect:BaryonicPhysics}. Common steps in analyzing a simulation and assessing its validity are presented in Sections \ref{sect:AnalyzingSimulations} and \ref{sect:AssessingSimulations}.

\subsection{Initial conditions}
\label{sect:InitialConditions}

Astrophysicists setting up cosmological simulations benefit from a well-understood statistical framework for the initial conditions (ICs). Observations of the first acoustic peak in the cosmic microwave background (CMB) power spectrum—obtained through balloon-borne experiments \citep{deBernardis+2000, Hanany+2000}, the Wilkinson Microwave Anisotropy Probe \citep{Spergel+2003, Hinshaw+2013}, and the Planck observatory \citep{Planck+2014, Planck+2020}—have provided tight constraints on the cosmological parameters that govern the expansion history of the Universe and the level of the initial density fluctuations that seeded the growth of structure over cosmic time.

The first step in setting up the ICs for a cosmological simulation involves generating a realization of a Gaussian random field with a power spectrum $P(k) = A_s k^{n_s} T^2(k)$. This random field encodes the fluctuations in the density of dark matter or cosmic gas and is scaled to the starting redshift $z_{\rm init}$ of the simulation using the linear growth factor. In this expression, $A_s$ defines the overall normalization of the amplitude of density fluctuations, while $A_s k^{n_s}$ describes the primordial power spectrum predicted by inflation theory. The transfer function $T(k)$ encapsulates the evolution of density fluctuations across the radiation- and matter-dominated epochs \citep[e.g.,][]{Bardeen+1986, 1998ApJ...496..605E}. It is obtained by solving the coupled equations of perturbations in the multi-species cosmic plasma using dedicated Boltzmann solvers such as LINGER \citep{Bertschinger+1995, Ma+1995}, CMBFast \citep{Seljak+1996}, CAMB \citep{Lewis+2000}, or CLASS \citep{Blas+2011}.

Next, the displacement field, which measures how far mass elements are shifted from their initially uniformly distribution at $t=0$, and the corresponding velocity field are calculated from the density fluctuations with the help of perturbation theory such as the Zel’dovich approximation \citep{Zeldovich1970}. More sophisticated methods, such as second-order Lagrangian perturbation theory (2LPT), are often used to achieve greater accuracy \citep{Scoccimarro+1998}. Subsequently, density fluctuations, displacements, and velocities are converted into the input formats required by the gravity and hydrodynamics solvers. Dark matter is often converted into equal mass point particles arranged into a nearly regular grid with particle positions and velocities inferred from the displacement and velocity fields. Lagrangian hydrodynamics codes follow an analogous procedure for the cosmic gas, while Eulerian fluid solvers discretize the density and velocity fields in form of a volumetric mesh.

Tools such as GRAFIC-2 \citep{Bertschinger+2001}, MUSIC \citep{Hahn+2011}, and MonofonIC \citep{Hahn+2021} are widely used for setting up ICs for cosmological simulations. These tools follow the steps outlined above for cosmological \emph{volume} simulation, i.e., for simulations that follow gas, stars, and dark matter in a comoving volume with periodic boundary conditions. In addition, MUSIC and GRAFIC-2 can generate ICs for cosmological \emph{zoom-in} simulations that resolve galaxies only in a sub-region of the original box, typically targeting a single galaxy or a small group of galaxies and their environment. To achieve this, these tools first create low-resolution ICs for the entire simulation volume and then progressively refine the resolution in the region of interest at $z_{\rm init}$. This targeted area, known as a Lagrangian patch, is often identified by running a low-resolution cosmological simulation, selecting a galaxy and its environment, identifying its constituent particles, and then calculating the convex hull that encloses these particles at the simulation's start.

\subsection{Numerical solvers}
\label{sect:NumericalSolvers}

Understanding galaxy formation and evolution requires sophisticated simulation techniques capable of modeling both large-scale cosmic structures and smaller-scale processes. Numerical solvers are the computational tools that handle the complex interactions and physical processes within the simulations. They are responsible for calculating the gravitational forces that shape the large-scale structure of the universe and the hydrodynamic behavior of baryonic matter. 

In cosmological simulations, the precision and reliability of these numerical methods directly influence the fidelity of the simulation results. This section provides an overview of the key numerical solvers used in modern cosmological simulations, emphasizing methods for solving gravitational and hydrodynamical equations. Understanding these tools is essential for grasping how they contribute to our ability to simulate and interpret the processes that drive galaxy formation and evolution.

\subsubsection{Solving for dark matter}

Dark matter is a fundamental component of the universe, comprising approximately 85\% of its total mass \citep{Planck+2020}. Within the framework of the $\Lambda{}$CDM paradigm, dark matter is understood to consist of non-relativistic, weakly interacting particles. Although it does not interact with electromagnetic forces and remains invisible to direct observation, dark matter exerts a gravitational influence that drives cosmic structure formation. Cosmological simulations of galaxies typically adopt this $\Lambda{}$CDM paradigm, treating dark matter as a collisionless fluid \citep[see][for a summary on alternative forms]{ReviewVogelsberger+2020}. In the continuum limit, its behavior is described by the collisionless Boltzmann equation, with the gravitational potential assumed to be Newtonian:
\begin{align}
\mathrm{\textbf{collisionless Boltzmann equation:}} & \quad \frac{\mathrm{d} f }{ \mathrm{d} t} = \frac{\partial f}{\partial t} + \boldsymbol{v} \frac{\partial f }{\partial \boldsymbol{r}} 
- \frac{\partial \phi}{\partial \boldsymbol{r}} \frac{\partial f}{\partial \boldsymbol{v}} = 0  \quad , \\
\mathrm{\textbf{Poisson's  equation:}} &\quad \nabla ^{2} \phi = 4 \pi G \int f \mathrm{d} \boldsymbol{v} \quad .
 \end{align}
Here $f(\boldsymbol{r}, \boldsymbol{v}, t)$ represents the phase-space distribution function, which describes the density of particles at position $\boldsymbol{r}$, velocity $\boldsymbol{v}$, and time $t$. The term $\phi(\boldsymbol{r}, t)$ is the gravitational potential, and $G$ is the gravitational constant. The first equation, the collisionless Boltzmann equation, governs the evolution of $f$ in the absence of collisions, while the second equation, Poisson’s equation, relates the gravitational potential $\phi$ to the density distribution of particles.

In most cosmological simulations, dark matter is typically represented by discrete particles that approximate its phase-space distribution. This particle-based approach provides a practical method for modeling the large-scale gravitational effects of dark matter, though other techniques exist \citep[see][for a summary]{ReviewVogelsberger+2020}.  In these simulations, gravity is typically assumed to act instantaneously across the cosmological box, without accounting for any delay due to the finite speed of light. To accurately compute gravitational interactions among these particles, simulations rely on gravity solvers, which enable the capture of the hierarchical structure of the Universe. Directly simulating these interactions would involve solving an $\mathcal{O} (N^2)$ problem, where each particle interacts with every other. To improve computational efficiency, cosmological simulations use a variety of techniques, with tree algorithms and particle-mesh (PM) methods being the most widely applied. These approaches allow for effective computation of gravitational forces over the vast scales present in cosmological models.

\smallskip

\noindent
\textbf{Tree Algorithms:} These algorithms are commonly used to efficiently approximate gravitational forces in simulations by hierarchically grouping distant particles \citep[see][]{Barnes+HutTree}. This grouping enables the summation of collective gravitational contributions, effectively solving Poisson's equation in its integral form. This approach reduces the computational complexity of force calculations to $\mathcal{O}(N \mathrm{log} N)$, making it feasible to simulate large cosmic volumes. Some implementations further optimize efficiency, approaching $\mathcal{O} (N)$ scaling, by using (fast) multipole expansions, which approximate gravitational forces based on the collective influence of particle groups rather than individual interactions \citep{MultipoleGrid1987, Multipole2000}. Codes such as \textsc{Gasoline} \citep{Gasoline2}, \textsc{ChaNGa} \citep{Changa}, and \textsc{PKDGRAV} \citep{Potter+2017} utilize such a tree algorithm approach. 

\smallskip

\noindent
\textbf{Particle-Mesh (PM) Techniques:} In this technique, Poisson's equation is solved on a fixed grid, leveraging Fourier transforms to efficiently compute the gravitational potential \citep{Brandt1977, Grid1981}. The forces are then derived by applying a finite difference scheme to the potential and subsequently interpolated to the particle positions. This approach achieves a computational scaling of $\mathcal{O}(N \log N)$, making it highly efficient for large-scale simulations. To improve resolution in regions requiring finer detail, PM methods often employ nested grids, where multiple levels of grid refinement capture small-scale structures more accurately. This approach enhances the overall resolution of the simulation, particularly in high-density regions, while maintaining computational efficiency over large scales. It is used by codes such as \textsc{Art}, \textsc{Ramses}, and \textsc{Enzo} \citep{Art, Ramses, Enzo}. 

\smallskip

\noindent
\textbf{Hybrid Techniques:} A range of methods combine short-range direct summation with Fourier techniques for long-range forces, providing a balance between computational efficiency and accuracy. One common example is the tree-particle-mesh (Tree-PM) method, which approximates short-range interactions using a hierarchical tree structure while applying Fourier transforms for efficient long-range force calculations. Some implementations further enhance this approach by employing multigrid methods on the coarsest grid to improve convergence speed and accuracy. These hybrid solvers are now standard in modern simulation codes such as \textsc{Gadget} \citep{Gadget4}, \textsc{Arepo} \citep{Springel2010}, \textsc{Gizmo} \citep{Hopkins2015}, and \textsc{Swift} \citep{Swift2024}, as they offer high computational accuracy as well as efficiency. 

\subsubsection{Solving for cosmic gas}

While dark matter drives the formation of large-scale structures, baryonic matter -- primarily in the form of cosmic gas -- undergoes more complex processes that ultimately lead to the formation of galaxies. 
Unlike cold dark matter, the gas component of baryonic matter is treated as a fluid and governed, in principle, by the Navier-Stokes equations. In practice, however, many astrophysical simulations approximate fluid behavior by solving the Euler equations, which describe the dynamics of an ideal gas under gravitational forces while neglecting viscous effects and heat conduction.
Accurately simulating this behavior requires advanced hydrodynamics solvers that can handle these equations effectively. The discretization schemes used in these simulations can be broadly classified into three categories: Lagrangian, Eulerian, and arbitrary Lagrange-Eulerian techniques. Each of these approaches will be discussed in more detail below.

\smallskip

\noindent
\textbf{Lagrangian Techniques (e.g., SPH):} Lagrangian methods, such as Smoothed Particle Hydrodynamics (SPH), represent fluid as discrete particles that move through space, following the equations of motion derived from the hydrodynamical equations \citep{Lucy1977, Gingold+1977}. This approach is especially effective for simulating flows with large density variations, like those present in galaxy formation. By naturally adapting to the fluid flow, Lagrangian techniques achieve high resolution in dense regions. Improved versions of SPH have been developed to address challenges in traditional SPH methods, such as difficulties with shock capturing, artificial viscosity, and inaccurate fluid mixing in low-density regions \citep[see][]{Agertz2007, Springel2010a, Price2012, Zhang+2022}. These advancements enable SPH-based codes to better capture shocks, turbulence, and complex fluid interactions \citep{Read+Hayfield2012, Hopkins+2013, Keller+2014}. Modern simulation codes implementing SPH include \textsc{Gadget-4} \citep{Gadget4}, \textsc{Gasoline} \citep{Gasoline2}, \textsc{ChaNGa} \citep{Changa}, and \textsc{Swift} \citep{Swift2024}.

\smallskip

\noindent
\textbf{Eulerian Techniques (e.g., AMR):} These methods divide the simulation volume into a grid, solving the hydrodynamical equations by discretizing them within each cell. Using finite difference, finite volume, or finite element techniques, they calculate fluxes of mass, momentum, and energy across cell boundaries \citep{Colella+Woodward1984}. While effective at capturing shocks and complex flows, traditional Eulerian methods face challenges such as advection errors, grid-locking, and imperfect angular momentum conservation, which can reduce the accuracy of fluid dynamics simulations. Modern codes address these issues by implementing improved numerical schemes and adaptive techniques. In galaxy formation simulations, the large dynamic range of processes requires balancing resolution across vastly different scales. To achieve this, modern codes for cosmological simulations employ Adaptive Mesh Refinement (AMR), dynamically refining the grid in regions of interest \citep{Berger+Oliger1984, Berger+Colella1989}. This approach allows for high-resolution simulations of fluid dynamics where needed, capturing shocks and complex flows. AMR-based codes include \textsc{Enzo} \citep{Enzo}, \textsc{Ramses} \citep{Ramses}, and \textsc{Art} \citep{Art}. 

\smallskip

\noindent
\textbf{Arbitrary Lagrangian-Eulerian Techniques (e.g., Moving Mesh):} Arbitrary Lagrangian-Eulerian techniques blend the adaptability of Lagrangian methods with the structured grid framework of Eulerian approaches. For astrophysical applications, such techniques have been realized using Voronoi tessellations of discrete mesh-generating points that move freely with the fluid. This approach adapts to the motion of the fluid by using Voronoi cells as dynamically adjusting computational volumes. Most importantly, due to the mathematical properties of the Voronoi tessellation, the mesh continuously deforms and updates as the points move, avoiding issues such as mesh tangling \citep{Springel2010, Duffell+2011, Vandenbroucke+2016}. Other mesh-free methods have extended these principles further by applying finite mass and finite volume techniques without relying on Voronoi-based frameworks. Modern simulation codes employing such techniques for cosmological simulations of galaxies include Arepo \citep{Springel2010}, which uses a Voronoi-based grid, and \textsc{Gizmo} \citep{Hopkins2015}, which implements meshless finite mass and finite volume methods.

\subsection{Modeling of baryonic physics}
\label{sect:BaryonicPhysics}

\begin{figure}[t]
\centering
\includegraphics[width=\textwidth]{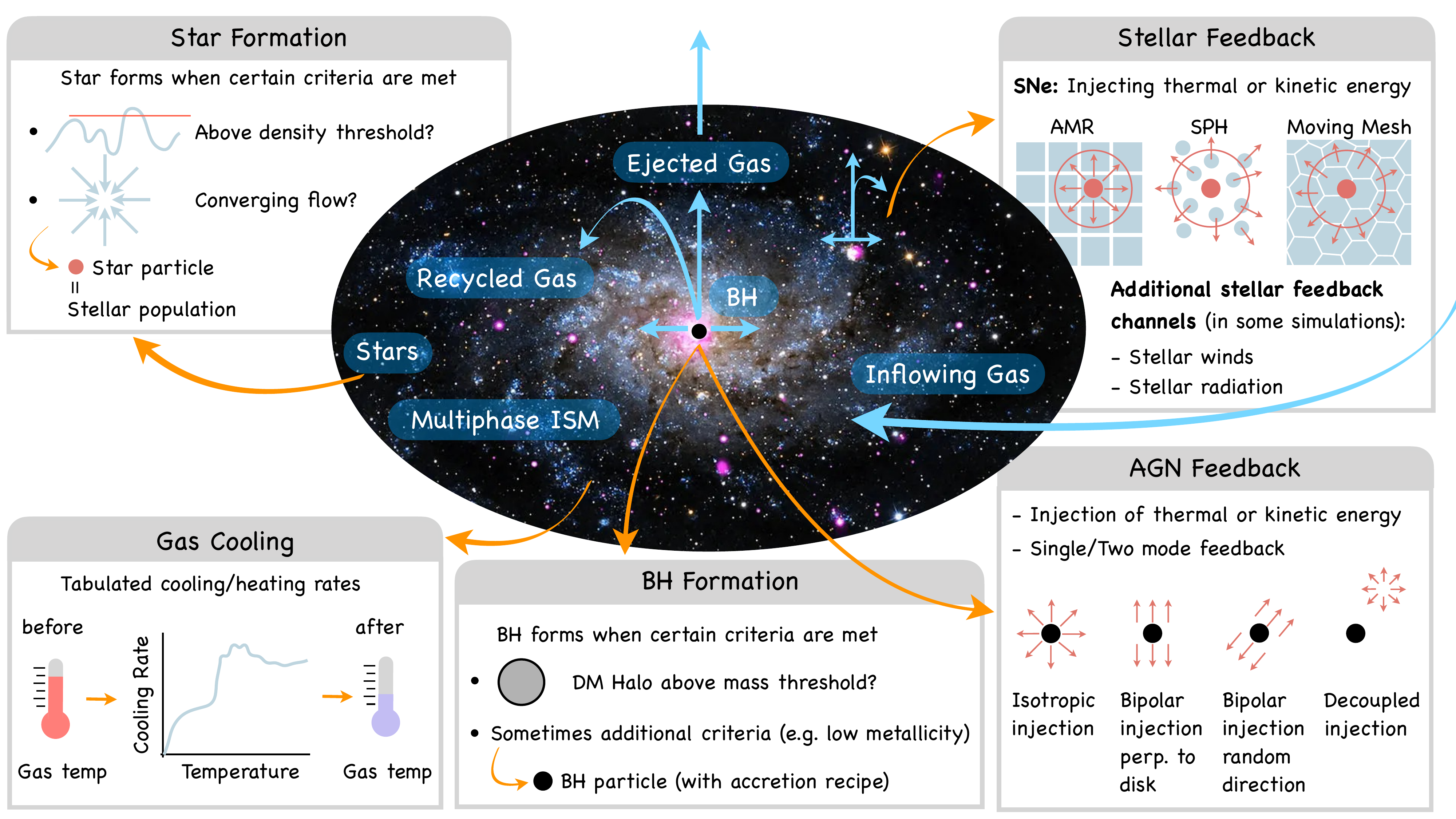}
\caption{Illustration of key baryonic processes in galaxies, using the Triangulum Galaxy (M33) as an example (Image credit: X-ray: NASA/CXC/SAO/P. Plucinsky et al.). The image highlights the cycle of gas in galaxies, showing processes such as gas accretion and recycling through feedback mechanisms. Gas flows into the galaxy from the CGM (not shown), fueling star formation and potentially feeding a black hole at the center (incidentally, M33 does not appear to host a visible black hole). Stellar feedback drives outflows, which enrich the surrounding medium and may later return as re-accreted gas, sustaining the cycle. Additionally, the black hole can influence the baryonic cycle by accreting gas and driving powerful outflows through active galactic nuclei (AGN) feedback, impacting both the interstellar and CGM. The boxes (starting from the top left and moving clockwise) highlight key aspects of how baryonic physics is implemented in simulations: Star Formation, Stellar Feedback, AGN Feedback, Black Hole Formation, and Gas Cooling. Each box provides an illustrative summary of possible implementations in simulations. For more details, see the descriptions in Section~\ref{sect:BaryonicPhysics}.}
\label{fig:BaryonicProcesses}
\end{figure}

Baryonic matter, which constitutes the visible matter in the Universe, plays a crucial role in the formation and evolution of galaxies. A key aspect of galaxy evolution is the baryonic cycle, which regulates gas flows into, through, and out of galaxies. Gas from the circumgalactic medium (CGM) accretes onto galaxies, replenishing the ISM and fueling star formation and feeding the black hole at the center of the galaxy. Stars and black holes then release energy and momentum through feedback, driving outflows that enrich the CGM. Over time, some of this material cools and accretes again, completing the cycle. Accurately modeling this cycle in cosmological simulations is essential, as it directly shapes observable galaxy properties such as luminosity, structure, and composition. 

This section provides an overview of the key baryonic physics components commonly implemented in current state-of-the-art simulations, including gas cooling, the ISM, star formation, stellar feedback, supermassive black holes, and active galactic nuclei (AGN) feedback, which together drive the flow of cosmic gas into and out of galaxies in a complex baryonic cycle. Fig.~\ref{fig:BaryonicProcesses} illustrates these key baryonic processes and their interconnected roles within galaxies. 
More recent developments incorporate additional processes that are potentially relevant, such as magnetohydrodynamics, cosmic rays, radiation hydrodynamics, thermal conduction, and viscosity. While each of these processes influences the thermal, kinetic, and chemical state of the gas, a detailed discussion of their treatment in simulations lies beyond the scope of this review. 

\smallskip

\noindent
\textbf{Gas Cooling and Heating:}  Cooling plays a vital role in galaxy formation, allowing gas to dissipate internal energy and condense into denser structures that eventually form stars. Cooling and heating processes involve several mechanisms, including collisional excitation, ionization, inverse Compton scattering, and free-free emission. In cosmological simulations, these cooling processes are typically coupled to the energy equation using cooling functions. Early simulations primarily relied on cooling rates assuming collisional ionization equilibrium, but more recent models account for the photoionization of metals by the metagalactic UV/X-ray background \citep[see][]{Haardt+Madau1996, Haardt+Madau2012, FaucherGiguere+2009, FaucherGiguere2020}. This background is generally assumed to be spatially uniform and time-dependent, influencing the cooling rates over cosmic time. To manage the complexities of cooling processes, simulations use detailed cooling rate tables that account for various conditions, including temperature, density, composition, and the effects of radiation fields \citep{Sutherland+Dopita1993, Wiersma+2009, Rosen+Bregman1995, Smith+2016}. These tables enable more accurate modeling of cooling processes, particularly in environments with significant variations in ionization states and metal abundances.

In simulations that resolve the cold phase of the ISM, cooling can occur at temperatures below $10^4$~K, where fine-structure and molecular cooling become significant. However, reaching the densities required for molecular cooling is challenging in large-scale cosmological simulations, leading many existing models to omit this process. Additionally, in dense regions, self-shielding becomes important, allowing gas to cool more efficiently by reducing the influence of external radiation. This is usually approximated with fitting functions calibrated using full radiation transport calculations \citep[e.g.,][]{Gnedin+2009, Gnedin+Hollon2012, Rahmati+2013}. A further layer of complexity in cooling processes comes from local radiation fields within the ISM, which interact with the gas differently than the uniform cosmological background \citep[see][for an implementation]{Ploeckinger+Schaye202}; for simulations that include both local sources and non-equilibrium cooling effects, see, for instance, \citet{Richings+2022}. While gas cooling is a direct physical process implemented in simulations without the need for sub-resolution models, accurately following all cooling processes requires sufficient numerical resolution to resolve the different gas phases. However, the computational demands of large-scale cosmological simulations often limit this resolution, making it challenging to model the detailed cooling dynamics, particularly within the ISM.

\smallskip

\noindent
\textbf{Interstellar Medium:} The space between stars within a galaxy is filled with a complex mixture of gas and dust which together make up the ISM. This medium includes distinct phases, such as hot ionized gas, warm neutral gas, and cold dense regions, each playing a critical role in star formation, the regulation of galactic dynamics, and the cycling of matter between stars and the galactic environment. The different phases are also interconnected through processes such as heating, cooling, turbulence, and magnetic fields. 

The thermal Jeans mass of the ISM defines the scale below which density perturbations in the ISM are stabilized against gravitational collapse by pressure forces. The necessity of resolving the thermal Jeans mass in simulations remains an area of debate. Achieving such resolution is particularly challenging in cosmological simulations due to the large dynamic range in density and temperature. To circumvent such difficulties, some simulations employ an effective polytropic equation of state or apply a pressure floor to the dense gas phase, reflecting the equilibrium of a two-phase ISM, to artificially raise the Jeans mass \citep{Springel+2003, Bournaud+2010, Agertz+2011, DallaVecchia+2012}. These approaches help maintain numerical stability and approximate the conditions within the ISM when full resolution is not feasible. Other research groups argue that the relevant scale is the much larger turbulent Jeans mass, emphasizing that resolving this mass is not strictly necessary. They argue that even without resolving this mass, simulations can still accurately capture the fragmentation hierarchy on resolved scales, though it is truncated at the resolution limit \citep[e.g.,][]{Hopkins+2018}. 

Resolving the relevant scales is only one facet of the complexities involved in modeling the ISM. Capturing the interplay between the various sources of feedback and the physical processes that shape its structure and energetics adds further layers of complexity. In high-resolution simulations, these processes can be directly modeled, while in lower-resolution simulations, they often need to be approximated through sub-grid physics. Gaining a deeper understanding of these processes represents a major theoretical challenge for future cosmological galaxy simulations.

\smallskip
 
\noindent
\textbf{Star Formation and Evolution:} Stars form in the dense regions of molecular clouds within the ISM, where gravitational forces become strong enough to overcome the internal pressure of the gas, causing it to collapse. In cosmological simulations, this process is modeled using sub-grid recipes, because star-forming regions and the relevant temporal scales cannot be resolved directly. 

In such sub-grid models, gas is converted into collisionless star particles with a certain efficiency, usually calibrated to reproduce observed global scaling relations, either at galaxy-wide scales or within giant molecular clouds. This efficiency can be implemented either deterministically, where a fixed fraction of gas forms stars, or stochastically, where a probability function governs star formation. Most modern simulations form stars stochastically, converting gas into star particles through probabilistic sampling of the local star formation rate. This approach reproduces the expected rate on average. While individual realizations introduce scatter, fluctuations average out over time and across regions, preserving the overall star formation history. A common approach follows the Schmidt law, where the local star formation rate density, $\dot{\rho}_{*}$, scales with the gas density $\rho_\mathrm{gas}$, over a time-scale, often expressed as $\dot{\rho}_{*} =  \epsilon \rho_{\mathrm{gas}} / t $, where $\epsilon$ is the local conversion efficiency, typically ranging from 0.01 to 1. The time-scale usually depends on local gas properties like the dynamical and/or gas cooling time \citep{Katz1992, Cen+Ostriker1992}. In addition, star formation generally only occurs when certain gas conditions as met, such as: \textit{gas density} or \textit{pressure threshold} (where the gas exceeds a set density or pressure threshold), \textit{gravitational instability} (where gas is locally self-gravitating and gravitational forces dominate over thermal or turbulent pressure), \textit{Jeans instability} (where the gas mass required for gravitational collapse is exceeded), or \textit{converging flows} (where the condition $\nabla \cdot \vec{v} < 0$ indicates gas is flowing inward). Sometimes, a combination of these criteria is also used to determine where star formation occurs. Some models also incorporate subgrid turbulence to account for unresolved turbulent motions within the gas, which can influence star formation efficiency. These models often use a combination of local gas properties, such as turbulent velocity dispersion or Mach number, to refine the criteria for star formation, capturing the role of turbulence in supporting or triggering gravitational collapse \citep[e.g.,][]{Schmidt+Federrath2011, Federrath+Klessen2012, Semenov+2016, Kretschmer+Teyssier2020, Semenov2024}. Note that in some simulations, the local star formation efficiency can differ from the galaxy-scale efficiency due to self-regulation by feedback processes \citep[see e.g.,][]{Hopkins+2018}.

The star particles created through the star formation mechanisms represent single-metallicity and single-age stellar populations, modeled with assumptions about their initial mass function and evolutionary processes. Star particles are created by taking the gas mass within a defined region and some codes impose a limit to ensure that a minimum amount of gas remains, preventing the formation of unphysically low-density regions. As these stars evolve, they return mass to the surrounding medium and enrich the gas with metals through processes such as Type II and Ia supernovae and asymptotic giant branch (AGB) stellar winds. Type II supernovae provide the bulk of this enrichment, while Type Ia supernovae contribute significantly to iron, and AGB stars are major sources of slow neutron capture elements. In cosmological simulations, these metals are ejected into the surrounding gas and then advected with the flow, following the hydrodynamic processes of the simulation. Metal yields are derived from stellar evolution models, but significant uncertainties -- particularly for low-metallicity and massive stars -- remain, often by about a factor of two or more. Binary evolution can also alter yields, adding further complexity to the predictions. These uncertainties can affect the predictions of metal abundances within simulations by at least a similar factor. The enriched gas can promote more efficient cooling and star formation in future generations of stars, which are more metal-rich, while also being transported into the CGM and intergalactic medium (IGM) through galaxy-scale outflows.

\smallskip

\noindent
\textbf{Stellar Feedback:} Massive stars are crucial to galaxy evolution through stellar feedback, which injects energy, momentum, and enriched material into the ISM, particularly in the later stages of stellar evolution. The primary mechanisms of stellar feedback include stellar winds, radiation, and supernova explosions. Stellar feedback regulates star formation by disrupting star-forming regions like molecular clouds, driving turbulence in the ISM, and launching large-scale outflows that expel gas from the galaxy into the CGM. Stellar feedback has been shown to be essential in cosmological simulations for reproducing key galaxy properties, especially in low-mass galaxies, such as the mass-metallicity relation and the galaxy stellar mass function.

In cosmological simulations, stellar feedback is modeled through sub-grid recipes, because the processes involved occur on spatial and temporal scales far smaller than can be resolved. At each time step, simulations sum the energy released by stellar particles representing a stellar population, injecting either thermal energy, kinetic energy, or both into the surrounding gas. 

Technically, a supernova event can be modeled as a point injection of a large amount of thermal energy, which should expand as a shock into its environment \citep[see e.g.,][]{Hu+2023}. In cosmological simulations, however, limited resolution in space and time causes energy injected as thermal energy to be radiated away too quickly, preventing the shock from properly expanding into the ISM -- also called the overcooling problem \citep[e.g.,][]{Katz1992, Smith+2018}. To address this limitation, numerous models have been developed that modify a pure thermal injection scheme including modified thermal injection schemes \citep[e.g.,][]{Murante+2010, DallaVecchia+2012, Tremmel+2017}, temporarily disabling radiative cooling after SNe events \citep[e.g.,][]{Stinson+2006, Teyssier+2013}, and modeling clusters of SNe as expanding superbubbles \citep[e.g.,][]{Keller+2014, ElBadry2019}. Other supernovae models compensate for resolution limitations by injecting the terminal momentum of supernovae ejecta, also accounting for environmental factors around the star \citep[see e.g.,][]{Murray+2005, Shetty+2008, Agertz+2013, Martizzi+2015, Kim+2015, Hopkins+2018, Pittard+2019, Marinacci+2019, Hopkins+2023, Bieri+2023}. However, this approach requires sufficiently high resolution to resolve the relevant physical scales, making it less applicable in lower-resolution setups. Another different strategy is to inject momentum into the gas around star-forming regions using hydrodynamically-decoupled wind particles. Decoupling these accelerated particles from the hydrodynamic scheme improves numerical convergence and enables simulations to model large-scale galactic outflows more effectively through non-local feedback \citep[e.g.,][]{Vogelsberger+2014, Pillepich+2018, Grand+2019, Dave+2019, Huang+2020, Smith+2024}.

We emphasize that cosmological simulations aim to model the results of stellar feedback rather than capture every detail of the underlying physics. In practice, models are assessed based on their ability to produce outcomes that match observations. These effective, albeit simplified, approaches capture the relative effects of stellar feedback without claiming to represent the full underlying physics.

Detailed studies of the ISM have shown that additional stellar feedback mechanisms, such as ionizing radiation and stellar winds impact the ISM structure by heating the gas in star forming regions, injecting momentum into the gas and driving local turbulence. This, in turn, has an effect on gas accretion onto the star forming regions and thus further regulates the local efficiency of star formation \citep{Geen+2015b, Rathjen+2021, Grudic+2022, Kim+2023}. Recent models have become more detailed, incorporating not only supernova feedback but also other stellar feedback mechanisms. They include energy and momentum injection from stellar winds, as well as the effects of photoionization and radiation pressure from young, massive stars \citep[e.g.,][]{Agertz+2013, Hopkins+2014, Smith+2019}. These additions have been shown to regulate and suppress star formation while, in some cases, enhancing galactic outflows. The sub-grid models used in those simulations rely, however, on a number of assumptions regarding, for instance, the coupling between the radiation and the gas such as the absorption of photons, mean free paths, optical depths, and shielding. It is, therefore, important to note that we still lack a clear understanding of how different stellar feedback processes couple to and impact the ISM locally and how the launched winds propagate to larger scales. This remains a key challenge in modern studies of galaxy formation and evolution.

\smallskip

\noindent
\textbf{Supermassive Black Hole Formation and Evolution:} Supermassive black holes (SMBHs) are thought to form in the early Universe and reside at the centers of galaxies with diverse properties. They appear in a broad range of systems, from the most massive galaxies to even lower-mass ones \citep{Gehren1984, Kormendy+1995, Filippenko+2003, Reines+2016}. They play a pivotal role in the evolution of massive galaxies through feedback processes. In cosmological simulations, SMBH formation is modeled through ``seeding'' mechanisms due to the inability to resolve their formation from first principles. Typically, SMBHs are seeded in cosmological simulations in dark matter haloes with masses of at least $10^{10} - 10^{11}$~M$_\odot$. A range of initial seed masses is used, from heavy seeds of $10^{4}$~M$_\odot$--$10^{6}$~M$_\odot$ \citep[e.g.,][]{Sijacki+2009, DiMatteo+2012, Hirschmann+2012, Sijacki+2015, Schaye+2015, Pillepich+2018, Dave+2019} to light seeds with masses around $\sim 100$~M$_\odot$ \citep{Wellons+2023}. Some models further refine seeding by placing black holes in dense regions, often with additional criteria such as low metallicity \citep[e.g.,][]{Bellovary+2010, Taylor+Kobayashi2014, Dubois+2014b, Volonteri+2016, Hopkins+2023_FIRE3}. 
Most cosmological simulations primarily focus on studying the growth of SMBHs and AGN feedback in massive galaxies rather than their initial formation. These simulations have successfully reproduced the AGN luminosity function, which is dominated by black holes with masses around $10^{8}$M$_\odot$, where the exact details of SMBH formation are less critical. Furthermore, beyond galaxies themselves, AGN feedback is thought to regulate the thermal state of the intragroup and intracluster medium, thereby reproducing the observed X-ray scaling relations in groups and clusters \citep[e.g.,][]{McCarthy+2010, LeBrun+2014}. However, it is important to note that most simulations are tuned to reproduce such BH-galaxy scaling relations, leading to broadly consistent predictions for these relations across different models. In contrast, the actual AGN luminosity functions and other AGN statistics can vary significantly between simulations, reflecting differences in the underlying assumptions and implementation of feedback and accretion processes. For a detailed comparison of these variations, see \citep{Habouzit+2021, Habouzit+2022}. In addition, recent studies have also started exploring the impact of black hole formation in smaller galaxies \citep[e.g.,][]{Habouzit+2017}. For a review on SMBH formation and seeding mechanisms, see \citet{Volonteri+2021}.

The evolution of SMBHs in simulations is governed by accretion of surrounding gas and mergers with other black holes. Accretion is often modeled using a Bondi-Hoyle-like accretion rate formula, capped by the Eddington limit. While many cosmological simulations adopt this Eddington limit, it is not a strict constraint on growth rates. High-resolution simulations of accretion flows have demonstrated that accretion rates can significantly exceed the Eddington limit, and some cosmological simulations do not enforce this cap. The accretion rate is often artificially enhanced in low-resolution regions of simulations to account for unresolved gas phases and to mitigate the suppression of Bondi accretion at low black hole masses, ensuring a realistic gas accretion rate. More recent advancements in modeling have moved beyond the classical Bondi-Hoyle prescription. Such approaches often replace the Bondi-Hoyle model with alternatives that account for distinct large-scale properties of the host galaxy, such as gravitational torques driving gas inflows \citep{Hopkins+Quataert2011, AnglesAlcazar+2017a, AnglesAlcazar+2017b, Dave+2019}, vorticity \citep{Krumholz+2004}, or the velocity dispersion of the bulge \citep{Hobbs+2011}. Alternatively, some models focus on capturing unresolved small-scale physics through accretion disk frameworks \citep{Power+2011, Dubois+2014b} or by incorporating processes such as circularization and viscous transfer of gas \citep{Debuhr+2010, Rosas-Guevara+2015}. See also \citet{Negri+Volonteri2017} for a detailed overview on accretion models. Furthermore, extreme accretion rates exceeding the Eddington limit -- such as those observed in high-redshift protogalaxies and during the growth of massive black hole seeds \citep{Inayoshi+2020, Lupi+2024} -- are increasingly modeled using super-Eddington scenarios \citep[e.g.,][]{Regan+2019, Massonnea+2023, Husvko+2024}. 

In addition to accretion, SMBHs grow through mergers. In cosmological simulations, when two black holes come within a close distance -- typically the numerical accretion radius -- their masses are combined. Some simulations assume an instantaneous merger while other simulations incorporate delayed merger processes, taking into account criteria like dynamical friction timescales. An additional challenge in these models is ensuring that the merged black hole remains at the center of the galaxy rather than wandering unrealistically at large radii due to finite resolution effects \citep[see e.g.,][]{Bartlett+2021, Bahe+2022}. To address this issue, many simulations simplify the problem by artificially pinning black holes to galaxy centers, effectively bypassing the difficulties introduced by finite resolution. Alternatively, cosmological simulations of galaxies often employ a ``dragging'' or repositioning force to keep the black hole at the center of the galaxy \citep[see e.g.,][]{Tremmel+2015, Tremmel+2017, Pfister+2019, Ma+2021, Dubois+2014, Damiano+2024}.

\smallskip

\noindent
\textbf{Active Galactic Nuclei (AGN) Feedback:} As gas accretes onto an SMBH, it releases immense energy, manifesting as an AGN. The resulting AGN feedback heats the surrounding gas, drives large-scale outflows, and regulates both star formation and SMBH growth. This is particularly important in massive galaxies, where stellar feedback alone struggles to overcome the deep gravitational wells, making AGN feedback a crucial component in their evolution \citep[e.g.,][]{Silk+Rees1998}. In cosmological simulations of galaxy formation, AGN feedback is implemented through sub-grid models, because the physical processes occur on scales smaller than the resolution of the simulations. These models emulate the impact of AGN energy injection, typically divided into `quasar-mode' and `radio-mode feedback' \citep[see e.g.,][]{Dubois+2016, Pillepich+2018a, Dave+2019}. However, some simulations instead implement a single-mode feedback model without explicitly distinguishing between these two modes \citep[e.g.,][]{Schaye+2015, Tremmel+2017}. For both modes, a broad range of model implementations exists in the literature, each varying in how energy is injected and interacts with the surrounding gas.

Quasar-mode feedback typically corresponds to high black hole accretion rates and is modeled by injecting thermal or kinetic energy -- or a combination of both -- into the surrounding gas, with the energy amount often scaled to the accretion rate. Some simulations inject the feedback energy locally \citep[e.g.,][]{Dubois+2012, Choi+2012, Costa+2014, Weinberger+2017}, while others implement hydro-decoupled winds that effectively transport the feedback energy into the CGM  \citep[e.g.,][]{Dave+2019}. Similarly, radio-mode models use various methods, often injecting energy thermally or kinetically through off-center bubbles \citep[e.g.,][]{Sijacki+2007, Schaye+2015, Grand+2017} or as large-scale bipolar winds. In some simulations, these bipolar winds are injected in random directions \citep[e.g.,][]{Weinberger+2017}, while others align the outflows with the angular momentum of the stellar disk \citep[e.g.,][]{Dubois+2012}. See also \citet[][particularly Figure~1]{Lagos+2024b} for a detailed overview of models used in three popular cosmological hydrodynamical simulations.

These simulations demonstrate that AGN feedback is essential for reproducing the observed properties of massive galaxies, including their stellar masses, star formation rates, and colors \citep[e.g.,][]{Martizzi+2012}. However, the success of these simulations relies heavily on effective AGN feedback models. An ongoing question is, therefore, how accurately these sub-grid models capture the true impact of AGN feedback on multi-phase gas dynamics and the resulting large-scale outflows. Recent simulations have incorporated more sophisticated AGN feedback models to better capture its role in galaxy formation across multiple scales. These models often derive the injection of kinetic or thermal energy from smaller-scale simulations and, in some cases, use observational data of large-scale winds to constrain the feedback properties, without explicitly tuning their feedback parameters to match more global observables \citep{Choi+2012, AnglesAlcazar+2017a, Dave+2019, Costa+2020}. Efforts such as those coupling multiple modes of AGN feedback, including mechanical, radiative, and cosmic rays, with a multi-phase ISM and multi-channel stellar feedback, reflect ongoing advancements in this area \citep{Wellons+2023, Mercedes-Feliz+2023, Byrne+2024}, which will remain a central research area for modern cosmological simulations of galaxies.

\subsection{Analyzing simulations}
\label{sect:AnalyzingSimulations}

Extracting scientific insights from a cosmological galaxy simulation typically requires extensive data-analysis of the simulation snapshots. During this post-processing step, information stored in the snapshots, such as particle positions, velocities, and masses, is converted into physical quantities, e.g., galaxy sizes, gas flow rates, or star formation histories. Alternatively, simulation data may be mapped directly into observables such as galaxy colors, spectra, or emission line luminosities via forward modeling. While the details of the post-processing analysis depend on the specific scientific goals, the following steps are often included.

\smallskip
\noindent
\textbf{Creation of the halo catalog:}
Various techniques have been developed to identify and characterize dark matter halos \citep[e.g.,][]{Knebe+2011, Onions+2012, Angulo+2022}. The friends-of-friends (FoF) algorithm identifies halos by grouping together particles that are within a specified linking length of each other \citep{Davis+1985}. Another option is to first locate the halo centers, e.g., by searching for peaks in the matter density, and then grow spheres around the halo centers such that the average density of (bound) particles within the sphere matches a given value \citep[e.g.,][]{Lacey+1994, Gill+2004}. Popular codes based on the first and second approach are SUBFIND \citep{Springel+2001} and the AMIGA halo finder \citep[AHF,][]{Knollmann+2009}, respectively. While originally developed for the analysis of collision-less (dark matter only) simulations, both approaches can be adopted to include also gas and stellar matter component \citep[e.g.,][]{Dolag+2009}. A further technique is to identify halos by linking particles in phase-space \citep{Diemand+2006}, which proves beneficial when dealing with halos that spatially overlap, e.g., during halo mergers or when considering sub-halos that reside within halos. The widely used halo finder ROCKSTAR \citep{Behroozi+2013a} implements this idea in a hierarchical fashion combining FoF in position space and in 6-dimensional phase-space.

\smallskip
\noindent
\textbf{Creation of the halo merger tree:}
The halo merger tree links identified halos across time, making it possible to study how halos move, grow in mass, and merge with each other. This identification is usually based on cross-matching the particles in halos at different time steps \citep[e.g.,][]{Kauffmann+1999, Springel+2005, Srisawat+2013}. Problems with the halo identification and tracking can be mitigated by evolving the positions of identified halos across snapshots based on their mutual gravitational attraction \citep{Behroozi+2013b}.

\smallskip
\noindent
\textbf{Global galaxy properties:}
Galaxies are commonly identified using the halo catalog and include all bound matter within a defined radius $R_{\rm gal}$ around a halo or sub-halo center. A typical choice for this galaxy radius is 10\% to 20\% of the virial radius of the halo \citep[e.g.,][]{Price+2017, Hopkins+2018}. The galaxy’s mass in stars, gas, metals, or dark matter is then calculated by summing the mass of the respective matter component within $R_{\rm gal}$. The star formation rate of the galaxy can be inferred from the recorded stellar ages of star particles by calculating the amount of stellar mass formed within $R_{\rm gal}$ in a given past time interval, often 5-100 Myr. The galaxy’s size is frequently characterized by the stellar half-mass radius, $R_{\rm half}$, defined as the radius enclosing half of the stellar mass within $R_{\rm gal}$. These intrinsic properties typically differ somewhat from the masses, sizes, and star formation rates, that would be derived if the galaxy were observed through a telescope.

\begin{figure}[t]
\centering
\includegraphics[width=.8\textwidth]{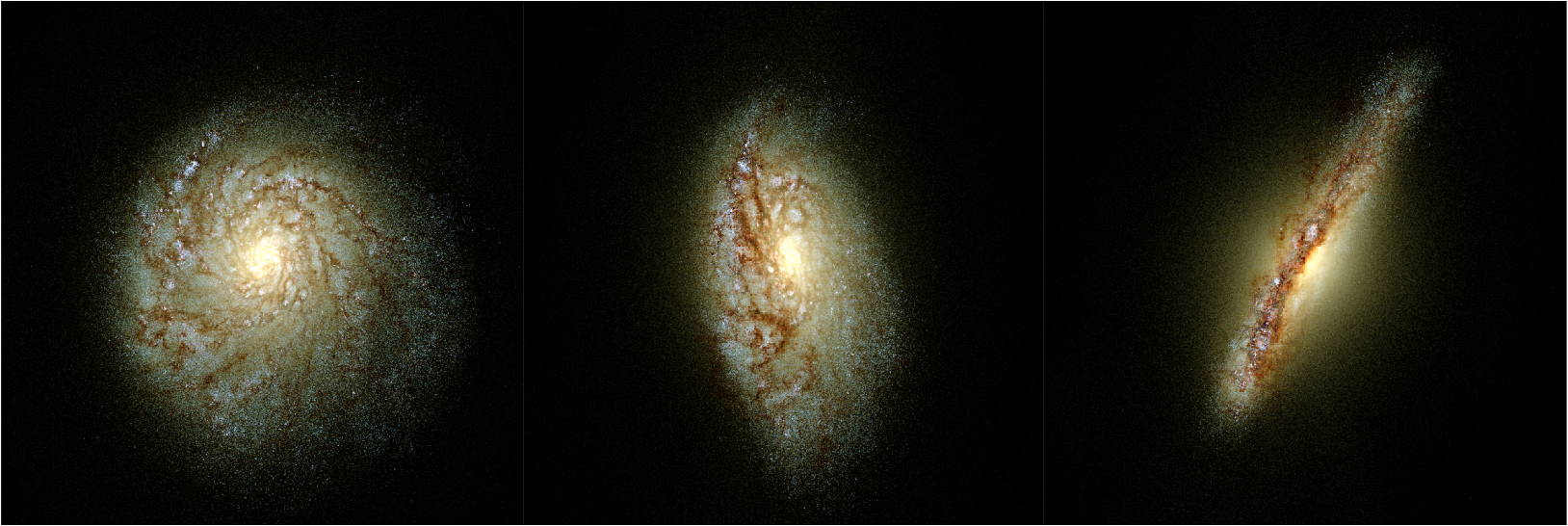}
\caption{Mock images of a Milky Way-like galaxy viewed from three different angles. The images are generated using dust radiative transfer, with red, green, and blue color channels representing luminosities at 0.3, 0.5, and 1 $\mu{}$m, respectively. Dust lanes appear as dark bands where starlight is significantly obscured, which is especially noticeable when the galaxy is viewed edge-on. Adapted from \cite{Narayanan+2021}.}
\label{fig:mock_images}
\end{figure}

\smallskip
\noindent
\textbf{Mock observations:}
Forward modeling aims to mitigate biases that arise when comparing intrinsic galaxy properties with observational data. To achieve this, simulation output is converted into mock observational data. Depending on the scientific goals, idealized mocks may be created by converting simulation data into observable quantities without accounting for all the physical and observational limitations, such as foreground noise, instrumental effects, and survey strategy, that would be included in a full forward modeling approach. Stellar population synthesis modeling is used to convert the masses, ages, and metallicities of star particles into mock spectra and luminosities \citep{Bruzual+2003}. A variety of codes and tabulated data are available for this purpose such a Starburst-99 \citep{Leitherer+1999}, the library of \cite{Bruzual+2003}, the Flexible Stellar Population Synthesis (FSPS, \citep{Conroy+2009}, and BPASS \citep{Eldridge+2009, Stanway+2018}. The light emitted by star particles can then be recorded by a virtual camera to create mock surface brightness maps or data cubes, using tools such as pnbody \citep{Revaz+2013} or py-ananke \citep{Thob+2024}. Three-dimensional dust radiative transfer codes and derived software packages, such as SUNRISE \citep{Jonsson2006}, SKIRT \citep{Baes+2011, Camps+2020}, HYPERION \citep{Robitaille+2011}, RADMC-3D \citep{Dullemond+2012}, and and POWDERDAY \citep{Narayanan+2021} further account for effects of dust extinction, reddening, scattering, and/or re-emission, see Fig.~\ref{fig:mock_images}. The intensity of nebular emission lines arising from ionized regions near massive stars can be calculated with codes such as MAPPINGS \citep{Dopita+1996, Dopita+2013} and Cloudy \citep{Ferland+1998, Ferland+2017}.

\subsection{Assessing simulation results}
\label{sect:AssessingSimulations}

Evaluating cosmological simulations requires comparing their results across various metrics to ensure that they align with theoretical expectations and observational data within their range of validity. This includes tests of internal consistency, comparisons with theoretical models, cross-simulation comparisons, and validation against observations. Each method offers insights into the accuracy of physical models and the numerical robustness. In this section, we will explore these evaluation methods in more detail.

\smallskip

\noindent
\textbf{Internal Comparison:} Cosmological simulations are often evaluated by examining their internal consistency. One common approach is \textit{convergence testing}, where simulations are run at different resolutions or with varying simulation domain sizes to check if key results remain stable under these changes. Adjusting resolution or domain size can impact physical processes like gas cooling, star formation, and feedback events. A larger simulation volume also captures a wider range of cosmic environments, adding diversity to the simulation. Generally, a simulation is considered converged when further increases in resolution or domain size no longer lead to significant changes in key outcomes. However, convergence testing becomes more challenging when hydrodynamical sub-grid models are calibrated at a specific resolution to reproduce observables such as the stellar mass function \citep[for a more in-depth discussion, see][]{Schaye+2015}. This issue remains an active subject of discussion and debate within the community.

In addition, \textit{Parameter variation tests} are crucial for evaluating the sensitivity of simulation results to chosen model parameters. In particular, uncertainties introduced by sub-grid models and stochastic processes - such as star formation and feedback mechanisms - can propagate through the simulation, leading to significant differences in final outcomes \citep{Keller+2019, Genel+2019, Davies+2021, Villaescusa-Navarro+2021, Davies+2022, Ni+2023}. This highlights the importance of thoroughly testing how variations in these parameters impact the stability and robustness of results.

\smallskip

\noindent
\textbf{Cross-Simulation Comparisons:} An important method to evaluate cosmological simulations is to compare the results across different simulation suites. Typically, each simulation uses distinct sub-grid models, feedback implementations, and numerical techniques which could all lead to varied predictions for galaxy properties. While many simulations now converge on key predictions (see above), discrepancies persist in other areas, particularly in the modeling of the CGM, as well as in reproducing the luminosities of lower-mass galaxies and low surface brightness features \citep[see][for a review]{Crain+VanDeVoort2023}. 

\begin{figure}[t]
\centering
\includegraphics[width=\textwidth]{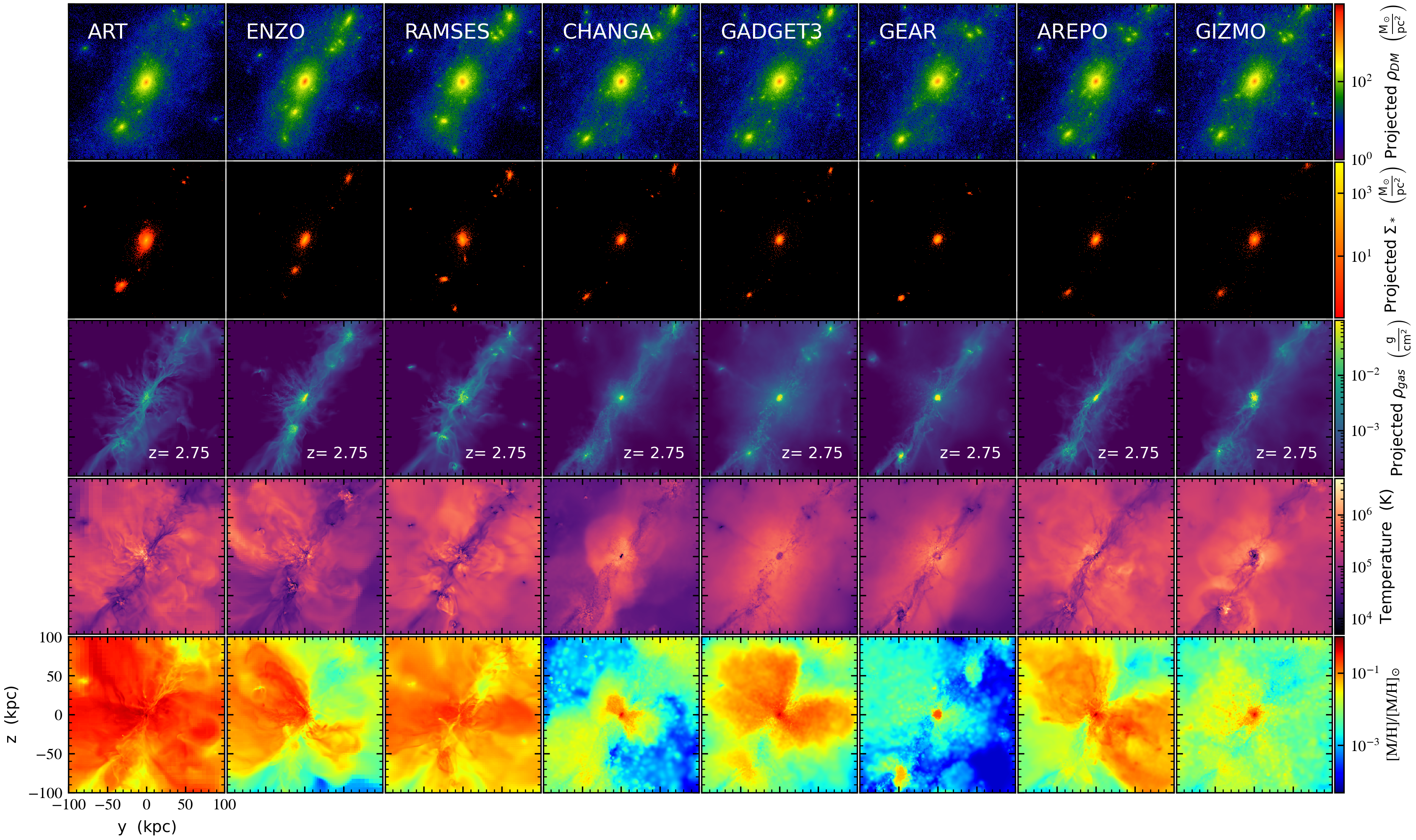}
\caption{Image from the AGORA project showing dark matter surface density, stellar surface density, gas density, density-weighted temperature, density-weighted metallicity projections at redshift $z=2.75$ (from top row to the bottom). Each column shows results from a different simulation code (as indicated), all based on identical initial conditions and physical parameters. Differences, especially in the thermal state and metallicity distribution of the CGM, highlight code-dependent variations in the simulation results (Image credit: S. Roca-Fàbrega).}
\label{fig:Agora}
\end{figure}

Addressing these discrepancies is one of the central goals of cross-simulation comparison projects such as Aquila \citep{Scannapieco+2012}, AGORA \citep{Kim+2014, AgoraCosmo2024}. These initiatives standardize comparisons by using the same initial conditions across different codes and implementing the same cooling tables and feedback mechanisms. This approach helps to understand how differences in numerical methods can lead to variations in the results of cosmological galaxy simulations, providing valuable insights into the robustness and range of validity of simulation predictions (see also Fig.~\ref{fig:Agora} for an illustration of such a comparison). 

\smallskip

\noindent
\textbf{Comparison with Theoretical Expectations:} Simulations are often compared with theoretical models that describe key galaxy formation processes. For example, gas regulator models and bathtub models provide frameworks for understanding how galaxies maintain equilibrium between star formation, inflows, and outflows \citep[e.g.,][]{Bouche+2010, Dave+2012, Lilly+2013, Feldmann2015, Bassini+2024}.
By comparing simulation results with such models, one can assess the extent to which simulations reproduce self-regulated star formation, where feedback mechanisms modulate star formation in response to the gas supply. Further comparisons may involve semi-analytic or semi-empirical models \citep{Kauffmann+1993, Somerville+1999, Bower+2006, Guo+2011, Moster+2013, Behroozi+2013, DeLucia+2024, Lagos+2024}, which provide a simplified framework for understanding galaxy evolution through approximations based on observed data or theoretical predictions. Comparison with such models helps to better understand whether the simulated evolution of galaxies reflect the underlying physical processes as predicted by theory. 

\smallskip

\noindent

\textbf{Observational Comparisons:} An essential approach to evaluating cosmological simulations is to compare their predictions with observations. Scaling relations provide critical benchmarks for determining whether simulations successfully capture key galaxy properties, such as star formation rates, gas content, mass growth, and SMBH activity. Examples include the Kennicutt-Schmidt relation, which links star formation rates to gas surface densities, and the stellar-to-halo mass relation that measures stellar mass growth within dark matter halos. The galaxy luminosity function, which describes the number density of galaxies of different luminosities, is also a key benchmark relation, alongside the mass-metallicity relation, which tracks the relationship between galaxy mass and chemical enrichment.
The AGN luminosity function and the observed stellar-to-black hole mass relation further test SMBH growth and feedback mechanisms. Additional tests focus on structural and dynamical properties, such as galaxy sizes and kinematics, including rotation curves and velocity dispersions, which reflect the influence of both dark matter and baryonic processes. For simulations that aim to capture realistic ISM properties it is important to ensure that they accurately model the transitions between gas phases, along with their mass and volume distributions. Comparing the simulated ISM properties with observations helps to validate whether the simulations can realistically capture the complex interplay between cooling, self-gravity, turbulence, shear, and heating from feedback processes during the evolution of the galaxy. Given the limited resolution of cosmological simulations, fully capturing these processes remains an essential challenge for modern galaxy formation models.

\section{Outlook: Toward next-generation cosmological simulations of galaxies}
\label{sec:Advancements}

Cosmological galaxy simulations have made tremendous strides in placing galaxy formation within the cosmological context, successfully reproducing many observed galaxy properties. However, significant obstacles remain, particularly in modeling the baryonic physics of galaxy formation. 

A core challenge is the limited resolution of cosmological simulations, which makes it difficult to capture crucial physical processes, such as the dynamics and phase structure of the ISM, the spatial and temporal clustering of star formation, and the coupling of feedback with gas in and around galaxies. One way to address this is by using ultra-high-resolution simulations of low mass galaxies, which allow detailed modeling of the ISM and feedback processes \citep{Agertz+2020, Gutcke+2021, Andersson+2024}. These simulations help uncover how gas flows, star formation, and feedback influence galaxy evolution at smaller mass scales; however, due to their extensive computational cost, they are currently limited to the simulation of low mass galaxies. An alternative approach involves extreme zoom-in simulations of specific regions, such as supermassive black holes at galaxy centers \citep{Beckmann+2019, Angles-Alcazar+2021, Hopkins+2024}, which achieve unparalleled resolution but are limited by the short time steps required in the highest resolved areas. A separate ansatz is to apply high resolution to extended regions of interest, such as the CGM and accretion flows \citep{Nelson+2016, VandeVoort+2019, Ramesh+2024}. This targeted approach enables simulations to better capture the relevant physics, such as cooling flows, turbulent mixing, and the phase structure of the gas, while managing computational costs by avoiding high-resolution requirements across the entire simulation volume. Focusing on early cosmic epochs can also mitigate the high computational costs, enabling such simulations to reach resolutions comparable to state-of-the-art zoom-in simulations run to $z=0$ (e.g., \citealt{Ceverino+2017, Feldmann+2025}). In addition, simulations targeting the Cosmic Dawn or the Epoch of Reionization may more accurately account for physical processes such as radiation (magneto-)hydrodynamics or the formation of Pop-III stars \citep{Rosdahl+2018, Trebitsch+2021, Kannan+2022, Bhagwat+2024}.

To account for processes occurring below the resolution limit, cosmological simulations usually resort to sub-grid models. Reliance on these approximate models can introduce uncertainties and biases stemming from model degeneracy and inadequacy. One promising approach forward involves leveraging high-resolution, smaller-scale simulations that capture the detailed physics of the multi-phase ISM and feedback processes and that are then used to refine the sub-grid models \citep[e.g.,][]{Motwani+2022, Bieri+2023, Smith+2024}. 
Furthermore, integrating additional physics—such as non-equilibrium cooling \citep[e.g.,][]{Lupi+2020, Katz2022, Richings+2022}, magnetic fields \citep[e.g.,][]{Pakmor+2017, Pillepich+2018a, Pakmor+2024}, cosmic rays \citep[e.g.,][]{Chan+2019, Buck+2020}, and thermal conduction \citep[e.g.,][]{Talbot+2024}—allows simulations to explore the impact of these processes on galaxy evolution \citep[e.g.,][]{Su+2017, Hopkins+2020}. Together, these approaches, while increasing computational demands, allow simulations to more accurately and consistently represent the formation and evolution of galaxies in a cosmological context.

Addressing these challenges requires computational innovations. Cosmological galaxy simulations are among the most demanding in astrophysics, necessitating supercomputers with massively parallel architectures. To meet these challenges, significant efforts are directed toward improving scalability, leveraging advanced hardware, and exploring innovative computational techniques. Hardware accelerators, such as Graphics processing unit (GPUs) and neural accelerators (e.g., Tensor Processing Units), deliver significant speed-ups for hydrodynamics and gravity solvers \citep[e.g.,][]{Schneider+2015, Potter+2017, Cavelan+2020, Wibkin+2022}, enabling higher-resolution simulations at reduced computational cost.

Alongside hardware advancements, artificial intelligence and machine learning (ML) are poised to play a transformative role in the next generation of simulations. Simulation suites such as CAMELS \citep{Villaescusa-Navarro+2021} and DREAMS \citep{Rose+2024} are specifically designed for ML applications, providing rich data sets for training emulators, performing parameter inference, and exploring the astrophysics underlying galaxy formation.
ML has shown promise in accelerating fundamental processes, such as predicting large-scale structure formation \citep[e.g.,][]{He+2019, Jamieson+2023} and enabling faster simulations of fluid dynamics while preserving accuracy in modeling gas flows and turbulence \citep[e.g.,][]{Kochkov+2021, Auddy+2024}. Another key application is the development of ML-based emulators, trained on high-fidelity simulations, to predict baryonic properties in less expensive $N$-body simulations. These include field-level emulators predicting baryonic density fields and maps \citep[e.g.,][]{Troester+2019, Wadekar+2021, Bernardini+2022, Hassan+2022, Horowitz+2022} and halo-based emulators forecasting baryonic properties such as galaxy masses within halos \citep[e.g.,][]{Agarwal+2018, Lovell+2022, Delgado+2022}.

Sub-grid modeling also benefits from ML innovations. ML can efficiently explore sub-grid parameter spaces, calibrating models against observations \citep[e.g.,][]{Oh+2022, Jo+2023, Kugel+2023}, or replace sub-grid models with emulators for specific physical processes such as supernova feedback \citep[e.g.,][]{Hirashima+2023, Hirashima+2024}. While many ML approaches rely on data-driven ``black box'' neural networks trained to interpolate results, emerging paradigms such as physics-informed neural networks incorporate physical laws directly into the learning process, ensuring physically consistent predictions \citep[e.g.,][]{Raissi+2019}. These advancements represent an exciting frontier for cosmological simulations, bridging the gap between computational efficiency and physical fidelity, and paving the way for more comprehensive and scalable models of galaxy formation.

\begin{ack}[Acknowledgments]

We thank Daniel Anglés-Alcázar, Claude-André Faucher-Giguère, Lucio Mayer, Rüdiger Pakmor, Santi Roca-Fàbrega, and Freeke van de Voort for providing valuable comments and suggestions that helped improve the manuscript. Both authors contributed equally to the design and writing of this chapter.

\end{ack}


\begin{thebibliography}{}
\makeatletter
\relax
\def\mn@urlcharsother{\let\do\@makeother \do\$\do\&\do\#\do\^\do\_\do\%\do\~}
\def\mn@doi{\begingroup\mn@urlcharsother \@ifnextchar [ {\mn@doi@}
  {\mn@doi@[]}}
\def\mn@doi@[#1]#2{\def\@tempa{#1}\ifx\@tempa\@empty \href
  {http://dx.doi.org/#2} {doi:#2}\else \href {http://dx.doi.org/#2} {#1}\fi
  \endgroup}
\def\mn@eprint#1#2{\mn@eprint@#1:#2::\@nil}
\def\mn@eprint@arXiv#1{\href {http://arxiv.org/abs/#1} {{\tt arXiv:#1}}}
\def\mn@eprint@dblp#1{\href {http://dblp.uni-trier.de/rec/bibtex/#1.xml}
  {dblp:#1}}
\def\mn@eprint@#1:#2:#3:#4\@nil{\def\@tempa {#1}\def\@tempb {#2}\def\@tempc
  {#3}\ifx \@tempc \@empty \let \@tempc \@tempb \let \@tempb \@tempa \fi \ifx
  \@tempb \@empty \def\@tempb {arXiv}\fi \@ifundefined
  {mn@eprint@\@tempb}{\@tempb:\@tempc}{\expandafter \expandafter \csname
  mn@eprint@\@tempb\endcsname \expandafter{\@tempc}}}

\bibitem[\protect\citeauthoryear{{Aarseth}, {Gott}  \& {Turner}}{{Aarseth}
  et~al.}{1979}]{Aarseth+1979}
{Aarseth} S.~J.,  {Gott} J.~R. I.,   {Turner} E.~L.,  1979, \mn@doi [\apj]
  {10.1086/156892}, \href
  {https://ui.adsabs.harvard.edu/abs/1979ApJ...228..664A} {228, 664}

\bibitem[\protect\citeauthoryear{{Abadi}, {Navarro}, {Steinmetz}  \&
  {Eke}}{{Abadi} et~al.}{2003}]{Abadi+2003}
{Abadi} M.~G.,  {Navarro} J.~F.,  {Steinmetz} M.,   {Eke} V.~R.,  2003, \mn@doi
  [\apj] {10.1086/375512}, \href
  {https://ui.adsabs.harvard.edu/abs/2003ApJ...591..499A} {591, 499}
  (\mn@eprint {arXiv} {astro-ph/0211331})

\bibitem[\protect\citeauthoryear{{Agarwal}, {Dav{\'e}}  \& {Bassett}}{{Agarwal}
  et~al.}{2018}]{Agarwal+2018}
{Agarwal} S.,  {Dav{\'e}} R.,   {Bassett} B.~A.,  2018, \mn@doi [\mnras]
  {10.1093/mnras/sty1169}, \href
  {https://ui.adsabs.harvard.edu/abs/2018MNRAS.478.3410A} {478, 3410}
  (\mn@eprint {arXiv} {1712.03255})

\bibitem[\protect\citeauthoryear{{Agertz} \& {Kravtsov}}{{Agertz} \&
  {Kravtsov}}{2016}]{Agertz+2016}
{Agertz} O.,  {Kravtsov} A.~V.,  2016, \mn@doi [\apj]
  {10.3847/0004-637X/824/2/79}, \href
  {https://ui.adsabs.harvard.edu/abs/2016ApJ...824...79A} {824, 79} (\mn@eprint
  {arXiv} {1509.00853})

\bibitem[\protect\citeauthoryear{{Agertz} et~al.,}{{Agertz}
  et~al.}{2007}]{Agertz2007}
{Agertz} O.,  et~al., 2007, \mn@doi [\mnras]
  {10.1111/j.1365-2966.2007.12183.x}, \href
  {https://ui.adsabs.harvard.edu/abs/2007MNRAS.380..963A} {380, 963}
  (\mn@eprint {arXiv} {astro-ph/0610051})

\bibitem[\protect\citeauthoryear{{Agertz}, {Teyssier}  \& {Moore}}{{Agertz}
  et~al.}{2011}]{Agertz+2011}
{Agertz} O.,  {Teyssier} R.,   {Moore} B.,  2011, \mn@doi [\mnras]
  {10.1111/j.1365-2966.2010.17530.x}, \href
  {https://ui.adsabs.harvard.edu/abs/2011MNRAS.410.1391A} {410, 1391}
  (\mn@eprint {arXiv} {1004.0005})

\bibitem[\protect\citeauthoryear{{Agertz}, {Kravtsov}, {Leitner}  \&
  {Gnedin}}{{Agertz} et~al.}{2013}]{Agertz+2013}
{Agertz} O.,  {Kravtsov} A.~V.,  {Leitner} S.~N.,   {Gnedin} N.~Y.,  2013,
  \mn@doi [\apj] {10.1088/0004-637X/770/1/25}, \href
  {https://ui.adsabs.harvard.edu/abs/2013ApJ...770...25A} {770, 25} (\mn@eprint
  {arXiv} {1210.4957})

\bibitem[\protect\citeauthoryear{{Agertz} et~al.,}{{Agertz}
  et~al.}{2020}]{Agertz+2020}
{Agertz} O.,  et~al., 2020, \mn@doi [\mnras] {10.1093/mnras/stz3053}, \href
  {https://ui.adsabs.harvard.edu/abs/2020MNRAS.491.1656A} {491, 1656}
  (\mn@eprint {arXiv} {1904.02723})

\bibitem[\protect\citeauthoryear{{Agertz} et~al.,}{{Agertz}
  et~al.}{2021}]{Agertz+2021}
{Agertz} O.,  et~al., 2021, \mn@doi [\mnras] {10.1093/mnras/stab322}, \href
  {https://ui.adsabs.harvard.edu/abs/2021MNRAS.503.5826A} {503, 5826}
  (\mn@eprint {arXiv} {2006.06008})

\bibitem[\protect\citeauthoryear{{Andersson}, {Rey}, {Pontzen}, {Cadiou},
  {Agertz}, {Read}  \& {Martin}}{{Andersson} et~al.}{2025}]{Andersson+2024}
{Andersson} E.~P.,  {Rey} M.~P.,  {Pontzen} A.,  {Cadiou} C.,  {Agertz} O.,
  {Read} J.~I.,   {Martin} N.~F.,  2025, \mn@doi [\apj]
  {10.3847/1538-4357/ad99d6}, \href
  {https://ui.adsabs.harvard.edu/abs/2025ApJ...978..129A} {978, 129}
  (\mn@eprint {arXiv} {2409.08073})

\bibitem[\protect\citeauthoryear{{Angl{\'e}s-Alc{\'a}zar}, {Dav{\'e}},
  {Faucher-Gigu{\`e}re}, {{\"O}zel}  \& {Hopkins}}{{Angl{\'e}s-Alc{\'a}zar}
  et~al.}{2017a}]{AnglesAlcazar+2017a}
{Angl{\'e}s-Alc{\'a}zar} D.,  {Dav{\'e}} R.,  {Faucher-Gigu{\`e}re} C.-A.,
  {{\"O}zel} F.,   {Hopkins} P.~F.,  2017a, \mn@doi [\mnras]
  {10.1093/mnras/stw2565}, \href
  {https://ui.adsabs.harvard.edu/abs/2017MNRAS.464.2840A} {464, 2840}
  (\mn@eprint {arXiv} {1603.08007})

\bibitem[\protect\citeauthoryear{{Angl{\'e}s-Alc{\'a}zar},
  {Faucher-Gigu{\`e}re}, {Kere{\v{s}}}, {Hopkins}, {Quataert}  \&
  {Murray}}{{Angl{\'e}s-Alc{\'a}zar} et~al.}{2017b}]{Angles-Alcazar+2017}
{Angl{\'e}s-Alc{\'a}zar} D.,  {Faucher-Gigu{\`e}re} C.-A.,  {Kere{\v{s}}} D.,
  {Hopkins} P.~F.,  {Quataert} E.,   {Murray} N.,  2017b, \mn@doi [\mnras]
  {10.1093/mnras/stx1517}, \href
  {https://ui.adsabs.harvard.edu/abs/2017MNRAS.470.4698A} {470, 4698}
  (\mn@eprint {arXiv} {1610.08523})

\bibitem[\protect\citeauthoryear{{Angl{\'e}s-Alc{\'a}zar},
  {Faucher-Gigu{\`e}re}, {Quataert}, {Hopkins}, {Feldmann}, {Torrey}, {Wetzel}
  \& {Kere{\v{s}}}}{{Angl{\'e}s-Alc{\'a}zar}
  et~al.}{2017c}]{AnglesAlcazar+2017b}
{Angl{\'e}s-Alc{\'a}zar} D.,  {Faucher-Gigu{\`e}re} C.-A.,  {Quataert} E.,
  {Hopkins} P.~F.,  {Feldmann} R.,  {Torrey} P.,  {Wetzel} A.,   {Kere{\v{s}}}
  D.,  2017c, \mn@doi [\mnras] {10.1093/mnrasl/slx161}, \href
  {https://ui.adsabs.harvard.edu/abs/2017MNRAS.472L.109A} {472, L109}
  (\mn@eprint {arXiv} {1707.03832})

\bibitem[\protect\citeauthoryear{{Angl{\'e}s-Alc{\'a}zar}
  et~al.,}{{Angl{\'e}s-Alc{\'a}zar} et~al.}{2021}]{Angles-Alcazar+2021}
{Angl{\'e}s-Alc{\'a}zar} D.,  et~al., 2021, \mn@doi [\apj]
  {10.3847/1538-4357/ac09e8}, \href
  {https://ui.adsabs.harvard.edu/abs/2021ApJ...917...53A} {917, 53} (\mn@eprint
  {arXiv} {2008.12303})

\bibitem[\protect\citeauthoryear{{Angulo} \& {Hahn}}{{Angulo} \&
  {Hahn}}{2022}]{Angulo+2022}
{Angulo} R.~E.,  {Hahn} O.,  2022, \mn@doi [Living Reviews in Computational
  Astrophysics] {10.1007/s41115-021-00013-z}, \href
  {https://ui.adsabs.harvard.edu/abs/2022LRCA....8....1A} {8, 1} (\mn@eprint
  {arXiv} {2112.05165})

\bibitem[\protect\citeauthoryear{{Auddy}, {Dey}, {Turner}  \& {Basu}}{{Auddy}
  et~al.}{2024}]{Auddy+2024}
{Auddy} S.,  {Dey} R.,  {Turner} N.~J.,   {Basu} S.,  2024, \mn@doi [Machine
  Learning: Science and Technology] {10.1088/2632-2153/ad3a32}, \href
  {https://ui.adsabs.harvard.edu/abs/2024MLS&T...5b5014A} {5, 025014}
  (\mn@eprint {arXiv} {2308.08010})

\bibitem[\protect\citeauthoryear{{Baes}, {Verstappen}, {De Looze}, {Fritz},
  {Saftly}, {Vidal P{\'e}rez}, {Stalevski}  \& {Valcke}}{{Baes}
  et~al.}{2011}]{Baes+2011}
{Baes} M.,  {Verstappen} J.,  {De Looze} I.,  {Fritz} J.,  {Saftly} W.,  {Vidal
  P{\'e}rez} E.,  {Stalevski} M.,   {Valcke} S.,  2011, \mn@doi [\apjs]
  {10.1088/0067-0049/196/2/22}, \href
  {https://ui.adsabs.harvard.edu/abs/2011ApJS..196...22B} {196, 22} (\mn@eprint
  {arXiv} {1108.5056})

\bibitem[\protect\citeauthoryear{{Bah{\'e}} et~al.,}{{Bah{\'e}}
  et~al.}{2022}]{Bahe+2022}
{Bah{\'e}} Y.~M.,  et~al., 2022, \mn@doi [\mnras] {10.1093/mnras/stac1339},
  \href {https://ui.adsabs.harvard.edu/abs/2022MNRAS.516..167B} {516, 167}
  (\mn@eprint {arXiv} {2109.01489})

\bibitem[\protect\citeauthoryear{{Bardeen}, {Bond}, {Kaiser}  \&
  {Szalay}}{{Bardeen} et~al.}{1986}]{Bardeen+1986}
{Bardeen} J.~M.,  {Bond} J.~R.,  {Kaiser} N.,   {Szalay} A.~S.,  1986, \mn@doi
  [\apj] {10.1086/164143}, \href
  {https://ui.adsabs.harvard.edu/abs/1986ApJ...304...15B} {304, 15}

\bibitem[\protect\citeauthoryear{{Barnes} \& {Hut}}{{Barnes} \&
  {Hut}}{1986}]{Barnes+HutTree}
{Barnes} J.,  {Hut} P.,  1986, \mn@doi [\nat] {10.1038/324446a0}, \href
  {https://ui.adsabs.harvard.edu/abs/1986Natur.324..446B} {324, 446}

\bibitem[\protect\citeauthoryear{{Bartlett}, {Desmond}, {Devriendt}, {Ferreira}
   \& {Slyz}}{{Bartlett} et~al.}{2021}]{Bartlett+2021}
{Bartlett} D.~J.,  {Desmond} H.,  {Devriendt} J.,  {Ferreira} P.~G.,   {Slyz}
  A.,  2021, \mn@doi [\mnras] {10.1093/mnras/staa3516}, \href
  {https://ui.adsabs.harvard.edu/abs/2021MNRAS.500.4639B} {500, 4639}
  (\mn@eprint {arXiv} {2007.01353})

\bibitem[\protect\citeauthoryear{{Bassini}, {Feldmann}, {Gensior}, {Hayward},
  {Faucher-Gigu{\`e}re}, {Cenci}, {Liang}  \& {Bernardini}}{{Bassini}
  et~al.}{2023}]{Bassini+2023}
{Bassini} L.,  {Feldmann} R.,  {Gensior} J.,  {Hayward} C.~C.,
  {Faucher-Gigu{\`e}re} C.-A.,  {Cenci} E.,  {Liang} L.,   {Bernardini} M.,
  2023, \mn@doi [\mnras] {10.1093/mnras/stad2617}, \href
  {https://ui.adsabs.harvard.edu/abs/2023MNRAS.525.5388B} {525, 5388}
  (\mn@eprint {arXiv} {2211.08423})

\bibitem[\protect\citeauthoryear{{Bassini}, {Feldmann}, {Gensior},
  {Faucher-Gigu{\`e}re}, {Cenci}, {Moreno}, {Bernardini}  \& {Liang}}{{Bassini}
  et~al.}{2024}]{Bassini+2024}
{Bassini} L.,  {Feldmann} R.,  {Gensior} J.,  {Faucher-Gigu{\`e}re} C.-A.,
  {Cenci} E.,  {Moreno} J.,  {Bernardini} M.,   {Liang} L.,  2024, \mn@doi
  [\mnras] {10.1093/mnrasl/slae036}, \href
  {https://ui.adsabs.harvard.edu/abs/2024MNRAS.532L..14B} {532, L14}
  (\mn@eprint {arXiv} {2401.13824})

\bibitem[\protect\citeauthoryear{{Beckmann}, {Devriendt}  \& {Slyz}}{{Beckmann}
  et~al.}{2019}]{Beckmann+2019}
{Beckmann} R.~S.,  {Devriendt} J.,   {Slyz} A.,  2019, \mn@doi [\mnras]
  {10.1093/mnras/sty2890}, \href
  {https://ui.adsabs.harvard.edu/abs/2019MNRAS.483.3488B} {483, 3488}
  (\mn@eprint {arXiv} {1810.01649})

\bibitem[\protect\citeauthoryear{{Behroozi}, {Wechsler}  \& {Wu}}{{Behroozi}
  et~al.}{2013a}]{Behroozi+2013a}
{Behroozi} P.~S.,  {Wechsler} R.~H.,   {Wu} H.-Y.,  2013a, \mn@doi [\apj]
  {10.1088/0004-637X/762/2/109}, \href
  {https://ui.adsabs.harvard.edu/abs/2013ApJ...762..109B} {762, 109}
  (\mn@eprint {arXiv} {1110.4372})

\bibitem[\protect\citeauthoryear{{Behroozi}, {Wechsler}, {Wu}, {Busha},
  {Klypin}  \& {Primack}}{{Behroozi} et~al.}{2013b}]{Behroozi+2013b}
{Behroozi} P.~S.,  {Wechsler} R.~H.,  {Wu} H.-Y.,  {Busha} M.~T.,  {Klypin}
  A.~A.,   {Primack} J.~R.,  2013b, \mn@doi [\apj]
  {10.1088/0004-637X/763/1/18}, \href
  {https://ui.adsabs.harvard.edu/abs/2013ApJ...763...18B} {763, 18} (\mn@eprint
  {arXiv} {1110.4370})

\bibitem[\protect\citeauthoryear{{Behroozi}, {Wechsler}  \&
  {Conroy}}{{Behroozi} et~al.}{2013c}]{Behroozi+2013}
{Behroozi} P.~S.,  {Wechsler} R.~H.,   {Conroy} C.,  2013c, \mn@doi [\apj]
  {10.1088/0004-637X/770/1/57}, \href
  {https://ui.adsabs.harvard.edu/abs/2013ApJ...770...57B} {770, 57} (\mn@eprint
  {arXiv} {1207.6105})

\bibitem[\protect\citeauthoryear{{Bellovary}, {Governato}, {Quinn}, {Wadsley},
  {Shen}  \& {Volonteri}}{{Bellovary} et~al.}{2010}]{Bellovary+2010}
{Bellovary} J.~M.,  {Governato} F.,  {Quinn} T.~R.,  {Wadsley} J.,  {Shen} S.,
   {Volonteri} M.,  2010, \mn@doi [\apjl] {10.1088/2041-8205/721/2/L148}, \href
  {https://ui.adsabs.harvard.edu/abs/2010ApJ...721L.148B} {721, L148}
  (\mn@eprint {arXiv} {1008.5147})

\bibitem[\protect\citeauthoryear{{Berger} \& {Colella}}{{Berger} \&
  {Colella}}{1989}]{Berger+Colella1989}
{Berger} M.~J.,  {Colella} P.,  1989, \mn@doi [Journal of Computational
  Physics] {10.1016/0021-9991(89)90035-1}, \href
  {https://ui.adsabs.harvard.edu/abs/1989JCoPh..82...64B} {82, 64}

\bibitem[\protect\citeauthoryear{{Berger} \& {Oliger}}{{Berger} \&
  {Oliger}}{1984}]{Berger+Oliger1984}
{Berger} M.~J.,  {Oliger} J.,  1984, \mn@doi [Journal of Computational Physics]
  {10.1016/0021-9991(84)90073-1}, \href
  {https://ui.adsabs.harvard.edu/abs/1984JCoPh..53..484B} {53, 484}

\bibitem[\protect\citeauthoryear{{Bernardini}, {Feldmann},
  {Angl{\'e}s-Alc{\'a}zar}, {Boylan-Kolchin}, {Bullock}, {Mayer}  \&
  {Stadel}}{{Bernardini} et~al.}{2022}]{Bernardini+2022}
{Bernardini} M.,  {Feldmann} R.,  {Angl{\'e}s-Alc{\'a}zar} D.,
  {Boylan-Kolchin} M.,  {Bullock} J.,  {Mayer} L.,   {Stadel} J.,  2022,
  \mn@doi [\mnras] {10.1093/mnras/stab3088}, \href
  {https://ui.adsabs.harvard.edu/abs/2022MNRAS.509.1323B} {509, 1323}
  (\mn@eprint {arXiv} {2110.11970})

\bibitem[\protect\citeauthoryear{{Bertschinger}}{{Bertschinger}}{1995}]{Bertschinger+1995}
{Bertschinger} E.,  1995, \mn@doi [arXiv e-prints]
  {10.48550/arXiv.astro-ph/9506070}, \href
  {https://ui.adsabs.harvard.edu/abs/1995astro.ph..6070B} {pp
  astro--ph/9506070} (\mn@eprint {arXiv} {astro-ph/9506070})

\bibitem[\protect\citeauthoryear{{Bertschinger}}{{Bertschinger}}{2001}]{Bertschinger+2001}
{Bertschinger} E.,  2001, \mn@doi [\apjs] {10.1086/322526}, \href
  {https://ui.adsabs.harvard.edu/abs/2001ApJS..137....1B} {137, 1} (\mn@eprint
  {arXiv} {astro-ph/0103301})

\bibitem[\protect\citeauthoryear{{Bhagwat}, {Costa}, {Ciardi}, {Pakmor}  \&
  {Garaldi}}{{Bhagwat} et~al.}{2024}]{Bhagwat+2024}
{Bhagwat} A.,  {Costa} T.,  {Ciardi} B.,  {Pakmor} R.,   {Garaldi} E.,  2024,
  \mn@doi [\mnras] {10.1093/mnras/stae1125}, \href
  {https://ui.adsabs.harvard.edu/abs/2024MNRAS.531.3406B} {531, 3406}
  (\mn@eprint {arXiv} {2310.16895})

\bibitem[\protect\citeauthoryear{{Bieri}, {Naab}, {Geen}, {Coles}, {Pakmor}  \&
  {Walch}}{{Bieri} et~al.}{2023}]{Bieri+2023}
{Bieri} R.,  {Naab} T.,  {Geen} S.,  {Coles} J.~P.,  {Pakmor} R.,   {Walch} S.,
   2023, \mn@doi [\mnras] {10.1093/mnras/stad1710}, \href
  {https://ui.adsabs.harvard.edu/abs/2023MNRAS.523.6336B} {523, 6336}
  (\mn@eprint {arXiv} {2209.06842})

\bibitem[\protect\citeauthoryear{{Blas}, {Lesgourgues}  \& {Tram}}{{Blas}
  et~al.}{2011}]{Blas+2011}
{Blas} D.,  {Lesgourgues} J.,   {Tram} T.,  2011, \mn@doi [\jcap]
  {10.1088/1475-7516/2011/07/034}, \href
  {https://ui.adsabs.harvard.edu/abs/2011JCAP...07..034B} {2011, 034}
  (\mn@eprint {arXiv} {1104.2933})

\bibitem[\protect\citeauthoryear{{Booth} \& {Schaye}}{{Booth} \&
  {Schaye}}{2009}]{Booth+2009}
{Booth} C.~M.,  {Schaye} J.,  2009, \mn@doi [\mnras]
  {10.1111/j.1365-2966.2009.15043.x}, \href
  {https://ui.adsabs.harvard.edu/abs/2009MNRAS.398...53B} {398, 53} (\mn@eprint
  {arXiv} {0904.2572})

\bibitem[\protect\citeauthoryear{{Bouch{\'e}} et~al.,}{{Bouch{\'e}}
  et~al.}{2010}]{Bouche+2010}
{Bouch{\'e}} N.,  et~al., 2010, \mn@doi [\apj] {10.1088/0004-637X/718/2/1001},
  \href {https://ui.adsabs.harvard.edu/abs/2010ApJ...718.1001B} {718, 1001}
  (\mn@eprint {arXiv} {0912.1858})

\bibitem[\protect\citeauthoryear{{Bournaud}, {Elmegreen}, {Teyssier}, {Block}
  \& {Puerari}}{{Bournaud} et~al.}{2010}]{Bournaud+2010}
{Bournaud} F.,  {Elmegreen} B.~G.,  {Teyssier} R.,  {Block} D.~L.,   {Puerari}
  I.,  2010, \mn@doi [\mnras] {10.1111/j.1365-2966.2010.17370.x}, \href
  {https://ui.adsabs.harvard.edu/abs/2010MNRAS.409.1088B} {409, 1088}
  (\mn@eprint {arXiv} {1007.2566})

\bibitem[\protect\citeauthoryear{{Bower}, {Benson}, {Malbon}, {Helly}, {Frenk},
  {Baugh}, {Cole}  \& {Lacey}}{{Bower} et~al.}{2006}]{Bower+2006}
{Bower} R.~G.,  {Benson} A.~J.,  {Malbon} R.,  {Helly} J.~C.,  {Frenk} C.~S.,
  {Baugh} C.~M.,  {Cole} S.,   {Lacey} C.~G.,  2006, \mn@doi [\mnras]
  {10.1111/j.1365-2966.2006.10519.x}, \href
  {https://ui.adsabs.harvard.edu/abs/2006MNRAS.370..645B} {370, 645}
  (\mn@eprint {arXiv} {astro-ph/0511338})

\bibitem[\protect\citeauthoryear{{Brandt}}{{Brandt}}{1977}]{Brandt1977}
{Brandt} A.,  1977, Mathematics of Computation, \href
  {https://doi.org/10.1090/S0025-5718-1977-0431719-X} {31, 333}

\bibitem[\protect\citeauthoryear{{Brooks}, {Kuhlen}, {Zolotov}  \&
  {Hooper}}{{Brooks} et~al.}{2013}]{Brooks+2013}
{Brooks} A.~M.,  {Kuhlen} M.,  {Zolotov} A.,   {Hooper} D.,  2013, \mn@doi
  [\apj] {10.1088/0004-637X/765/1/22}, \href
  {https://ui.adsabs.harvard.edu/abs/2013ApJ...765...22B} {765, 22} (\mn@eprint
  {arXiv} {1209.5394})

\bibitem[\protect\citeauthoryear{{Bruzual} \& {Charlot}}{{Bruzual} \&
  {Charlot}}{2003}]{Bruzual+2003}
{Bruzual} G.,  {Charlot} S.,  2003, \mn@doi [\mnras]
  {10.1046/j.1365-8711.2003.06897.x}, \href
  {https://ui.adsabs.harvard.edu/abs/2003MNRAS.344.1000B} {344, 1000}
  (\mn@eprint {arXiv} {astro-ph/0309134})

\bibitem[\protect\citeauthoryear{{Bryan} et~al.,}{{Bryan} et~al.}{2014}]{Enzo}
{Bryan} G.~L.,  et~al., 2014, \mn@doi [\apjs] {10.1088/0067-0049/211/2/19},
  \href {https://ui.adsabs.harvard.edu/abs/2014ApJS..211...19B} {211, 19}
  (\mn@eprint {arXiv} {1307.2265})

\bibitem[\protect\citeauthoryear{{Buck}, {Pfrommer}, {Pakmor}, {Grand}  \&
  {Springel}}{{Buck} et~al.}{2020}]{Buck+2020}
{Buck} T.,  {Pfrommer} C.,  {Pakmor} R.,  {Grand} R. J.~J.,   {Springel} V.,
  2020, \mn@doi [\mnras] {10.1093/mnras/staa1960}, \href
  {https://ui.adsabs.harvard.edu/abs/2020MNRAS.497.1712B} {497, 1712}
  (\mn@eprint {arXiv} {1911.00019})

\bibitem[\protect\citeauthoryear{{Bullock} \& {Boylan-Kolchin}}{{Bullock} \&
  {Boylan-Kolchin}}{2017}]{Bullock+2017}
{Bullock} J.~S.,  {Boylan-Kolchin} M.,  2017, \mn@doi [\araa]
  {10.1146/annurev-astro-091916-055313}, \href
  {https://ui.adsabs.harvard.edu/abs/2017ARA&A..55..343B} {55, 343} (\mn@eprint
  {arXiv} {1707.04256})

\bibitem[\protect\citeauthoryear{{Byrne} et~al.,}{{Byrne}
  et~al.}{2024}]{Byrne+2024}
{Byrne} L.,  et~al., 2024, \mn@doi [\apj] {10.3847/1538-4357/ad67ca}, \href
  {https://ui.adsabs.harvard.edu/abs/2024ApJ...973..149B} {973, 149}
  (\mn@eprint {arXiv} {2310.16086})

\bibitem[\protect\citeauthoryear{{Camps} \& {Baes}}{{Camps} \&
  {Baes}}{2020}]{Camps+2020}
{Camps} P.,  {Baes} M.,  2020, \mn@doi [Astronomy and Computing]
  {10.1016/j.ascom.2020.100381}, \href
  {https://ui.adsabs.harvard.edu/abs/2020A&C....3100381C} {31, 100381}
  (\mn@eprint {arXiv} {2003.00721})

\bibitem[\protect\citeauthoryear{{Cavelan}, {Cabez{\'o}n}, {Grabarczyk}  \&
  {Ciorba}}{{Cavelan} et~al.}{2020}]{Cavelan+2020}
{Cavelan} A.,  {Cabez{\'o}n} R.~M.,  {Grabarczyk} M.,   {Ciorba} F.~M.,  2020,
  in PASC '20: Proceedings of the Platform for Advanced Scientific Computing
  ConferenceJune 2020. p.~11 (\mn@eprint {arXiv} {2005.02656}),
  \mn@doi{10.1145/3394277.3401855}

\bibitem[\protect\citeauthoryear{{Cen} \& {Ostriker}}{{Cen} \&
  {Ostriker}}{1992}]{Cen+Ostriker1992}
{Cen} R.,  {Ostriker} J.,  1992, \mn@doi [\apj] {10.1086/171482}, \href
  {https://ui.adsabs.harvard.edu/abs/1992ApJ...393...22C} {393, 22}

\bibitem[\protect\citeauthoryear{{Ceverino}, {Glover}  \& {Klessen}}{{Ceverino}
  et~al.}{2017}]{Ceverino+2017}
{Ceverino} D.,  {Glover} S. C.~O.,   {Klessen} R.~S.,  2017, \mn@doi [\mnras]
  {10.1093/mnras/stx1386}, \href
  {https://ui.adsabs.harvard.edu/abs/2017MNRAS.470.2791C} {470, 2791}
  (\mn@eprint {arXiv} {1703.02913})

\bibitem[\protect\citeauthoryear{{Ceverino}, {Mandelker}, {Snyder}, {Lapiner},
  {Dekel}, {Primack}, {Ginzburg}  \& {Larkin}}{{Ceverino}
  et~al.}{2023}]{Ceverino+2023}
{Ceverino} D.,  {Mandelker} N.,  {Snyder} G.~F.,  {Lapiner} S.,  {Dekel} A.,
  {Primack} J.,  {Ginzburg} O.,   {Larkin} S.,  2023, \mn@doi [\mnras]
  {10.1093/mnras/stad1255}, \href
  {https://ui.adsabs.harvard.edu/abs/2023MNRAS.522.3912C} {522, 3912}
  (\mn@eprint {arXiv} {2210.15372})

\bibitem[\protect\citeauthoryear{{Chan}, {Kere{\v{s}}}, {O{\~n}orbe},
  {Hopkins}, {Muratov}, {Faucher-Gigu{\`e}re}  \& {Quataert}}{{Chan}
  et~al.}{2015}]{Chan+2015}
{Chan} T.~K.,  {Kere{\v{s}}} D.,  {O{\~n}orbe} J.,  {Hopkins} P.~F.,  {Muratov}
  A.~L.,  {Faucher-Gigu{\`e}re} C.~A.,   {Quataert} E.,  2015, \mn@doi [\mnras]
  {10.1093/mnras/stv2165}, \href
  {https://ui.adsabs.harvard.edu/abs/2015MNRAS.454.2981C} {454, 2981}
  (\mn@eprint {arXiv} {1507.02282})

\bibitem[\protect\citeauthoryear{{Chan}, {Kere{\v{s}}}, {Hopkins}, {Quataert},
  {Su}, {Hayward}  \& {Faucher-Gigu{\`e}re}}{{Chan} et~al.}{2019}]{Chan+2019}
{Chan} T.~K.,  {Kere{\v{s}}} D.,  {Hopkins} P.~F.,  {Quataert} E.,  {Su} K.~Y.,
   {Hayward} C.~C.,   {Faucher-Gigu{\`e}re} C.~A.,  2019, \mn@doi [\mnras]
  {10.1093/mnras/stz1895}, \href
  {https://ui.adsabs.harvard.edu/abs/2019MNRAS.488.3716C} {488, 3716}
  (\mn@eprint {arXiv} {1812.10496})

\bibitem[\protect\citeauthoryear{{Choi}, {Ostriker}, {Naab}  \&
  {Johansson}}{{Choi} et~al.}{2012}]{Choi+2012}
{Choi} E.,  {Ostriker} J.~P.,  {Naab} T.,   {Johansson} P.~H.,  2012, \mn@doi
  [\apj] {10.1088/0004-637X/754/2/125}, \href
  {https://ui.adsabs.harvard.edu/abs/2012ApJ...754..125C} {754, 125}
  (\mn@eprint {arXiv} {1205.2082})

\bibitem[\protect\citeauthoryear{{Colella} \& {Woodward}}{{Colella} \&
  {Woodward}}{1984}]{Colella+Woodward1984}
{Colella} P.,  {Woodward} P.~R.,  1984, \mn@doi [Journal of Computational
  Physics] {10.1016/0021-9991(84)90143-8}, \href
  {https://ui.adsabs.harvard.edu/abs/1984JCoPh..54..174C} {54, 174}

\bibitem[\protect\citeauthoryear{{Conroy}, {Gunn}  \& {White}}{{Conroy}
  et~al.}{2009}]{Conroy+2009}
{Conroy} C.,  {Gunn} J.~E.,   {White} M.,  2009, \mn@doi [\apj]
  {10.1088/0004-637X/699/1/486}, \href
  {https://ui.adsabs.harvard.edu/abs/2009ApJ...699..486C} {699, 486}
  (\mn@eprint {arXiv} {0809.4261})

\bibitem[\protect\citeauthoryear{{Correa}, {Schaye}, {Clauwens}, {Bower},
  {Crain}, {Schaller}, {Theuns}  \& {Thob}}{{Correa}
  et~al.}{2017}]{Correa+2017}
{Correa} C.~A.,  {Schaye} J.,  {Clauwens} B.,  {Bower} R.~G.,  {Crain} R.~A.,
  {Schaller} M.,  {Theuns} T.,   {Thob} A. C.~R.,  2017, \mn@doi [\mnras]
  {10.1093/mnrasl/slx133}, \href
  {https://ui.adsabs.harvard.edu/abs/2017MNRAS.472L..45C} {472, L45}
  (\mn@eprint {arXiv} {1704.06283})

\bibitem[\protect\citeauthoryear{{Costa}, {Sijacki}  \& {Haehnelt}}{{Costa}
  et~al.}{2014}]{Costa+2014}
{Costa} T.,  {Sijacki} D.,   {Haehnelt} M.~G.,  2014, \mn@doi [\mnras]
  {10.1093/mnras/stu1632}, \href
  {https://ui.adsabs.harvard.edu/abs/2014MNRAS.444.2355C} {444, 2355}
  (\mn@eprint {arXiv} {1406.2691})

\bibitem[\protect\citeauthoryear{{Costa}, {Pakmor}  \& {Springel}}{{Costa}
  et~al.}{2020}]{Costa+2020}
{Costa} T.,  {Pakmor} R.,   {Springel} V.,  2020, \mn@doi [\mnras]
  {10.1093/mnras/staa2321}, \href
  {https://ui.adsabs.harvard.edu/abs/2020MNRAS.497.5229C} {497, 5229}
  (\mn@eprint {arXiv} {2006.05997})

\bibitem[\protect\citeauthoryear{{Crain} \& {van de Voort}}{{Crain} \& {van de
  Voort}}{2023}]{Crain+VanDeVoort2023}
{Crain} R.~A.,  {van de Voort} F.,  2023, \mn@doi [\araa]
  {10.1146/annurev-astro-041923-043618}, \href
  {https://ui.adsabs.harvard.edu/abs/2023ARA&A..61..473C} {61, 473} (\mn@eprint
  {arXiv} {2309.17075})

\bibitem[\protect\citeauthoryear{{Dalla Vecchia} \& {Schaye}}{{Dalla Vecchia}
  \& {Schaye}}{2012}]{DallaVecchia+2012}
{Dalla Vecchia} C.,  {Schaye} J.,  2012, \mn@doi [\mnras]
  {10.1111/j.1365-2966.2012.21704.x}, \href
  {https://ui.adsabs.harvard.edu/abs/2012MNRAS.426..140D} {426, 140}
  (\mn@eprint {arXiv} {1203.5667})

\bibitem[\protect\citeauthoryear{{Damiano}, {Valentini}, {Borgani},
  {Tornatore}, {Murante}, {Ragagnin}, {Ragone-Figueroa}  \& {Dolag}}{{Damiano}
  et~al.}{2024}]{Damiano+2024}
{Damiano} A.,  {Valentini} M.,  {Borgani} S.,  {Tornatore} L.,  {Murante} G.,
  {Ragagnin} A.,  {Ragone-Figueroa} C.,   {Dolag} K.,  2024, \mn@doi [\aap]
  {10.1051/0004-6361/202450021}, \href
  {https://ui.adsabs.harvard.edu/abs/2024A&A...692A..81D} {692, A81}
  (\mn@eprint {arXiv} {2403.12600})

\bibitem[\protect\citeauthoryear{{Dav{\'e}}, {Finlator}  \&
  {Oppenheimer}}{{Dav{\'e}} et~al.}{2012}]{Dave+2012}
{Dav{\'e}} R.,  {Finlator} K.,   {Oppenheimer} B.~D.,  2012, \mn@doi [\mnras]
  {10.1111/j.1365-2966.2011.20148.x}, \href
  {https://ui.adsabs.harvard.edu/abs/2012MNRAS.421...98D} {421, 98} (\mn@eprint
  {arXiv} {1108.0426})

\bibitem[\protect\citeauthoryear{{Dav{\'e}}, {Angl{\'e}s-Alc{\'a}zar},
  {Narayanan}, {Li}, {Rafieferantsoa}  \& {Appleby}}{{Dav{\'e}}
  et~al.}{2019}]{Dave+2019}
{Dav{\'e}} R.,  {Angl{\'e}s-Alc{\'a}zar} D.,  {Narayanan} D.,  {Li} Q.,
  {Rafieferantsoa} M.~H.,   {Appleby} S.,  2019, \mn@doi [\mnras]
  {10.1093/mnras/stz937}, \href
  {https://ui.adsabs.harvard.edu/abs/2019MNRAS.486.2827D} {486, 2827}
  (\mn@eprint {arXiv} {1901.10203})

\bibitem[\protect\citeauthoryear{{Davies}, {Crain}  \& {Pontzen}}{{Davies}
  et~al.}{2021}]{Davies+2021}
{Davies} J.~J.,  {Crain} R.~A.,   {Pontzen} A.,  2021, \mn@doi [\mnras]
  {10.1093/mnras/staa3643}, \href
  {https://ui.adsabs.harvard.edu/abs/2021MNRAS.501..236D} {501, 236}
  (\mn@eprint {arXiv} {2006.13221})

\bibitem[\protect\citeauthoryear{{Davies}, {Pontzen}  \& {Crain}}{{Davies}
  et~al.}{2022}]{Davies+2022}
{Davies} J.~J.,  {Pontzen} A.,   {Crain} R.~A.,  2022, \mn@doi [\mnras]
  {10.1093/mnras/stac1742}, \href
  {https://ui.adsabs.harvard.edu/abs/2022MNRAS.515.1430D} {515, 1430}
  (\mn@eprint {arXiv} {2203.08157})

\bibitem[\protect\citeauthoryear{{Davis}, {Efstathiou}, {Frenk}  \&
  {White}}{{Davis} et~al.}{1985}]{Davis+1985}
{Davis} M.,  {Efstathiou} G.,  {Frenk} C.~S.,   {White} S.~D.~M.,  1985,
  \mn@doi [\apj] {10.1086/163168}, \href
  {https://ui.adsabs.harvard.edu/abs/1985ApJ...292..371D} {292, 371}

\bibitem[\protect\citeauthoryear{{De Lucia}, {Fontanot}, {Xie}  \&
  {Hirschmann}}{{De Lucia} et~al.}{2024}]{DeLucia+2024}
{De Lucia} G.,  {Fontanot} F.,  {Xie} L.,   {Hirschmann} M.,  2024, \mn@doi
  [\aap] {10.1051/0004-6361/202349045}, \href
  {https://ui.adsabs.harvard.edu/abs/2024A&A...687A..68D} {687, A68}
  (\mn@eprint {arXiv} {2401.06211})

\bibitem[\protect\citeauthoryear{{Debuhr}, {Quataert}, {Ma}  \&
  {Hopkins}}{{Debuhr} et~al.}{2010}]{Debuhr+2010}
{Debuhr} J.,  {Quataert} E.,  {Ma} C.-P.,   {Hopkins} P.,  2010, \mn@doi
  [\mnras] {10.1111/j.1745-3933.2010.00881.x}, \href
  {https://ui.adsabs.harvard.edu/abs/2010MNRAS.406L..55D} {406, L55}
  (\mn@eprint {arXiv} {0909.2872})

\bibitem[\protect\citeauthoryear{{Dehnen}}{{Dehnen}}{2000}]{Multipole2000}
{Dehnen} W.,  2000, \mn@doi [\apjl] {10.1086/312724}, \href
  {https://ui.adsabs.harvard.edu/abs/2000ApJ...536L..39D} {536, L39}
  (\mn@eprint {arXiv} {astro-ph/0003209})

\bibitem[\protect\citeauthoryear{{Dekel} et~al.,}{{Dekel}
  et~al.}{2009}]{Dekel+2009}
{Dekel} A.,  et~al., 2009, \mn@doi [\nat] {10.1038/nature07648}, \href
  {https://ui.adsabs.harvard.edu/abs/2009Natur.457..451D} {457, 451}
  (\mn@eprint {arXiv} {0808.0553})

\bibitem[\protect\citeauthoryear{{Delgado}, {Wadekar}, {Hadzhiyska}, {Bose},
  {Hernquist}  \& {Ho}}{{Delgado} et~al.}{2022}]{Delgado+2022}
{Delgado} A.~M.,  {Wadekar} D.,  {Hadzhiyska} B.,  {Bose} S.,  {Hernquist} L.,
   {Ho} S.,  2022, \mn@doi [\mnras] {10.1093/mnras/stac1951}, \href
  {https://ui.adsabs.harvard.edu/abs/2022MNRAS.515.2733D} {515, 2733}
  (\mn@eprint {arXiv} {2111.02422})

\bibitem[\protect\citeauthoryear{{Di Matteo}, {Khandai}, {DeGraf}, {Feng},
  {Croft}, {Lopez}  \& {Springel}}{{Di Matteo} et~al.}{2012}]{DiMatteo+2012}
{Di Matteo} T.,  {Khandai} N.,  {DeGraf} C.,  {Feng} Y.,  {Croft} R.~A.~C.,
  {Lopez} J.,   {Springel} V.,  2012, \mn@doi [\apjl]
  {10.1088/2041-8205/745/2/L29}, \href
  {https://ui.adsabs.harvard.edu/abs/2012ApJ...745L..29D} {745, L29}
  (\mn@eprint {arXiv} {1107.1253})

\bibitem[\protect\citeauthoryear{{Diemand}, {Kuhlen}  \& {Madau}}{{Diemand}
  et~al.}{2006}]{Diemand+2006}
{Diemand} J.,  {Kuhlen} M.,   {Madau} P.,  2006, \mn@doi [\apj]
  {10.1086/506377}, \href
  {https://ui.adsabs.harvard.edu/abs/2006ApJ...649....1D} {649, 1} (\mn@eprint
  {arXiv} {astro-ph/0603250})

\bibitem[\protect\citeauthoryear{{Dolag}, {Borgani}, {Murante}  \&
  {Springel}}{{Dolag} et~al.}{2009}]{Dolag+2009}
{Dolag} K.,  {Borgani} S.,  {Murante} G.,   {Springel} V.,  2009, \mn@doi
  [\mnras] {10.1111/j.1365-2966.2009.15034.x}, \href
  {https://ui.adsabs.harvard.edu/abs/2009MNRAS.399..497D} {399, 497}
  (\mn@eprint {arXiv} {0808.3401})

\bibitem[\protect\citeauthoryear{{Dopita} \& {Sutherland}}{{Dopita} \&
  {Sutherland}}{1996}]{Dopita+1996}
{Dopita} M.~A.,  {Sutherland} R.~S.,  1996, \mn@doi [\apjs] {10.1086/192255},
  \href {https://ui.adsabs.harvard.edu/abs/1996ApJS..102..161D} {102, 161}

\bibitem[\protect\citeauthoryear{{Dopita}, {Sutherland}, {Nicholls}, {Kewley}
  \& {Vogt}}{{Dopita} et~al.}{2013}]{Dopita+2013}
{Dopita} M.~A.,  {Sutherland} R.~S.,  {Nicholls} D.~C.,  {Kewley} L.~J.,
  {Vogt} F. P.~A.,  2013, \mn@doi [\apjs] {10.1088/0067-0049/208/1/10}, \href
  {https://ui.adsabs.harvard.edu/abs/2013ApJS..208...10D} {208, 10} (\mn@eprint
  {arXiv} {1307.5950})

\bibitem[\protect\citeauthoryear{{Dubois}, {Devriendt}, {Slyz}  \&
  {Teyssier}}{{Dubois} et~al.}{2012}]{Dubois+2012}
{Dubois} Y.,  {Devriendt} J.,  {Slyz} A.,   {Teyssier} R.,  2012, \mn@doi
  [\mnras] {10.1111/j.1365-2966.2011.20236.x}, \href
  {https://ui.adsabs.harvard.edu/abs/2012MNRAS.420.2662D} {420, 2662}
  (\mn@eprint {arXiv} {1108.0110})

\bibitem[\protect\citeauthoryear{{Dubois}, {Volonteri}, {Silk}, {Devriendt}  \&
  {Slyz}}{{Dubois} et~al.}{2014a}]{Dubois+2014b}
{Dubois} Y.,  {Volonteri} M.,  {Silk} J.,  {Devriendt} J.,   {Slyz} A.,  2014a,
  \mn@doi [\mnras] {10.1093/mnras/stu425}, \href
  {https://ui.adsabs.harvard.edu/abs/2014MNRAS.440.2333D} {440, 2333}
  (\mn@eprint {arXiv} {1401.1220})

\bibitem[\protect\citeauthoryear{{Dubois} et~al.,}{{Dubois}
  et~al.}{2014b}]{Dubois+2014}
{Dubois} Y.,  et~al., 2014b, \mn@doi [\mnras] {10.1093/mnras/stu1227}, \href
  {https://ui.adsabs.harvard.edu/abs/2014MNRAS.444.1453D} {444, 1453}
  (\mn@eprint {arXiv} {1402.1165})

\bibitem[\protect\citeauthoryear{{Dubois}, {Peirani}, {Pichon}, {Devriendt},
  {Gavazzi}, {Welker}  \& {Volonteri}}{{Dubois} et~al.}{2016}]{Dubois+2016}
{Dubois} Y.,  {Peirani} S.,  {Pichon} C.,  {Devriendt} J.,  {Gavazzi} R.,
  {Welker} C.,   {Volonteri} M.,  2016, \mn@doi [\mnras]
  {10.1093/mnras/stw2265}, \href
  {https://ui.adsabs.harvard.edu/abs/2016MNRAS.463.3948D} {463, 3948}
  (\mn@eprint {arXiv} {1606.03086})

\bibitem[\protect\citeauthoryear{{Dubois} et~al.,}{{Dubois}
  et~al.}{2021}]{Dubois+2021}
{Dubois} Y.,  et~al., 2021, \mn@doi [\aap] {10.1051/0004-6361/202039429}, \href
  {https://ui.adsabs.harvard.edu/abs/2021A&A...651A.109D} {651, A109}
  (\mn@eprint {arXiv} {2009.10578})

\bibitem[\protect\citeauthoryear{{Duffell} \& {MacFadyen}}{{Duffell} \&
  {MacFadyen}}{2011}]{Duffell+2011}
{Duffell} P.~C.,  {MacFadyen} A.~I.,  2011, \mn@doi [\apjs]
  {10.1088/0067-0049/197/2/15}, \href
  {https://ui.adsabs.harvard.edu/abs/2011ApJS..197...15D} {197, 15} (\mn@eprint
  {arXiv} {1104.3562})

\bibitem[\protect\citeauthoryear{{Dullemond}, {Juhasz}, {Pohl}, {Sereshti},
  {Shetty}, {Peters}, {Commercon}  \& {Flock}}{{Dullemond}
  et~al.}{2012}]{Dullemond+2012}
{Dullemond} C.~P.,  {Juhasz} A.,  {Pohl} A.,  {Sereshti} F.,  {Shetty} R.,
  {Peters} T.,  {Commercon} B.,   {Flock} M.,  2012, {RADMC-3D: A multi-purpose
  radiative transfer tool}, Astrophysics Source Code Library, record
  ascl:1202.015

\bibitem[\protect\citeauthoryear{{Dutton}, {Macci{\`o}}, {Frings}, {Wang},
  {Stinson}, {Penzo}  \& {Kang}}{{Dutton} et~al.}{2016}]{Dutton+2016}
{Dutton} A.~A.,  {Macci{\`o}} A.~V.,  {Frings} J.,  {Wang} L.,  {Stinson}
  G.~S.,  {Penzo} C.,   {Kang} X.,  2016, \mn@doi [\mnras]
  {10.1093/mnrasl/slv193}, \href
  {https://ui.adsabs.harvard.edu/abs/2016MNRAS.457L..74D} {457, L74}
  (\mn@eprint {arXiv} {1512.00453})

\bibitem[\protect\citeauthoryear{{Efstathiou} \& {Eastwood}}{{Efstathiou} \&
  {Eastwood}}{1981}]{Efstathiou+1981}
{Efstathiou} G.,  {Eastwood} J.~W.,  1981, \mn@doi [\mnras]
  {10.1093/mnras/194.3.503}, \href
  {https://ui.adsabs.harvard.edu/abs/1981MNRAS.194..503E} {194, 503}

\bibitem[\protect\citeauthoryear{{Eisenstein} \& {Hu}}{{Eisenstein} \&
  {Hu}}{1998}]{1998ApJ...496..605E}
{Eisenstein} D.~J.,  {Hu} W.,  1998, \mn@doi [\apj] {10.1086/305424}, \href
  {https://ui.adsabs.harvard.edu/abs/1998ApJ...496..605E} {496, 605}
  (\mn@eprint {arXiv} {astro-ph/9709112})

\bibitem[\protect\citeauthoryear{{El-Badry} et~al.,}{{El-Badry}
  et~al.}{2018}]{El-Badry+2018}
{El-Badry} K.,  et~al., 2018, \mn@doi [\mnras] {10.1093/mnras/stx2482}, \href
  {https://ui.adsabs.harvard.edu/abs/2018MNRAS.473.1930E} {473, 1930}
  (\mn@eprint {arXiv} {1705.10321})

\bibitem[\protect\citeauthoryear{{El-Badry}, {Ostriker}, {Kim}, {Quataert}  \&
  {Weisz}}{{El-Badry} et~al.}{2019}]{ElBadry2019}
{El-Badry} K.,  {Ostriker} E.~C.,  {Kim} C.-G.,  {Quataert} E.,   {Weisz}
  D.~R.,  2019, \mn@doi [\mnras] {10.1093/mnras/stz2773}, \href
  {https://ui.adsabs.harvard.edu/abs/2019MNRAS.490.1961E} {490, 1961}
  (\mn@eprint {arXiv} {1902.09547})

\bibitem[\protect\citeauthoryear{{Eldridge} \& {Stanway}}{{Eldridge} \&
  {Stanway}}{2009}]{Eldridge+2009}
{Eldridge} J.~J.,  {Stanway} E.~R.,  2009, \mn@doi [\mnras]
  {10.1111/j.1365-2966.2009.15514.x}, \href
  {https://ui.adsabs.harvard.edu/abs/2009MNRAS.400.1019E} {400, 1019}
  (\mn@eprint {arXiv} {0908.1386})

\bibitem[\protect\citeauthoryear{{Engler} et~al.,}{{Engler}
  et~al.}{2021}]{Engler+2021}
{Engler} C.,  et~al., 2021, \mn@doi [\mnras] {10.1093/mnras/stab2437}, \href
  {https://ui.adsabs.harvard.edu/abs/2021MNRAS.507.4211E} {507, 4211}
  (\mn@eprint {arXiv} {2101.12215})

\bibitem[\protect\citeauthoryear{{Faucher-Gigu{\`e}re}}{{Faucher-Gigu{\`e}re}}{2020}]{FaucherGiguere2020}
{Faucher-Gigu{\`e}re} C.-A.,  2020, \mn@doi [\mnras] {10.1093/mnras/staa302},
  \href {https://ui.adsabs.harvard.edu/abs/2020MNRAS.493.1614F} {493, 1614}
  (\mn@eprint {arXiv} {1903.08657})

\bibitem[\protect\citeauthoryear{{Faucher-Gigu{\`e}re} \&
  {Oh}}{{Faucher-Gigu{\`e}re} \& {Oh}}{2023}]{FaucherGiguere+Peng2023}
{Faucher-Gigu{\`e}re} C.-A.,  {Oh} S.~P.,  2023, \mn@doi [\araa]
  {10.1146/annurev-astro-052920-125203}, \href
  {https://ui.adsabs.harvard.edu/abs/2023ARA&A..61..131F} {61, 131} (\mn@eprint
  {arXiv} {2301.10253})

\bibitem[\protect\citeauthoryear{{Faucher-Gigu{\`e}re}, {Lidz}, {Zaldarriaga}
  \& {Hernquist}}{{Faucher-Gigu{\`e}re} et~al.}{2009}]{FaucherGiguere+2009}
{Faucher-Gigu{\`e}re} C.-A.,  {Lidz} A.,  {Zaldarriaga} M.,   {Hernquist} L.,
  2009, \mn@doi [\apj] {10.1088/0004-637X/703/2/1416}, \href
  {https://ui.adsabs.harvard.edu/abs/2009ApJ...703.1416F} {703, 1416}
  (\mn@eprint {arXiv} {0901.4554})

\bibitem[\protect\citeauthoryear{{Federrath} \& {Klessen}}{{Federrath} \&
  {Klessen}}{2012}]{Federrath+Klessen2012}
{Federrath} C.,  {Klessen} R.~S.,  2012, \mn@doi [\apj]
  {10.1088/0004-637X/761/2/156}, \href
  {https://ui.adsabs.harvard.edu/abs/2012ApJ...761..156F} {761, 156}
  (\mn@eprint {arXiv} {1209.2856})

\bibitem[\protect\citeauthoryear{{Feldmann}}{{Feldmann}}{2015}]{Feldmann2015}
{Feldmann} R.,  2015, \mn@doi [\mnras] {10.1093/mnras/stv552}, \href
  {https://ui.adsabs.harvard.edu/abs/2015MNRAS.449.3274F} {449, 3274}
  (\mn@eprint {arXiv} {1412.2755})

\bibitem[\protect\citeauthoryear{{Feldmann}, {Carollo}  \& {Mayer}}{{Feldmann}
  et~al.}{2011}]{Feldmann+2011}
{Feldmann} R.,  {Carollo} C.~M.,   {Mayer} L.,  2011, \mn@doi [\apj]
  {10.1088/0004-637X/736/2/88}, \href
  {https://ui.adsabs.harvard.edu/abs/2011ApJ...736...88F} {736, 88} (\mn@eprint
  {arXiv} {1008.3386})

\bibitem[\protect\citeauthoryear{{Feldmann}, {Hopkins}, {Quataert},
  {Faucher-Gigu{\`e}re}  \& {Kere{\v{s}}}}{{Feldmann}
  et~al.}{2016}]{Feldmann+2016}
{Feldmann} R.,  {Hopkins} P.~F.,  {Quataert} E.,  {Faucher-Gigu{\`e}re} C.-A.,
   {Kere{\v{s}}} D.,  2016, \mn@doi [\mnras] {10.1093/mnrasl/slw014}, \href
  {https://ui.adsabs.harvard.edu/abs/2016MNRAS.458L..14F} {458, L14}
  (\mn@eprint {arXiv} {1601.04704})

\bibitem[\protect\citeauthoryear{{Feldmann} et~al.,}{{Feldmann}
  et~al.}{2023}]{Feldmann+2023}
{Feldmann} R.,  et~al., 2023, \mn@doi [\mnras] {10.1093/mnras/stad1205}, \href
  {https://ui.adsabs.harvard.edu/abs/2023MNRAS.522.3831F} {522, 3831}
  (\mn@eprint {arXiv} {2205.15325})

\bibitem[\protect\citeauthoryear{{Feldmann} et~al.,}{{Feldmann}
  et~al.}{2025}]{Feldmann+2025}
{Feldmann} R.,  et~al., 2025, \mn@doi [\mnras] {10.1093/mnras/stae2633}, \href
  {https://ui.adsabs.harvard.edu/abs/2025MNRAS.536..988F} {536, 988}
  (\mn@eprint {arXiv} {2407.02674})

\bibitem[\protect\citeauthoryear{{Ferland}, {Korista}, {Verner}, {Ferguson},
  {Kingdon}  \& {Verner}}{{Ferland} et~al.}{1998}]{Ferland+1998}
{Ferland} G.~J.,  {Korista} K.~T.,  {Verner} D.~A.,  {Ferguson} J.~W.,
  {Kingdon} J.~B.,   {Verner} E.~M.,  1998, \mn@doi [\pasp] {10.1086/316190},
  \href {https://ui.adsabs.harvard.edu/abs/1998PASP..110..761F} {110, 761}

\bibitem[\protect\citeauthoryear{{Ferland} et~al.,}{{Ferland}
  et~al.}{2017}]{Ferland+2017}
{Ferland} G.~J.,  et~al., 2017, \mn@doi [\rmxaa] {10.48550/arXiv.1705.10877},
  \href {https://ui.adsabs.harvard.edu/abs/2017RMxAA..53..385F} {53, 385}
  (\mn@eprint {arXiv} {1705.10877})

\bibitem[\protect\citeauthoryear{{Filippenko} \& {Ho}}{{Filippenko} \&
  {Ho}}{2003}]{Filippenko+2003}
{Filippenko} A.~V.,  {Ho} L.~C.,  2003, \mn@doi [\apjl] {10.1086/375361}, \href
  {https://ui.adsabs.harvard.edu/abs/2003ApJ...588L..13F} {588, L13}
  (\mn@eprint {arXiv} {astro-ph/0303429})

\bibitem[\protect\citeauthoryear{{Garrison-Kimmel} et~al.,}{{Garrison-Kimmel}
  et~al.}{2019}]{Garrison-Kimmel2019}
{Garrison-Kimmel} S.,  et~al., 2019, \mn@doi [\mnras] {10.1093/mnras/stz1317},
  \href {https://ui.adsabs.harvard.edu/abs/2019MNRAS.487.1380G} {487, 1380}
  (\mn@eprint {arXiv} {1806.04143})

\bibitem[\protect\citeauthoryear{{Geen}, {Hennebelle}, {Tremblin}  \&
  {Rosdahl}}{{Geen} et~al.}{2015}]{Geen+2015b}
{Geen} S.,  {Hennebelle} P.,  {Tremblin} P.,   {Rosdahl} J.,  2015, \mn@doi
  [\mnras] {10.1093/mnras/stv2272}, \href
  {https://ui.adsabs.harvard.edu/abs/2015MNRAS.454.4484G} {454, 4484}
  (\mn@eprint {arXiv} {1507.02981})

\bibitem[\protect\citeauthoryear{{Gehren}, {Fried}, {Wehinger}  \&
  {Wyckoff}}{{Gehren} et~al.}{1984}]{Gehren1984}
{Gehren} T.,  {Fried} J.,  {Wehinger} P.~A.,   {Wyckoff} S.,  1984, \mn@doi
  [\apj] {10.1086/161763}, \href
  {https://ui.adsabs.harvard.edu/abs/1984ApJ...278...11G} {278, 11}

\bibitem[\protect\citeauthoryear{{Genel} et~al.,}{{Genel}
  et~al.}{2019}]{Genel+2019}
{Genel} S.,  et~al., 2019, \mn@doi [\apj] {10.3847/1538-4357/aaf4bb}, \href
  {https://ui.adsabs.harvard.edu/abs/2019ApJ...871...21G} {871, 21} (\mn@eprint
  {arXiv} {1807.07084})

\bibitem[\protect\citeauthoryear{{Gill}, {Knebe}  \& {Gibson}}{{Gill}
  et~al.}{2004}]{Gill+2004}
{Gill} S. P.~D.,  {Knebe} A.,   {Gibson} B.~K.,  2004, \mn@doi [\mnras]
  {10.1111/j.1365-2966.2004.07786.x}, \href
  {https://ui.adsabs.harvard.edu/abs/2004MNRAS.351..399G} {351, 399}
  (\mn@eprint {arXiv} {astro-ph/0404258})

\bibitem[\protect\citeauthoryear{{Gingold} \& {Monaghan}}{{Gingold} \&
  {Monaghan}}{1977}]{Gingold+1977}
{Gingold} R.~A.,  {Monaghan} J.~J.,  1977, \mn@doi [\mnras]
  {10.1093/mnras/181.3.375}, \href
  {https://ui.adsabs.harvard.edu/abs/1977MNRAS.181..375G} {181, 375}

\bibitem[\protect\citeauthoryear{{Gnedin} \& {Hollon}}{{Gnedin} \&
  {Hollon}}{2012}]{Gnedin+Hollon2012}
{Gnedin} N.~Y.,  {Hollon} N.,  2012, \mn@doi [\apjs]
  {10.1088/0067-0049/202/2/13}, \href
  {https://ui.adsabs.harvard.edu/abs/2012ApJS..202...13G} {202, 13} (\mn@eprint
  {arXiv} {1201.5116})

\bibitem[\protect\citeauthoryear{{Gnedin}, {Tassis}  \& {Kravtsov}}{{Gnedin}
  et~al.}{2009}]{Gnedin+2009}
{Gnedin} N.~Y.,  {Tassis} K.,   {Kravtsov} A.~V.,  2009, \mn@doi [\apj]
  {10.1088/0004-637X/697/1/55}, \href
  {https://ui.adsabs.harvard.edu/abs/2009ApJ...697...55G} {697, 55} (\mn@eprint
  {arXiv} {0810.4148})

\bibitem[\protect\citeauthoryear{{Grand} et~al.,}{{Grand}
  et~al.}{2017}]{Grand+2017}
{Grand} R. J.~J.,  et~al., 2017, \mn@doi [\mnras] {10.1093/mnras/stx071}, \href
  {https://ui.adsabs.harvard.edu/abs/2017MNRAS.467..179G} {467, 179}
  (\mn@eprint {arXiv} {1610.01159})

\bibitem[\protect\citeauthoryear{{Grand} et~al.,}{{Grand}
  et~al.}{2019}]{Grand+2019}
{Grand} R. J.~J.,  et~al., 2019, \mn@doi [\mnras] {10.1093/mnras/stz2928},
  \href {https://ui.adsabs.harvard.edu/abs/2019MNRAS.490.4786G} {490, 4786}
  (\mn@eprint {arXiv} {1909.04038})

\bibitem[\protect\citeauthoryear{{Greengard} \& {Rokhlin}}{{Greengard} \&
  {Rokhlin}}{1987}]{MultipoleGrid1987}
{Greengard} L.,  {Rokhlin} V.,  1987, \mn@doi [Journal of Computational
  Physics] {10.1016/0021-9991(87)90140-9}, \href
  {https://ui.adsabs.harvard.edu/abs/1987JCoPh..73..325G} {73, 325}

\bibitem[\protect\citeauthoryear{{Grudi{\'c}}, {Guszejnov}, {Offner}, {Rosen},
  {Raju}, {Faucher-Gigu{\`e}re}  \& {Hopkins}}{{Grudi{\'c}}
  et~al.}{2022}]{Grudic+2022}
{Grudi{\'c}} M.~Y.,  {Guszejnov} D.,  {Offner} S. S.~R.,  {Rosen} A.~L.,
  {Raju} A.~N.,  {Faucher-Gigu{\`e}re} C.-A.,   {Hopkins} P.~F.,  2022, \mn@doi
  [\mnras] {10.1093/mnras/stac526}, \href
  {https://ui.adsabs.harvard.edu/abs/2022MNRAS.512..216G} {512, 216}
  (\mn@eprint {arXiv} {2201.00882})

\bibitem[\protect\citeauthoryear{{Guedes}, {Callegari}, {Madau}  \&
  {Mayer}}{{Guedes} et~al.}{2011}]{Guesdes+2011}
{Guedes} J.,  {Callegari} S.,  {Madau} P.,   {Mayer} L.,  2011, \mn@doi [\apj]
  {10.1088/0004-637X/742/2/76}, \href
  {https://ui.adsabs.harvard.edu/abs/2011ApJ...742...76G} {742, 76} (\mn@eprint
  {arXiv} {1103.6030})

\bibitem[\protect\citeauthoryear{{Guo} et~al.,}{{Guo} et~al.}{2011}]{Guo+2011}
{Guo} Q.,  et~al., 2011, \mn@doi [\mnras] {10.1111/j.1365-2966.2010.18114.x},
  \href {https://ui.adsabs.harvard.edu/abs/2011MNRAS.413..101G} {413, 101}
  (\mn@eprint {arXiv} {1006.0106})

\bibitem[\protect\citeauthoryear{{Gutcke}, {Pakmor}, {Naab}  \&
  {Springel}}{{Gutcke} et~al.}{2021}]{Gutcke+2021}
{Gutcke} T.~A.,  {Pakmor} R.,  {Naab} T.,   {Springel} V.,  2021, \mn@doi
  [\mnras] {10.1093/mnras/staa3875}, \href
  {https://ui.adsabs.harvard.edu/abs/2021MNRAS.501.5597G} {501, 5597}
  (\mn@eprint {arXiv} {2010.07311})

\bibitem[\protect\citeauthoryear{{Haardt} \& {Madau}}{{Haardt} \&
  {Madau}}{1996}]{Haardt+Madau1996}
{Haardt} F.,  {Madau} P.,  1996, \mn@doi [\apj] {10.1086/177035}, \href
  {https://ui.adsabs.harvard.edu/abs/1996ApJ...461...20H} {461, 20} (\mn@eprint
  {arXiv} {astro-ph/9509093})

\bibitem[\protect\citeauthoryear{{Haardt} \& {Madau}}{{Haardt} \&
  {Madau}}{2012}]{Haardt+Madau2012}
{Haardt} F.,  {Madau} P.,  2012, \mn@doi [\apj] {10.1088/0004-637X/746/2/125},
  \href {https://ui.adsabs.harvard.edu/abs/2012ApJ...746..125H} {746, 125}
  (\mn@eprint {arXiv} {1105.2039})

\bibitem[\protect\citeauthoryear{{Habouzit}, {Volonteri}  \&
  {Dubois}}{{Habouzit} et~al.}{2017}]{Habouzit+2017}
{Habouzit} M.,  {Volonteri} M.,   {Dubois} Y.,  2017, \mn@doi [\mnras]
  {10.1093/mnras/stx666}, \href
  {https://ui.adsabs.harvard.edu/abs/2017MNRAS.468.3935H} {468, 3935}
  (\mn@eprint {arXiv} {1605.09394})

\bibitem[\protect\citeauthoryear{{Habouzit} et~al.,}{{Habouzit}
  et~al.}{2021}]{Habouzit+2021}
{Habouzit} M.,  et~al., 2021, \mn@doi [\mnras] {10.1093/mnras/stab496}, \href
  {https://ui.adsabs.harvard.edu/abs/2021MNRAS.503.1940H} {503, 1940}
  (\mn@eprint {arXiv} {2006.10094})

\bibitem[\protect\citeauthoryear{{Habouzit} et~al.,}{{Habouzit}
  et~al.}{2022}]{Habouzit+2022}
{Habouzit} M.,  et~al., 2022, \mn@doi [\mnras] {10.1093/mnras/stab3147}, \href
  {https://ui.adsabs.harvard.edu/abs/2022MNRAS.509.3015H} {509, 3015}
  (\mn@eprint {arXiv} {2111.01802})

\bibitem[\protect\citeauthoryear{{Hafen} et~al.,}{{Hafen}
  et~al.}{2019}]{Hafen+2019}
{Hafen} Z.,  et~al., 2019, \mn@doi [\mnras] {10.1093/mnras/stz1773}, \href
  {https://ui.adsabs.harvard.edu/abs/2019MNRAS.488.1248H} {488, 1248}
  (\mn@eprint {arXiv} {1811.11753})

\bibitem[\protect\citeauthoryear{{Hahn} \& {Abel}}{{Hahn} \&
  {Abel}}{2011}]{Hahn+2011}
{Hahn} O.,  {Abel} T.,  2011, \mn@doi [\mnras]
  {10.1111/j.1365-2966.2011.18820.x}, \href
  {https://ui.adsabs.harvard.edu/abs/2011MNRAS.415.2101H} {415, 2101}
  (\mn@eprint {arXiv} {1103.6031})

\bibitem[\protect\citeauthoryear{{Hahn}, {Rampf}  \& {Uhlemann}}{{Hahn}
  et~al.}{2021}]{Hahn+2021}
{Hahn} O.,  {Rampf} C.,   {Uhlemann} C.,  2021, \mn@doi [\mnras]
  {10.1093/mnras/staa3773}, \href
  {https://ui.adsabs.harvard.edu/abs/2021MNRAS.503..426H} {503, 426}
  (\mn@eprint {arXiv} {2008.09124})

\bibitem[\protect\citeauthoryear{{Hanany} et~al.,}{{Hanany}
  et~al.}{2000}]{Hanany+2000}
{Hanany} S.,  et~al., 2000, \mn@doi [\apjl] {10.1086/317322}, \href
  {https://ui.adsabs.harvard.edu/abs/2000ApJ...545L...5H} {545, L5} (\mn@eprint
  {arXiv} {astro-ph/0005123})

\bibitem[\protect\citeauthoryear{{Hassan} et~al.,}{{Hassan}
  et~al.}{2022}]{Hassan+2022}
{Hassan} S.,  et~al., 2022, \mn@doi [\apj] {10.3847/1538-4357/ac8b09}, \href
  {https://ui.adsabs.harvard.edu/abs/2022ApJ...937...83H} {937, 83} (\mn@eprint
  {arXiv} {2110.02983})

\bibitem[\protect\citeauthoryear{{He}, {Li}, {Feng}, {Ho}, {Ravanbakhsh},
  {Chen}  \& {P{\'o}czos}}{{He} et~al.}{2019}]{He+2019}
{He} S.,  {Li} Y.,  {Feng} Y.,  {Ho} S.,  {Ravanbakhsh} S.,  {Chen} W.,
  {P{\'o}czos} B.,  2019, \mn@doi [Proceedings of the National Academy of
  Science] {10.1073/pnas.1821458116}, \href
  {https://ui.adsabs.harvard.edu/abs/2019PNAS..11613825H} {116, 13825}
  (\mn@eprint {arXiv} {1811.06533})

\bibitem[\protect\citeauthoryear{{Hinshaw} et~al.,}{{Hinshaw}
  et~al.}{2013}]{Hinshaw+2013}
{Hinshaw} G.,  et~al., 2013, \mn@doi [\apjs] {10.1088/0067-0049/208/2/19},
  \href {https://ui.adsabs.harvard.edu/abs/2013ApJS..208...19H} {208, 19}
  (\mn@eprint {arXiv} {1212.5226})

\bibitem[\protect\citeauthoryear{{Hirashima}, {Moriwaki}, {Fujii}, {Hirai},
  {Saitoh}  \& {Makino}}{{Hirashima} et~al.}{2023}]{Hirashima+2023}
{Hirashima} K.,  {Moriwaki} K.,  {Fujii} M.~S.,  {Hirai} Y.,  {Saitoh} T.~R.,
  {Makino} J.,  2023, \mn@doi [\mnras] {10.1093/mnras/stad2864}, \href
  {https://ui.adsabs.harvard.edu/abs/2023MNRAS.526.4054H} {526, 4054}
  (\mn@eprint {arXiv} {2302.00026})

\bibitem[\protect\citeauthoryear{Hirashima, Moriwaki, Fujii, Hirai, Saitoh,
  Makino, Steinwandel  \& Ho}{Hirashima et~al.}{2025}]{Hirashima+2024}
Hirashima K.,  Moriwaki K.,  Fujii M.~S.,  Hirai Y.,  Saitoh T.~R.,  Makino J.,
   Steinwandel U.~P.,   Ho S.,  2025, \mn@doi [The Astrophysical Journal]
  {10.3847/1538-4357/add689}, 987, 86 (\mn@eprint {arXiv} {2410.23346})

\bibitem[\protect\citeauthoryear{{Hirschmann}, {Somerville}, {Naab}  \&
  {Burkert}}{{Hirschmann} et~al.}{2012}]{Hirschmann+2012}
{Hirschmann} M.,  {Somerville} R.~S.,  {Naab} T.,   {Burkert} A.,  2012,
  \mn@doi [\mnras] {10.1111/j.1365-2966.2012.21626.x}, \href
  {https://ui.adsabs.harvard.edu/abs/2012MNRAS.426..237H} {426, 237}
  (\mn@eprint {arXiv} {1206.6112})

\bibitem[\protect\citeauthoryear{{Hirschmann}, {Dolag}, {Saro}, {Bachmann},
  {Borgani}  \& {Burkert}}{{Hirschmann} et~al.}{2014}]{Hirschmann+2014}
{Hirschmann} M.,  {Dolag} K.,  {Saro} A.,  {Bachmann} L.,  {Borgani} S.,
  {Burkert} A.,  2014, \mn@doi [\mnras] {10.1093/mnras/stu1023}, \href
  {https://ui.adsabs.harvard.edu/abs/2014MNRAS.442.2304H} {442, 2304}
  (\mn@eprint {arXiv} {1308.0333})

\bibitem[\protect\citeauthoryear{{Hobbs}, {Nayakshin}, {Power}  \&
  {King}}{{Hobbs} et~al.}{2011}]{Hobbs+2011}
{Hobbs} A.,  {Nayakshin} S.,  {Power} C.,   {King} A.,  2011, \mn@doi [\mnras]
  {10.1111/j.1365-2966.2011.18333.x}, \href
  {https://ui.adsabs.harvard.edu/abs/2011MNRAS.413.2633H} {413, 2633}
  (\mn@eprint {arXiv} {1001.3883})

\bibitem[\protect\citeauthoryear{{Hockney} \& {Eastwood}}{{Hockney} \&
  {Eastwood}}{1988}]{Grid1981}
{Hockney} R.~W.,  {Eastwood} J.~W.,  1988, {Computer Simulation Using
  Particles}, 1st edn.
CRC Press, Boca Raton, \mn@doi{10.1201/9780367806934}

\bibitem[\protect\citeauthoryear{{Holmberg}}{{Holmberg}}{1941}]{Holmberg1941}
{Holmberg} E.,  1941, \mn@doi [\apj] {10.1086/144344}, \href
  {https://ui.adsabs.harvard.edu/abs/1941ApJ....94..385H} {94, 385}

\bibitem[\protect\citeauthoryear{{Hopkins}}{{Hopkins}}{2013}]{Hopkins+2013}
{Hopkins} P.~F.,  2013, \mn@doi [\mnras] {10.1093/mnras/sts210}, \href
  {https://ui.adsabs.harvard.edu/abs/2013MNRAS.428.2840H} {428, 2840}
  (\mn@eprint {arXiv} {1206.5006})

\bibitem[\protect\citeauthoryear{{Hopkins}}{{Hopkins}}{2015}]{Hopkins2015}
{Hopkins} P.~F.,  2015, \mn@doi [\mnras] {10.1093/mnras/stv195}, \href
  {https://ui.adsabs.harvard.edu/abs/2015MNRAS.450...53H} {450, 53} (\mn@eprint
  {arXiv} {1409.7395})

\bibitem[\protect\citeauthoryear{{Hopkins} \& {Quataert}}{{Hopkins} \&
  {Quataert}}{2011}]{Hopkins+Quataert2011}
{Hopkins} P.~F.,  {Quataert} E.,  2011, \mn@doi [\mnras]
  {10.1111/j.1365-2966.2011.18542.x}, \href
  {https://ui.adsabs.harvard.edu/abs/2011MNRAS.415.1027H} {415, 1027}
  (\mn@eprint {arXiv} {1007.2647})

\bibitem[\protect\citeauthoryear{{Hopkins}, {Kere{\v{s}}}, {O{\~n}orbe},
  {Faucher-Gigu{\`e}re}, {Quataert}, {Murray}  \& {Bullock}}{{Hopkins}
  et~al.}{2014}]{Hopkins+2014}
{Hopkins} P.~F.,  {Kere{\v{s}}} D.,  {O{\~n}orbe} J.,  {Faucher-Gigu{\`e}re}
  C.-A.,  {Quataert} E.,  {Murray} N.,   {Bullock} J.~S.,  2014, \mn@doi
  [\mnras] {10.1093/mnras/stu1738}, \href
  {https://ui.adsabs.harvard.edu/abs/2014MNRAS.445..581H} {445, 581}
  (\mn@eprint {arXiv} {1311.2073})

\bibitem[\protect\citeauthoryear{{Hopkins} et~al.,}{{Hopkins}
  et~al.}{2018}]{Hopkins+2018}
{Hopkins} P.~F.,  et~al., 2018, \mn@doi [\mnras] {10.1093/mnras/sty1690}, \href
  {https://ui.adsabs.harvard.edu/abs/2018MNRAS.480..800H} {480, 800}
  (\mn@eprint {arXiv} {1702.06148})

\bibitem[\protect\citeauthoryear{{Hopkins} et~al.,}{{Hopkins}
  et~al.}{2020}]{Hopkins+2020}
{Hopkins} P.~F.,  et~al., 2020, \mn@doi [\mnras] {10.1093/mnras/stz3321}, \href
  {https://ui.adsabs.harvard.edu/abs/2020MNRAS.492.3465H} {492, 3465}
  (\mn@eprint {arXiv} {1905.04321})

\bibitem[\protect\citeauthoryear{{Hopkins} et~al.,}{{Hopkins}
  et~al.}{2023a}]{Hopkins+2023_FIRE3}
{Hopkins} P.~F.,  et~al., 2023a, \mn@doi [\mnras] {10.1093/mnras/stac3489},
  \href {https://ui.adsabs.harvard.edu/abs/2023MNRAS.519.3154H} {519, 3154}
  (\mn@eprint {arXiv} {2203.00040})

\bibitem[\protect\citeauthoryear{{Hopkins} et~al.,}{{Hopkins}
  et~al.}{2023b}]{Hopkins+2023}
{Hopkins} P.~F.,  et~al., 2023b, \mn@doi [\mnras] {10.1093/mnras/stad1902},
  \href {https://ui.adsabs.harvard.edu/abs/2023MNRAS.525.2241H} {525, 2241}
  (\mn@eprint {arXiv} {2301.08263})

\bibitem[\protect\citeauthoryear{{Hopkins} et~al.,}{{Hopkins}
  et~al.}{2024}]{Hopkins+2024}
{Hopkins} P.~F.,  et~al., 2024, \mn@doi [The Open Journal of Astrophysics]
  {10.21105/astro.2309.13115}, \href
  {https://ui.adsabs.harvard.edu/abs/2024OJAp....7E..18H} {7, 18} (\mn@eprint
  {arXiv} {2309.13115})

\bibitem[\protect\citeauthoryear{{Horowitz}, {Dornfest}, {Luki{\'c}}  \&
  {Harrington}}{{Horowitz} et~al.}{2022}]{Horowitz+2022}
{Horowitz} B.,  {Dornfest} M.,  {Luki{\'c}} Z.,   {Harrington} P.,  2022,
  \mn@doi [\apj] {10.3847/1538-4357/ac9ea7}, \href
  {https://ui.adsabs.harvard.edu/abs/2022ApJ...941...42H} {941, 42} (\mn@eprint
  {arXiv} {2106.12675})

\bibitem[\protect\citeauthoryear{{Hu} et~al.,}{{Hu} et~al.}{2023}]{Hu+2023}
{Hu} C.-Y.,  et~al., 2023, \mn@doi [\apj] {10.3847/1538-4357/accf9e}, \href
  {https://ui.adsabs.harvard.edu/abs/2023ApJ...950..132H} {950, 132}
  (\mn@eprint {arXiv} {2208.10528})

\bibitem[\protect\citeauthoryear{{Huang}, {Katz}, {Scannapieco}, {Cottle},
  {Dav{\'e}}, {Weinberg}, {Peeples}  \& {Br{\"u}ggen}}{{Huang}
  et~al.}{2020}]{Huang+2020}
{Huang} S.,  {Katz} N.,  {Scannapieco} E.,  {Cottle} J.,  {Dav{\'e}} R.,
  {Weinberg} D.~H.,  {Peeples} M.~S.,   {Br{\"u}ggen} M.,  2020, \mn@doi
  [\mnras] {10.1093/mnras/staa1978}, \href
  {https://ui.adsabs.harvard.edu/abs/2020MNRAS.497.2586H} {497, 2586}
  (\mn@eprint {arXiv} {2005.13585})

\bibitem[\protect\citeauthoryear{{Hu{\v{s}}ko}, {Lacey}, {Roper}, {Schaye},
  {Briggs}  \& {Schaller}}{{Hu{\v{s}}ko} et~al.}{2025}]{Husvko+2024}
{Hu{\v{s}}ko} F.,  {Lacey} C.~G.,  {Roper} W.~J.,  {Schaye} J.,  {Briggs}
  J.~M.,   {Schaller} M.,  2025, \mn@doi [\mnras] {10.1093/mnras/staf146},
  \href {https://ui.adsabs.harvard.edu/abs/2025MNRAS.537.2559H} {537, 2559}
  (\mn@eprint {arXiv} {2410.09450})

\bibitem[\protect\citeauthoryear{{Inayoshi}, {Visbal}  \& {Haiman}}{{Inayoshi}
  et~al.}{2020}]{Inayoshi+2020}
{Inayoshi} K.,  {Visbal} E.,   {Haiman} Z.,  2020, \mn@doi [\araa]
  {10.1146/annurev-astro-120419-014455}, \href
  {https://ui.adsabs.harvard.edu/abs/2020ARA&A..58...27I} {58, 27} (\mn@eprint
  {arXiv} {1911.05791})

\bibitem[\protect\citeauthoryear{{Jamieson}, {Li}, {de Oliveira},
  {Villaescusa-Navarro}, {Ho}  \& {Spergel}}{{Jamieson}
  et~al.}{2023}]{Jamieson+2023}
{Jamieson} D.,  {Li} Y.,  {de Oliveira} R.~A.,  {Villaescusa-Navarro} F.,  {Ho}
  S.,   {Spergel} D.~N.,  2023, \mn@doi [\apj] {10.3847/1538-4357/acdb6c},
  \href {https://ui.adsabs.harvard.edu/abs/2023ApJ...952..145J} {952, 145}
  (\mn@eprint {arXiv} {2206.04594})

\bibitem[\protect\citeauthoryear{{Jo} et~al.,}{{Jo} et~al.}{2023}]{Jo+2023}
{Jo} Y.,  et~al., 2023, \mn@doi [\apj] {10.3847/1538-4357/aca8fe}, \href
  {https://ui.adsabs.harvard.edu/abs/2023ApJ...944...67J} {944, 67} (\mn@eprint
  {arXiv} {2211.16461})

\bibitem[\protect\citeauthoryear{{Jonsson}}{{Jonsson}}{2006}]{Jonsson2006}
{Jonsson} P.,  2006, \mn@doi [\mnras] {10.1111/j.1365-2966.2006.10884.x}, \href
  {https://ui.adsabs.harvard.edu/abs/2006MNRAS.372....2J} {372, 2} (\mn@eprint
  {arXiv} {astro-ph/0604118})

\bibitem[\protect\citeauthoryear{{Jung} et~al.,}{{Jung}
  et~al.}{2024}]{Jung+2024}
{Jung} M.,  et~al., 2024, \mn@doi [\apj] {10.3847/1538-4357/ad245b}, \href
  {https://ui.adsabs.harvard.edu/abs/2024ApJ...964..123J} {964, 123}
  (\mn@eprint {arXiv} {2402.05392})

\bibitem[\protect\citeauthoryear{{Kannan}, {Garaldi}, {Smith}, {Pakmor},
  {Springel}, {Vogelsberger}  \& {Hernquist}}{{Kannan}
  et~al.}{2022}]{Kannan+2022}
{Kannan} R.,  {Garaldi} E.,  {Smith} A.,  {Pakmor} R.,  {Springel} V.,
  {Vogelsberger} M.,   {Hernquist} L.,  2022, \mn@doi [\mnras]
  {10.1093/mnras/stab3710}, \href
  {https://ui.adsabs.harvard.edu/abs/2022MNRAS.511.4005K} {511, 4005}
  (\mn@eprint {arXiv} {2110.00584})

\bibitem[\protect\citeauthoryear{{Katz}}{{Katz}}{1992}]{Katz1992}
{Katz} N.,  1992, \mn@doi [\apj] {10.1086/171366}, \href
  {https://ui.adsabs.harvard.edu/abs/1992ApJ...391..502K} {391, 502}

\bibitem[\protect\citeauthoryear{{Katz}}{{Katz}}{2022}]{Katz2022}
{Katz} H.,  2022, \mn@doi [\mnras] {10.1093/mnras/stac423}, \href
  {https://ui.adsabs.harvard.edu/abs/2022MNRAS.512..348K} {512, 348}
  (\mn@eprint {arXiv} {2202.04083})

\bibitem[\protect\citeauthoryear{{Katz} \& {Gunn}}{{Katz} \&
  {Gunn}}{1991}]{Katz+1991}
{Katz} N.,  {Gunn} J.~E.,  1991, \mn@doi [\apj] {10.1086/170367}, \href
  {https://ui.adsabs.harvard.edu/abs/1991ApJ...377..365K} {377, 365}

\bibitem[\protect\citeauthoryear{{Katz}, {Weinberg}  \& {Hernquist}}{{Katz}
  et~al.}{1996}]{Katz+1996}
{Katz} N.,  {Weinberg} D.~H.,   {Hernquist} L.,  1996, \mn@doi [\apjs]
  {10.1086/192305}, \href
  {https://ui.adsabs.harvard.edu/abs/1996ApJS..105...19K} {105, 19} (\mn@eprint
  {arXiv} {astro-ph/9509107})

\bibitem[\protect\citeauthoryear{{Kauffmann}, {White}  \&
  {Guiderdoni}}{{Kauffmann} et~al.}{1993}]{Kauffmann+1993}
{Kauffmann} G.,  {White} S.~D.~M.,   {Guiderdoni} B.,  1993, \mn@doi [\mnras]
  {10.1093/mnras/264.1.201}, \href
  {https://ui.adsabs.harvard.edu/abs/1993MNRAS.264..201K} {264, 201}

\bibitem[\protect\citeauthoryear{{Kauffmann}, {Colberg}, {Diaferio}  \&
  {White}}{{Kauffmann} et~al.}{1999}]{Kauffmann+1999}
{Kauffmann} G.,  {Colberg} J.~M.,  {Diaferio} A.,   {White} S. D.~M.,  1999,
  \mn@doi [\mnras] {10.1046/j.1365-8711.1999.02202.x}, \href
  {https://ui.adsabs.harvard.edu/abs/1999MNRAS.303..188K} {303, 188}
  (\mn@eprint {arXiv} {astro-ph/9805283})

\bibitem[\protect\citeauthoryear{{Keller}, {Wadsley}, {Benincasa}  \&
  {Couchman}}{{Keller} et~al.}{2014}]{Keller+2014}
{Keller} B.~W.,  {Wadsley} J.,  {Benincasa} S.~M.,   {Couchman} H.~M.~P.,
  2014, \mn@doi [\mnras] {10.1093/mnras/stu1058}, \href
  {https://ui.adsabs.harvard.edu/abs/2014MNRAS.442.3013K} {442, 3013}
  (\mn@eprint {arXiv} {1405.2625})

\bibitem[\protect\citeauthoryear{{Keller}, {Wadsley}, {Wang}  \&
  {Kruijssen}}{{Keller} et~al.}{2019}]{Keller+2019}
{Keller} B.~W.,  {Wadsley} J.~W.,  {Wang} L.,   {Kruijssen} J.~M.~D.,  2019,
  \mn@doi [\mnras] {10.1093/mnras/sty2859}, \href
  {https://ui.adsabs.harvard.edu/abs/2019MNRAS.482.2244K} {482, 2244}
  (\mn@eprint {arXiv} {1803.05445})

\bibitem[\protect\citeauthoryear{{Kere{\v{s}}}, {Katz}, {Weinberg}  \&
  {Dav{\'e}}}{{Kere{\v{s}}} et~al.}{2005}]{Keres+2005}
{Kere{\v{s}}} D.,  {Katz} N.,  {Weinberg} D.~H.,   {Dav{\'e}} R.,  2005,
  \mn@doi [\mnras] {10.1111/j.1365-2966.2005.09451.x}, \href
  {https://ui.adsabs.harvard.edu/abs/2005MNRAS.363....2K} {363, 2} (\mn@eprint
  {arXiv} {astro-ph/0407095})

\bibitem[\protect\citeauthoryear{{Kim} \& {Ostriker}}{{Kim} \&
  {Ostriker}}{2015}]{Kim+2015}
{Kim} C.-G.,  {Ostriker} E.~C.,  2015, \mn@doi [\apj]
  {10.1088/0004-637X/802/2/99}, \href
  {https://ui.adsabs.harvard.edu/abs/2015ApJ...802...99K} {802, 99} (\mn@eprint
  {arXiv} {1410.1537})

\bibitem[\protect\citeauthoryear{{Kim} et~al.,}{{Kim} et~al.}{2014}]{Kim+2014}
{Kim} J.-h.,  et~al., 2014, \mn@doi [\apjs] {10.1088/0067-0049/210/1/14}, \href
  {https://ui.adsabs.harvard.edu/abs/2014ApJS..210...14K} {210, 14} (\mn@eprint
  {arXiv} {1308.2669})

\bibitem[\protect\citeauthoryear{{Kim}, {Kim}, {Gong}  \& {Ostriker}}{{Kim}
  et~al.}{2023}]{Kim+2023}
{Kim} C.-G.,  {Kim} J.-G.,  {Gong} M.,   {Ostriker} E.~C.,  2023, \mn@doi
  [\apj] {10.3847/1538-4357/acbd3a}, \href
  {https://ui.adsabs.harvard.edu/abs/2023ApJ...946....3K} {946, 3} (\mn@eprint
  {arXiv} {2211.13293})

\bibitem[\protect\citeauthoryear{{Klypin}, {Kravtsov}, {Valenzuela}  \&
  {Prada}}{{Klypin} et~al.}{1999}]{Klypin+1999}
{Klypin} A.,  {Kravtsov} A.~V.,  {Valenzuela} O.,   {Prada} F.,  1999, \mn@doi
  [\apj] {10.1086/307643}, \href
  {https://ui.adsabs.harvard.edu/abs/1999ApJ...522...82K} {522, 82} (\mn@eprint
  {arXiv} {astro-ph/9901240})

\bibitem[\protect\citeauthoryear{{Knebe} et~al.,}{{Knebe}
  et~al.}{2011}]{Knebe+2011}
{Knebe} A.,  et~al., 2011, \mn@doi [\mnras] {10.1111/j.1365-2966.2011.18858.x},
  \href {https://ui.adsabs.harvard.edu/abs/2011MNRAS.415.2293K} {415, 2293}
  (\mn@eprint {arXiv} {1104.0949})

\bibitem[\protect\citeauthoryear{{Knollmann} \& {Knebe}}{{Knollmann} \&
  {Knebe}}{2009}]{Knollmann+2009}
{Knollmann} S.~R.,  {Knebe} A.,  2009, \mn@doi [\apjs]
  {10.1088/0067-0049/182/2/608}, \href
  {https://ui.adsabs.harvard.edu/abs/2009ApJS..182..608K} {182, 608}
  (\mn@eprint {arXiv} {0904.3662})

\bibitem[\protect\citeauthoryear{{Kochkov}, {Smith}, {Alieva}, {Wang},
  {Brenner}  \& {Hoyer}}{{Kochkov} et~al.}{2021}]{Kochkov+2021}
{Kochkov} D.,  {Smith} J.~A.,  {Alieva} A.,  {Wang} Q.,  {Brenner} M.~P.,
  {Hoyer} S.,  2021, \mn@doi [Proceedings of the National Academy of Science]
  {10.1073/pnas.2101784118}, \href
  {https://ui.adsabs.harvard.edu/abs/2021PNAS..11801784K} {118, e2101784118}
  (\mn@eprint {arXiv} {2102.01010})

\bibitem[\protect\citeauthoryear{{Kormendy} \& {Richstone}}{{Kormendy} \&
  {Richstone}}{1995}]{Kormendy+1995}
{Kormendy} J.,  {Richstone} D.,  1995, \mn@doi [\araa]
  {10.1146/annurev.aa.33.090195.003053}, \href
  {https://ui.adsabs.harvard.edu/abs/1995ARA&A..33..581K} {33, 581}

\bibitem[\protect\citeauthoryear{{Kravtsov}, {Klypin}  \&
  {Khokhlov}}{{Kravtsov} et~al.}{1997}]{Art}
{Kravtsov} A.~V.,  {Klypin} A.~A.,   {Khokhlov} A.~M.,  1997, \mn@doi [\apjs]
  {10.1086/313015}, \href
  {https://ui.adsabs.harvard.edu/abs/1997ApJS..111...73K} {111, 73} (\mn@eprint
  {arXiv} {astro-ph/9701195})

\bibitem[\protect\citeauthoryear{{Kravtsov}, {Gnedin}  \& {Klypin}}{{Kravtsov}
  et~al.}{2004}]{Kravtsov+2004}
{Kravtsov} A.~V.,  {Gnedin} O.~Y.,   {Klypin} A.~A.,  2004, \mn@doi [\apj]
  {10.1086/421322}, \href
  {https://ui.adsabs.harvard.edu/abs/2004ApJ...609..482K} {609, 482}
  (\mn@eprint {arXiv} {astro-ph/0401088})

\bibitem[\protect\citeauthoryear{{Kretschmer} \& {Teyssier}}{{Kretschmer} \&
  {Teyssier}}{2020}]{Kretschmer+Teyssier2020}
{Kretschmer} M.,  {Teyssier} R.,  2020, \mn@doi [\mnras]
  {10.1093/mnras/stz3495}, \href
  {https://ui.adsabs.harvard.edu/abs/2020MNRAS.492.1385K} {492, 1385}
  (\mn@eprint {arXiv} {1906.11836})

\bibitem[\protect\citeauthoryear{{Krumholz}, {McKee}  \& {Klein}}{{Krumholz}
  et~al.}{2004}]{Krumholz+2004}
{Krumholz} M.~R.,  {McKee} C.~F.,   {Klein} R.~I.,  2004, \mn@doi [\apj]
  {10.1086/421935}, \href
  {https://ui.adsabs.harvard.edu/abs/2004ApJ...611..399K} {611, 399}
  (\mn@eprint {arXiv} {astro-ph/0312612})

\bibitem[\protect\citeauthoryear{{Kugel} et~al.,}{{Kugel}
  et~al.}{2023}]{Kugel+2023}
{Kugel} R.,  et~al., 2023, \mn@doi [\mnras] {10.1093/mnras/stad2540}, \href
  {https://ui.adsabs.harvard.edu/abs/2023MNRAS.526.6103K} {526, 6103}
  (\mn@eprint {arXiv} {2306.05492})

\bibitem[\protect\citeauthoryear{{Lacey} \& {Cole}}{{Lacey} \&
  {Cole}}{1994}]{Lacey+1994}
{Lacey} C.,  {Cole} S.,  1994, \mn@doi [\mnras] {10.1093/mnras/271.3.676},
  \href {https://ui.adsabs.harvard.edu/abs/1994MNRAS.271..676L} {271, 676}
  (\mn@eprint {arXiv} {astro-ph/9402069})

\bibitem[\protect\citeauthoryear{{Lagos} et~al.,}{{Lagos}
  et~al.}{2024}]{Lagos+2024}
{Lagos} C. d.~P.,  et~al., 2024, \mn@doi [\mnras] {10.1093/mnras/stae1024},
  \href {https://ui.adsabs.harvard.edu/abs/2024MNRAS.531.3551L} {531, 3551}
  (\mn@eprint {arXiv} {2309.02310})

\bibitem[\protect\citeauthoryear{{Lagos} et~al.,}{{Lagos}
  et~al.}{2025}]{Lagos+2024b}
{Lagos} C. d.~P.,  et~al., 2025, \mn@doi [\mnras] {10.1093/mnras/stae2626},
  \href {https://ui.adsabs.harvard.edu/abs/2025MNRAS.536.2324L} {536, 2324}
  (\mn@eprint {arXiv} {2409.16916})

\bibitem[\protect\citeauthoryear{{Le Brun}, {McCarthy}, {Schaye}  \&
  {Ponman}}{{Le Brun} et~al.}{2014}]{LeBrun+2014}
{Le Brun} A. M.~C.,  {McCarthy} I.~G.,  {Schaye} J.,   {Ponman} T.~J.,  2014,
  \mn@doi [\mnras] {10.1093/mnras/stu608}, \href
  {https://ui.adsabs.harvard.edu/abs/2014MNRAS.441.1270L} {441, 1270}
  (\mn@eprint {arXiv} {1312.5462})

\bibitem[\protect\citeauthoryear{{Leitherer} et~al.,}{{Leitherer}
  et~al.}{1999}]{Leitherer+1999}
{Leitherer} C.,  et~al., 1999, \mn@doi [\apjs] {10.1086/313233}, \href
  {https://ui.adsabs.harvard.edu/abs/1999ApJS..123....3L} {123, 3} (\mn@eprint
  {arXiv} {astro-ph/9902334})

\bibitem[\protect\citeauthoryear{{Lewis}, {Challinor}  \& {Lasenby}}{{Lewis}
  et~al.}{2000}]{Lewis+2000}
{Lewis} A.,  {Challinor} A.,   {Lasenby} A.,  2000, \mn@doi [\apj]
  {10.1086/309179}, \href
  {https://ui.adsabs.harvard.edu/abs/2000ApJ...538..473L} {538, 473}
  (\mn@eprint {arXiv} {astro-ph/9911177})

\bibitem[\protect\citeauthoryear{{Lilly}, {Carollo}, {Pipino}, {Renzini}  \&
  {Peng}}{{Lilly} et~al.}{2013}]{Lilly+2013}
{Lilly} S.~J.,  {Carollo} C.~M.,  {Pipino} A.,  {Renzini} A.,   {Peng} Y.,
  2013, \mn@doi [\apj] {10.1088/0004-637X/772/2/119}, \href
  {https://ui.adsabs.harvard.edu/abs/2013ApJ...772..119L} {772, 119}
  (\mn@eprint {arXiv} {1303.5059})

\bibitem[\protect\citeauthoryear{{Lovell}, {Wilkins}, {Thomas}, {Schaller},
  {Baugh}, {Fabbian}  \& {Bah{\'e}}}{{Lovell} et~al.}{2022}]{Lovell+2022}
{Lovell} C.~C.,  {Wilkins} S.~M.,  {Thomas} P.~A.,  {Schaller} M.,  {Baugh}
  C.~M.,  {Fabbian} G.,   {Bah{\'e}} Y.,  2022, \mn@doi [\mnras]
  {10.1093/mnras/stab3221}, \href
  {https://ui.adsabs.harvard.edu/abs/2022MNRAS.509.5046L} {509, 5046}
  (\mn@eprint {arXiv} {2106.04980})

\bibitem[\protect\citeauthoryear{{Lucy}}{{Lucy}}{1977}]{Lucy1977}
{Lucy} L.~B.,  1977, \mn@doi [\aj] {10.1086/112164}, \href
  {https://ui.adsabs.harvard.edu/abs/1977AJ.....82.1013L} {82, 1013}

\bibitem[\protect\citeauthoryear{{Lupi}, {Pallottini}, {Ferrara}, {Bovino},
  {Carniani}  \& {Vallini}}{{Lupi} et~al.}{2020}]{Lupi+2020}
{Lupi} A.,  {Pallottini} A.,  {Ferrara} A.,  {Bovino} S.,  {Carniani} S.,
  {Vallini} L.,  2020, \mn@doi [\mnras] {10.1093/mnras/staa1842}, \href
  {https://ui.adsabs.harvard.edu/abs/2020MNRAS.496.5160L} {496, 5160}
  (\mn@eprint {arXiv} {2004.06118})

\bibitem[\protect\citeauthoryear{{Lupi}, {Trinca}, {Volonteri}, {Dotti}  \&
  {Mazzucchelli}}{{Lupi} et~al.}{2024}]{Lupi+2024}
{Lupi} A.,  {Trinca} A.,  {Volonteri} M.,  {Dotti} M.,   {Mazzucchelli} C.,
  2024, \mn@doi [\aap] {10.1051/0004-6361/202451249}, \href
  {https://ui.adsabs.harvard.edu/abs/2024A&A...689A.128L} {689, A128}
  (\mn@eprint {arXiv} {2406.17847})

\bibitem[\protect\citeauthoryear{{Ma} \& {Bertschinger}}{{Ma} \&
  {Bertschinger}}{1995}]{Ma+1995}
{Ma} C.-P.,  {Bertschinger} E.,  1995, \mn@doi [\apj] {10.1086/176550}, \href
  {https://ui.adsabs.harvard.edu/abs/1995ApJ...455....7M} {455, 7} (\mn@eprint
  {arXiv} {astro-ph/9506072})

\bibitem[\protect\citeauthoryear{{Ma}, {Hopkins}, {Ma},
  {Angl{\'e}s-Alc{\'a}zar}, {Faucher-Gigu{\`e}re}  \& {Kelley}}{{Ma}
  et~al.}{2021}]{Ma+2021}
{Ma} L.,  {Hopkins} P.~F.,  {Ma} X.,  {Angl{\'e}s-Alc{\'a}zar} D.,
  {Faucher-Gigu{\`e}re} C.-A.,   {Kelley} L.~Z.,  2021, \mn@doi [\mnras]
  {10.1093/mnras/stab2713}, \href
  {https://ui.adsabs.harvard.edu/abs/2021MNRAS.508.1973M} {508, 1973}
  (\mn@eprint {arXiv} {2101.02727})

\bibitem[\protect\citeauthoryear{{Marinacci}, {Sales}, {Vogelsberger}, {Torrey}
   \& {Springel}}{{Marinacci} et~al.}{2019}]{Marinacci+2019}
{Marinacci} F.,  {Sales} L.~V.,  {Vogelsberger} M.,  {Torrey} P.,   {Springel}
  V.,  2019, \mn@doi [\mnras] {10.1093/mnras/stz2391}, \href
  {https://ui.adsabs.harvard.edu/abs/2019MNRAS.489.4233M} {489, 4233}
  (\mn@eprint {arXiv} {1905.08806})

\bibitem[\protect\citeauthoryear{{Martizzi}, {Teyssier}, {Moore}  \&
  {Wentz}}{{Martizzi} et~al.}{2012}]{Martizzi+2012}
{Martizzi} D.,  {Teyssier} R.,  {Moore} B.,   {Wentz} T.,  2012, \mn@doi
  [\mnras] {10.1111/j.1365-2966.2012.20879.x}, \href
  {https://ui.adsabs.harvard.edu/abs/2012MNRAS.422.3081M} {422, 3081}
  (\mn@eprint {arXiv} {1112.2752})

\bibitem[\protect\citeauthoryear{{Martizzi}, {Faucher-Gigu{\`e}re}  \&
  {Quataert}}{{Martizzi} et~al.}{2015}]{Martizzi+2015}
{Martizzi} D.,  {Faucher-Gigu{\`e}re} C.-A.,   {Quataert} E.,  2015, \mn@doi
  [\mnras] {10.1093/mnras/stv562}, \href
  {https://ui.adsabs.harvard.edu/abs/2015MNRAS.450..504M} {450, 504}
  (\mn@eprint {arXiv} {1409.4425})

\bibitem[\protect\citeauthoryear{{Massonneau}, {Volonteri}, {Dubois}  \&
  {Beckmann}}{{Massonneau} et~al.}{2023}]{Massonnea+2023}
{Massonneau} W.,  {Volonteri} M.,  {Dubois} Y.,   {Beckmann} R.~S.,  2023,
  \mn@doi [\aap] {10.1051/0004-6361/202243170}, \href
  {https://ui.adsabs.harvard.edu/abs/2023A&A...670A.180M} {670, A180}
  (\mn@eprint {arXiv} {2201.08766})

\bibitem[\protect\citeauthoryear{{Mayer}, {Governato}  \& {Kaufmann}}{{Mayer}
  et~al.}{2008}]{Mayer+2008}
{Mayer} L.,  {Governato} F.,   {Kaufmann} T.,  2008, \mn@doi [Advanced Science
  Letters] {10.48550/arXiv.0801.3845}, \href
  {https://ui.adsabs.harvard.edu/abs/2008ASL.....1....7M} {1, 7} (\mn@eprint
  {arXiv} {0801.3845})

\bibitem[\protect\citeauthoryear{{McCarthy} et~al.,}{{McCarthy}
  et~al.}{2010}]{McCarthy+2010}
{McCarthy} I.~G.,  et~al., 2010, \mn@doi [\mnras]
  {10.1111/j.1365-2966.2010.16750.x}, \href
  {https://ui.adsabs.harvard.edu/abs/2010MNRAS.406..822M} {406, 822}
  (\mn@eprint {arXiv} {0911.2641})

\bibitem[\protect\citeauthoryear{{Menon}, {Wesolowski}, {Zheng}, {Jetley},
  {Kale}, {Quinn}  \& {Governato}}{{Menon} et~al.}{2015}]{Changa}
{Menon} H.,  {Wesolowski} L.,  {Zheng} G.,  {Jetley} P.,  {Kale} L.,  {Quinn}
  T.,   {Governato} F.,  2015, \mn@doi [Computational Astrophysics and
  Cosmology] {10.1186/s40668-015-0007-9}, \href
  {https://ui.adsabs.harvard.edu/abs/2015ComAC...2....1M} {2, 1} (\mn@eprint
  {arXiv} {1409.1929})

\bibitem[\protect\citeauthoryear{{Mercedes-Feliz} et~al.,}{{Mercedes-Feliz}
  et~al.}{2023}]{Mercedes-Feliz+2023}
{Mercedes-Feliz} J.,  et~al., 2023, \mn@doi [\mnras] {10.1093/mnras/stad2079},
  \href {https://ui.adsabs.harvard.edu/abs/2023MNRAS.524.3446M} {524, 3446}
  (\mn@eprint {arXiv} {2301.01784})

\bibitem[\protect\citeauthoryear{{Mitchell}, {Schaye}, {Bower}  \&
  {Crain}}{{Mitchell} et~al.}{2020}]{Mitchell+2020}
{Mitchell} P.~D.,  {Schaye} J.,  {Bower} R.~G.,   {Crain} R.~A.,  2020, \mn@doi
  [\mnras] {10.1093/mnras/staa938}, \href
  {https://ui.adsabs.harvard.edu/abs/2020MNRAS.494.3971M} {494, 3971}
  (\mn@eprint {arXiv} {1910.09566})

\bibitem[\protect\citeauthoryear{{Moore}, {Ghigna}, {Governato}, {Lake},
  {Quinn}, {Stadel}  \& {Tozzi}}{{Moore} et~al.}{1999}]{Moore+1999}
{Moore} B.,  {Ghigna} S.,  {Governato} F.,  {Lake} G.,  {Quinn} T.,  {Stadel}
  J.,   {Tozzi} P.,  1999, \mn@doi [\apjl] {10.1086/312287}, \href
  {https://ui.adsabs.harvard.edu/abs/1999ApJ...524L..19M} {524, L19}
  (\mn@eprint {arXiv} {astro-ph/9907411})

\bibitem[\protect\citeauthoryear{{Moster}, {Naab}  \& {White}}{{Moster}
  et~al.}{2013}]{Moster+2013}
{Moster} B.~P.,  {Naab} T.,   {White} S. D.~M.,  2013, \mn@doi [\mnras]
  {10.1093/mnras/sts261}, \href
  {https://ui.adsabs.harvard.edu/abs/2013MNRAS.428.3121M} {428, 3121}
  (\mn@eprint {arXiv} {1205.5807})

\bibitem[\protect\citeauthoryear{{Motwani} et~al.,}{{Motwani}
  et~al.}{2022}]{Motwani+2022}
{Motwani} B.,  et~al., 2022, \mn@doi [\apj] {10.3847/1538-4357/ac3d2d}, \href
  {https://ui.adsabs.harvard.edu/abs/2022ApJ...926..139M} {926, 139}
  (\mn@eprint {arXiv} {2006.16314})

\bibitem[\protect\citeauthoryear{{Murante}, {Monaco}, {Giovalli}, {Borgani}  \&
  {Diaferio}}{{Murante} et~al.}{2010}]{Murante+2010}
{Murante} G.,  {Monaco} P.,  {Giovalli} M.,  {Borgani} S.,   {Diaferio} A.,
  2010, \mn@doi [\mnras] {10.1111/j.1365-2966.2010.16567.x}, \href
  {https://ui.adsabs.harvard.edu/abs/2010MNRAS.405.1491M} {405, 1491}
  (\mn@eprint {arXiv} {1002.4122})

\bibitem[\protect\citeauthoryear{{Muratov}, {Kere{\v{s}}},
  {Faucher-Gigu{\`e}re}, {Hopkins}, {Quataert}  \& {Murray}}{{Muratov}
  et~al.}{2015}]{Muratov+2015}
{Muratov} A.~L.,  {Kere{\v{s}}} D.,  {Faucher-Gigu{\`e}re} C.-A.,  {Hopkins}
  P.~F.,  {Quataert} E.,   {Murray} N.,  2015, \mn@doi [\mnras]
  {10.1093/mnras/stv2126}, \href
  {https://ui.adsabs.harvard.edu/abs/2015MNRAS.454.2691M} {454, 2691}
  (\mn@eprint {arXiv} {1501.03155})

\bibitem[\protect\citeauthoryear{{Murray}, {Quataert}  \& {Thompson}}{{Murray}
  et~al.}{2005}]{Murray+2005}
{Murray} N.,  {Quataert} E.,   {Thompson} T.~A.,  2005, \mn@doi [\apj]
  {10.1086/426067}, \href
  {https://ui.adsabs.harvard.edu/abs/2005ApJ...618..569M} {618, 569}
  (\mn@eprint {arXiv} {astro-ph/0406070})

\bibitem[\protect\citeauthoryear{{Naab} \& {Ostriker}}{{Naab} \&
  {Ostriker}}{2017}]{Naab+Ostriker2017}
{Naab} T.,  {Ostriker} J.~P.,  2017, \mn@doi [\araa]
  {10.1146/annurev-astro-081913-040019}, \href
  {https://ui.adsabs.harvard.edu/abs/2017ARA&A..55...59N} {55, 59} (\mn@eprint
  {arXiv} {1612.06891})

\bibitem[\protect\citeauthoryear{{Naab}, {Johansson}, {Ostriker}  \&
  {Efstathiou}}{{Naab} et~al.}{2007}]{Naab+2007}
{Naab} T.,  {Johansson} P.~H.,  {Ostriker} J.~P.,   {Efstathiou} G.,  2007,
  \mn@doi [\apj] {10.1086/510841}, \href
  {https://ui.adsabs.harvard.edu/abs/2007ApJ...658..710N} {658, 710}
  (\mn@eprint {arXiv} {astro-ph/0512235})

\bibitem[\protect\citeauthoryear{{Narayanan} et~al.,}{{Narayanan}
  et~al.}{2021}]{Narayanan+2021}
{Narayanan} D.,  et~al., 2021, \mn@doi [\apjs] {10.3847/1538-4365/abc487},
  \href {https://ui.adsabs.harvard.edu/abs/2021ApJS..252...12N} {252, 12}
  (\mn@eprint {arXiv} {2006.10757})

\bibitem[\protect\citeauthoryear{{Navarro} \& {Steinmetz}}{{Navarro} \&
  {Steinmetz}}{2000}]{Navarro+2000}
{Navarro} J.~F.,  {Steinmetz} M.,  2000, \mn@doi [\apj] {10.1086/309175}, \href
  {https://ui.adsabs.harvard.edu/abs/2000ApJ...538..477N} {538, 477}
  (\mn@eprint {arXiv} {astro-ph/0001003})

\bibitem[\protect\citeauthoryear{{Navarro} \& {White}}{{Navarro} \&
  {White}}{1994}]{Navarro+1994}
{Navarro} J.~F.,  {White} S. D.~M.,  1994, \mn@doi [\mnras]
  {10.1093/mnras/267.2.401}, \href
  {https://ui.adsabs.harvard.edu/abs/1994MNRAS.267..401N} {267, 401}

\bibitem[\protect\citeauthoryear{{Negri} \& {Volonteri}}{{Negri} \&
  {Volonteri}}{2017}]{Negri+Volonteri2017}
{Negri} A.,  {Volonteri} M.,  2017, \mn@doi [\mnras] {10.1093/mnras/stx362},
  \href {https://ui.adsabs.harvard.edu/abs/2017MNRAS.467.3475N} {467, 3475}
  (\mn@eprint {arXiv} {1610.04753})

\bibitem[\protect\citeauthoryear{{Nelson}, {Vogelsberger}, {Genel}, {Sijacki},
  {Kere{\v{s}}}, {Springel}  \& {Hernquist}}{{Nelson}
  et~al.}{2013}]{Nelson+2013}
{Nelson} D.,  {Vogelsberger} M.,  {Genel} S.,  {Sijacki} D.,  {Kere{\v{s}}} D.,
   {Springel} V.,   {Hernquist} L.,  2013, \mn@doi [\mnras]
  {10.1093/mnras/sts595}, \href
  {https://ui.adsabs.harvard.edu/abs/2013MNRAS.429.3353N} {429, 3353}
  (\mn@eprint {arXiv} {1301.6753})

\bibitem[\protect\citeauthoryear{{Nelson}, {Genel}, {Pillepich},
  {Vogelsberger}, {Springel}  \& {Hernquist}}{{Nelson}
  et~al.}{2016}]{Nelson+2016}
{Nelson} D.,  {Genel} S.,  {Pillepich} A.,  {Vogelsberger} M.,  {Springel} V.,
   {Hernquist} L.,  2016, \mn@doi [\mnras] {10.1093/mnras/stw1191}, \href
  {https://ui.adsabs.harvard.edu/abs/2016MNRAS.460.2881N} {460, 2881}
  (\mn@eprint {arXiv} {1503.02665})

\bibitem[\protect\citeauthoryear{{Ni} et~al.,}{{Ni} et~al.}{2023}]{Ni+2023}
{Ni} Y.,  et~al., 2023, \mn@doi [\apj] {10.3847/1538-4357/ad022a}, \href
  {https://ui.adsabs.harvard.edu/abs/2023ApJ...959..136N} {959, 136}
  (\mn@eprint {arXiv} {2304.02096})

\bibitem[\protect\citeauthoryear{{Ocvirk}, {Pichon}  \& {Teyssier}}{{Ocvirk}
  et~al.}{2008}]{Ocvirk+2008}
{Ocvirk} P.,  {Pichon} C.,   {Teyssier} R.,  2008, \mn@doi [\mnras]
  {10.1111/j.1365-2966.2008.13763.x}, \href
  {https://ui.adsabs.harvard.edu/abs/2008MNRAS.390.1326O} {390, 1326}
  (\mn@eprint {arXiv} {0803.4506})

\bibitem[\protect\citeauthoryear{{Ocvirk} et~al.,}{{Ocvirk}
  et~al.}{2020}]{Ocvirk+2020}
{Ocvirk} P.,  et~al., 2020, \mn@doi [\mnras] {10.1093/mnras/staa1266}, \href
  {https://ui.adsabs.harvard.edu/abs/2020MNRAS.496.4087O} {496, 4087}
  (\mn@eprint {arXiv} {1811.11192})

\bibitem[\protect\citeauthoryear{{Oh}, {An}, {Shin}, {Kim}  \& {Hong}}{{Oh}
  et~al.}{2022}]{Oh+2022}
{Oh} B.~K.,  {An} H.,  {Shin} E.-j.,  {Kim} J.-h.,   {Hong} S.~E.,  2022,
  \mn@doi [\mnras] {10.1093/mnras/stac1614}, \href
  {https://ui.adsabs.harvard.edu/abs/2022MNRAS.515..693O} {515, 693}
  (\mn@eprint {arXiv} {2203.06914})

\bibitem[\protect\citeauthoryear{{Onions} et~al.,}{{Onions}
  et~al.}{2012}]{Onions+2012}
{Onions} J.,  et~al., 2012, \mn@doi [\mnras]
  {10.1111/j.1365-2966.2012.20947.x}, \href
  {https://ui.adsabs.harvard.edu/abs/2012MNRAS.423.1200O} {423, 1200}
  (\mn@eprint {arXiv} {1203.3695})

\bibitem[\protect\citeauthoryear{{Pakmor} et~al.,}{{Pakmor}
  et~al.}{2017}]{Pakmor+2017}
{Pakmor} R.,  et~al., 2017, \mn@doi [\mnras] {10.1093/mnras/stx1074}, \href
  {https://ui.adsabs.harvard.edu/abs/2017MNRAS.469.3185P} {469, 3185}
  (\mn@eprint {arXiv} {1701.07028})

\bibitem[\protect\citeauthoryear{{Pakmor} et~al.,}{{Pakmor}
  et~al.}{2023}]{Pakmor+2023}
{Pakmor} R.,  et~al., 2023, \mn@doi [\mnras] {10.1093/mnras/stac3620}, \href
  {https://ui.adsabs.harvard.edu/abs/2023MNRAS.524.2539P} {524, 2539}
  (\mn@eprint {arXiv} {2210.10060})

\bibitem[\protect\citeauthoryear{{Pakmor} et~al.,}{{Pakmor}
  et~al.}{2024}]{Pakmor+2024}
{Pakmor} R.,  et~al., 2024, \mn@doi [\mnras] {10.1093/mnras/stae112}, \href
  {https://ui.adsabs.harvard.edu/abs/2024MNRAS.528.2308P} {528, 2308}
  (\mn@eprint {arXiv} {2309.13104})

\bibitem[\protect\citeauthoryear{{Pallottini} et~al.,}{{Pallottini}
  et~al.}{2022}]{Pallottini+2022}
{Pallottini} A.,  et~al., 2022, \mn@doi [\mnras] {10.1093/mnras/stac1281},
  \href {https://ui.adsabs.harvard.edu/abs/2022MNRAS.513.5621P} {513, 5621}
  (\mn@eprint {arXiv} {2201.02636})

\bibitem[\protect\citeauthoryear{{Pandya} et~al.,}{{Pandya}
  et~al.}{2021}]{Pandya+2021}
{Pandya} V.,  et~al., 2021, \mn@doi [\mnras] {10.1093/mnras/stab2714}, \href
  {https://ui.adsabs.harvard.edu/abs/2021MNRAS.508.2979P} {508, 2979}
  (\mn@eprint {arXiv} {2103.06891})

\bibitem[\protect\citeauthoryear{{Perivolaropoulos} \&
  {Skara}}{{Perivolaropoulos} \& {Skara}}{2022}]{Perivolaropoulos+2022}
{Perivolaropoulos} L.,  {Skara} F.,  2022, \mn@doi [\nar]
  {10.1016/j.newar.2022.101659}, \href
  {https://ui.adsabs.harvard.edu/abs/2022NewAR..9501659P} {95, 101659}
  (\mn@eprint {arXiv} {2105.05208})

\bibitem[\protect\citeauthoryear{{Pfister}, {Volonteri}, {Dubois}, {Dotti}  \&
  {Colpi}}{{Pfister} et~al.}{2019}]{Pfister+2019}
{Pfister} H.,  {Volonteri} M.,  {Dubois} Y.,  {Dotti} M.,   {Colpi} M.,  2019,
  \mn@doi [\mnras] {10.1093/mnras/stz822}, \href
  {https://ui.adsabs.harvard.edu/abs/2019MNRAS.486..101P} {486, 101}
  (\mn@eprint {arXiv} {1902.01297})

\bibitem[\protect\citeauthoryear{{Pillepich} et~al.,}{{Pillepich}
  et~al.}{2018a}]{Pillepich+2018a}
{Pillepich} A.,  et~al., 2018a, \mn@doi [\mnras] {10.1093/mnras/stx2656}, \href
  {https://ui.adsabs.harvard.edu/abs/2018MNRAS.473.4077P} {473, 4077}
  (\mn@eprint {arXiv} {1703.02970})

\bibitem[\protect\citeauthoryear{{Pillepich} et~al.,}{{Pillepich}
  et~al.}{2018b}]{Pillepich+2018}
{Pillepich} A.,  et~al., 2018b, \mn@doi [\mnras] {10.1093/mnras/stx3112}, \href
  {https://ui.adsabs.harvard.edu/abs/2018MNRAS.475..648P} {475, 648}
  (\mn@eprint {arXiv} {1707.03406})

\bibitem[\protect\citeauthoryear{{Pillepich} et~al.,}{{Pillepich}
  et~al.}{2019}]{Pillepich+2019}
{Pillepich} A.,  et~al., 2019, \mn@doi [\mnras] {10.1093/mnras/stz2338}, \href
  {https://ui.adsabs.harvard.edu/abs/2019MNRAS.490.3196P} {490, 3196}
  (\mn@eprint {arXiv} {1902.05553})

\bibitem[\protect\citeauthoryear{{Pittard}}{{Pittard}}{2019}]{Pittard+2019}
{Pittard} J.~M.,  2019, \mn@doi [\mnras] {10.1093/mnras/stz1885}, \href
  {https://ui.adsabs.harvard.edu/abs/2019MNRAS.488.3376P} {488, 3376}
  (\mn@eprint {arXiv} {1907.03519})

\bibitem[\protect\citeauthoryear{{Planck Collaboration} et~al.,}{{Planck
  Collaboration} et~al.}{2014}]{Planck+2014}
{Planck Collaboration} et~al., 2014, \mn@doi [\aap]
  {10.1051/0004-6361/201321591}, \href
  {https://ui.adsabs.harvard.edu/abs/2014A&A...571A..16P} {571, A16}
  (\mn@eprint {arXiv} {1303.5076})

\bibitem[\protect\citeauthoryear{{Planck Collaboration} et~al.,}{{Planck
  Collaboration} et~al.}{2020}]{Planck+2020}
{Planck Collaboration} et~al., 2020, \mn@doi [\aap]
  {10.1051/0004-6361/201833910}, \href
  {https://ui.adsabs.harvard.edu/abs/2020A&A...641A...6P} {641, A6} (\mn@eprint
  {arXiv} {1807.06209})

\bibitem[\protect\citeauthoryear{{Ploeckinger} \& {Schaye}}{{Ploeckinger} \&
  {Schaye}}{2020}]{Ploeckinger+Schaye202}
{Ploeckinger} S.,  {Schaye} J.,  2020, \mn@doi [\mnras]
  {10.1093/mnras/staa2172}, \href
  {https://ui.adsabs.harvard.edu/abs/2020MNRAS.497.4857P} {497, 4857}
  (\mn@eprint {arXiv} {2006.14322})

\bibitem[\protect\citeauthoryear{{Potter}, {Stadel}  \& {Teyssier}}{{Potter}
  et~al.}{2017}]{Potter+2017}
{Potter} D.,  {Stadel} J.,   {Teyssier} R.,  2017, \mn@doi [Computational
  Astrophysics and Cosmology] {10.1186/s40668-017-0021-1}, \href
  {https://ui.adsabs.harvard.edu/abs/2017ComAC...4....2P} {4, 2} (\mn@eprint
  {arXiv} {1609.08621})

\bibitem[\protect\citeauthoryear{{Power}, {Nayakshin}  \& {King}}{{Power}
  et~al.}{2011}]{Power+2011}
{Power} C.,  {Nayakshin} S.,   {King} A.,  2011, \mn@doi [\mnras]
  {10.1111/j.1365-2966.2010.17901.x}, \href
  {https://ui.adsabs.harvard.edu/abs/2011MNRAS.412..269P} {412, 269}
  (\mn@eprint {arXiv} {1003.0605})

\bibitem[\protect\citeauthoryear{{Price}}{{Price}}{2012}]{Price2012}
{Price} D.~J.,  2012, \mn@doi [Journal of Computational Physics]
  {10.1016/j.jcp.2010.12.011}, \href
  {https://ui.adsabs.harvard.edu/abs/2012JCoPh.231..759P} {231, 759}
  (\mn@eprint {arXiv} {1012.1885})

\bibitem[\protect\citeauthoryear{{Price}, {Kriek}, {Feldmann}, {Quataert},
  {Hopkins}, {Faucher-Gigu{\`e}re}, {Kere{\v{s}}}  \& {Barro}}{{Price}
  et~al.}{2017}]{Price+2017}
{Price} S.~H.,  {Kriek} M.,  {Feldmann} R.,  {Quataert} E.,  {Hopkins} P.~F.,
  {Faucher-Gigu{\`e}re} C.-A.,  {Kere{\v{s}}} D.,   {Barro} G.,  2017, \mn@doi
  [\apjl] {10.3847/2041-8213/aa7d4b}, \href
  {https://ui.adsabs.harvard.edu/abs/2017ApJ...844L...6P} {844, L6} (\mn@eprint
  {arXiv} {1707.01094})

\bibitem[\protect\citeauthoryear{{Rahmati}, {Pawlik}, {Rai{\v{c}}evi{\'c}}  \&
  {Schaye}}{{Rahmati} et~al.}{2013}]{Rahmati+2013}
{Rahmati} A.,  {Pawlik} A.~H.,  {Rai{\v{c}}evi{\'c}} M.,   {Schaye} J.,  2013,
  \mn@doi [\mnras] {10.1093/mnras/stt066}, \href
  {https://ui.adsabs.harvard.edu/abs/2013MNRAS.430.2427R} {430, 2427}
  (\mn@eprint {arXiv} {1210.7808})

\bibitem[\protect\citeauthoryear{{Raissi}, {Perdikaris}  \&
  {Karniadakis}}{{Raissi} et~al.}{2019}]{Raissi+2019}
{Raissi} M.,  {Perdikaris} P.,   {Karniadakis} G.~E.,  2019, \mn@doi [Journal
  of Computational Physics] {10.1016/j.jcp.2018.10.045}, \href
  {https://ui.adsabs.harvard.edu/abs/2019JCoPh.378..686R} {378, 686}

\bibitem[\protect\citeauthoryear{{Ramesh} \& {Nelson}}{{Ramesh} \&
  {Nelson}}{2024}]{Ramesh+2024}
{Ramesh} R.,  {Nelson} D.,  2024, \mn@doi [\mnras] {10.1093/mnras/stae237},
  \href {https://ui.adsabs.harvard.edu/abs/2024MNRAS.528.3320R} {528, 3320}
  (\mn@eprint {arXiv} {2307.11143})

\bibitem[\protect\citeauthoryear{{Rathjen} et~al.,}{{Rathjen}
  et~al.}{2021}]{Rathjen+2021}
{Rathjen} T.-E.,  et~al., 2021, \mn@doi [\mnras] {10.1093/mnras/stab900}, \href
  {https://ui.adsabs.harvard.edu/abs/2021MNRAS.504.1039R} {504, 1039}
  (\mn@eprint {arXiv} {2103.14128})

\bibitem[\protect\citeauthoryear{{Read} \& {Hayfield}}{{Read} \&
  {Hayfield}}{2012}]{Read+Hayfield2012}
{Read} J.~I.,  {Hayfield} T.,  2012, \mn@doi [\mnras]
  {10.1111/j.1365-2966.2012.20819.x}, \href
  {https://ui.adsabs.harvard.edu/abs/2012MNRAS.422.3037R} {422, 3037}
  (\mn@eprint {arXiv} {1111.6985})

\bibitem[\protect\citeauthoryear{{Regan}, {Downes}, {Volonteri}, {Beckmann},
  {Lupi}, {Trebitsch}  \& {Dubois}}{{Regan} et~al.}{2019}]{Regan+2019}
{Regan} J.~A.,  {Downes} T.~P.,  {Volonteri} M.,  {Beckmann} R.,  {Lupi} A.,
  {Trebitsch} M.,   {Dubois} Y.,  2019, \mn@doi [\mnras]
  {10.1093/mnras/stz1045}, \href
  {https://ui.adsabs.harvard.edu/abs/2019MNRAS.486.3892R} {486, 3892}
  (\mn@eprint {arXiv} {1811.04953})

\bibitem[\protect\citeauthoryear{{Reines}, {Reynolds}, {Miller}, {Sivakoff},
  {Greene}, {Hickox}  \& {Johnson}}{{Reines} et~al.}{2016}]{Reines+2016}
{Reines} A.~E.,  {Reynolds} M.~T.,  {Miller} J.~M.,  {Sivakoff} G.~R.,
  {Greene} J.~E.,  {Hickox} R.~C.,   {Johnson} K.~E.,  2016, \mn@doi [\apjl]
  {10.3847/2041-8205/830/2/L35}, \href
  {https://ui.adsabs.harvard.edu/abs/2016ApJ...830L..35R} {830, L35}
  (\mn@eprint {arXiv} {1610.01598})

\bibitem[\protect\citeauthoryear{{Revaz}}{{Revaz}}{2013}]{Revaz+2013}
{Revaz} Y.,  2013, {pNbody: A python parallelized N-body reduction toolbox},
  Astrophysics Source Code Library, record ascl:1302.004

\bibitem[\protect\citeauthoryear{{Revaz} \& {Jablonka}}{{Revaz} \&
  {Jablonka}}{2018}]{Revaz2018}
{Revaz} Y.,  {Jablonka} P.,  2018, \mn@doi [\aap]
  {10.1051/0004-6361/201832669}, \href
  {https://ui.adsabs.harvard.edu/abs/2018A&A...616A..96R} {616, A96}
  (\mn@eprint {arXiv} {1801.06222})

\bibitem[\protect\citeauthoryear{{Richings}, {Faucher-Gigu{\`e}re}, {Gurvich},
  {Schaye}  \& {Hayward}}{{Richings} et~al.}{2022}]{Richings+2022}
{Richings} A.~J.,  {Faucher-Gigu{\`e}re} C.-A.,  {Gurvich} A.~B.,  {Schaye} J.,
    {Hayward} C.~C.,  2022, \mn@doi [\mnras] {10.1093/mnras/stac2338}, \href
  {https://ui.adsabs.harvard.edu/abs/2022MNRAS.517.1557R} {517, 1557}
  (\mn@eprint {arXiv} {2208.02288})

\bibitem[\protect\citeauthoryear{{Robitaille}}{{Robitaille}}{2011}]{Robitaille+2011}
{Robitaille} T.~P.,  2011, \mn@doi [\aap] {10.1051/0004-6361/201117150}, \href
  {https://ui.adsabs.harvard.edu/abs/2011A&A...536A..79R} {536, A79}
  (\mn@eprint {arXiv} {1112.1071})

\bibitem[\protect\citeauthoryear{{Roca-F{\`a}brega} et~al.,}{{Roca-F{\`a}brega}
  et~al.}{2019}]{Roca-Fabrega+2019}
{Roca-F{\`a}brega} S.,  et~al., 2019, \mn@doi [\mnras] {10.1093/mnras/stz063},
  \href {https://ui.adsabs.harvard.edu/abs/2019MNRAS.484.3625R} {484, 3625}
  (\mn@eprint {arXiv} {1808.09973})

\bibitem[\protect\citeauthoryear{{Roca-F{\`a}brega} et~al.,}{{Roca-F{\`a}brega}
  et~al.}{2024}]{AgoraCosmo2024}
{Roca-F{\`a}brega} S.,  et~al., 2024, \mn@doi [\apj]
  {10.3847/1538-4357/ad43de}, \href
  {https://ui.adsabs.harvard.edu/abs/2024ApJ...968..125R} {968, 125}
  (\mn@eprint {arXiv} {2402.06202})

\bibitem[\protect\citeauthoryear{{Rodriguez-Gomez} et~al.,}{{Rodriguez-Gomez}
  et~al.}{2019}]{Rodriguez-Gomez+2019}
{Rodriguez-Gomez} V.,  et~al., 2019, \mn@doi [\mnras] {10.1093/mnras/sty3345},
  \href {https://ui.adsabs.harvard.edu/abs/2019MNRAS.483.4140R} {483, 4140}
  (\mn@eprint {arXiv} {1809.08239})

\bibitem[\protect\citeauthoryear{{Rosas-Guevara} et~al.,}{{Rosas-Guevara}
  et~al.}{2015}]{Rosas-Guevara+2015}
{Rosas-Guevara} Y.~M.,  et~al., 2015, \mn@doi [\mnras] {10.1093/mnras/stv2056},
  \href {https://ui.adsabs.harvard.edu/abs/2015MNRAS.454.1038R} {454, 1038}
  (\mn@eprint {arXiv} {1312.0598})

\bibitem[\protect\citeauthoryear{{Rosdahl} et~al.,}{{Rosdahl}
  et~al.}{2018}]{Rosdahl+2018}
{Rosdahl} J.,  et~al., 2018, \mn@doi [\mnras] {10.1093/mnras/sty1655}, \href
  {https://ui.adsabs.harvard.edu/abs/2018MNRAS.479..994R} {479, 994}
  (\mn@eprint {arXiv} {1801.07259})

\bibitem[\protect\citeauthoryear{{Rose} et~al.,}{{Rose}
  et~al.}{2025}]{Rose+2024}
{Rose} J.~C.,  et~al., 2025, \mn@doi [\apj] {10.3847/1538-4357/adb8e5}, \href
  {https://ui.adsabs.harvard.edu/abs/2025ApJ...982...68R} {982, 68} (\mn@eprint
  {arXiv} {2405.00766})

\bibitem[\protect\citeauthoryear{{Rosen} \& {Bregman}}{{Rosen} \&
  {Bregman}}{1995}]{Rosen+Bregman1995}
{Rosen} A.,  {Bregman} J.~N.,  1995, \mn@doi [\apj] {10.1086/175303}, \href
  {https://ui.adsabs.harvard.edu/abs/1995ApJ...440..634R} {440, 634}

\bibitem[\protect\citeauthoryear{{Sales}, {Wetzel}  \& {Fattahi}}{{Sales}
  et~al.}{2022}]{Sales+2022}
{Sales} L.~V.,  {Wetzel} A.,   {Fattahi} A.,  2022, \mn@doi [Nature Astronomy]
  {10.1038/s41550-022-01689-w}, \href
  {https://ui.adsabs.harvard.edu/abs/2022NatAs...6..897S} {6, 897} (\mn@eprint
  {arXiv} {2206.05295})

\bibitem[\protect\citeauthoryear{{Sawala} et~al.,}{{Sawala}
  et~al.}{2016}]{Sawala+2016}
{Sawala} T.,  et~al., 2016, \mn@doi [\mnras] {10.1093/mnras/stw145}, \href
  {https://ui.adsabs.harvard.edu/abs/2016MNRAS.457.1931S} {457, 1931}
  (\mn@eprint {arXiv} {1511.01098})

\bibitem[\protect\citeauthoryear{{Scannapieco} et~al.,}{{Scannapieco}
  et~al.}{2012}]{Scannapieco+2012}
{Scannapieco} C.,  et~al., 2012, \mn@doi [\mnras]
  {10.1111/j.1365-2966.2012.20993.x}, \href
  {https://ui.adsabs.harvard.edu/abs/2012MNRAS.423.1726S} {423, 1726}
  (\mn@eprint {arXiv} {1112.0315})

\bibitem[\protect\citeauthoryear{{Schaller} et~al.,}{{Schaller}
  et~al.}{2024}]{Swift2024}
{Schaller} M.,  et~al., 2024, \mn@doi [\mnras] {10.1093/mnras/stae922}, \href
  {https://ui.adsabs.harvard.edu/abs/2024MNRAS.530.2378S} {530, 2378}
  (\mn@eprint {arXiv} {2305.13380})

\bibitem[\protect\citeauthoryear{{Schaller}, {Schaye}, {Kugel}, {Broxterman}
  \& {van Daalen}}{{Schaller} et~al.}{2025}]{Schaller+2024}
{Schaller} M.,  {Schaye} J.,  {Kugel} R.,  {Broxterman} J.~C.,   {van Daalen}
  M.~P.,  2025, \mn@doi [\mnras] {10.1093/mnras/staf569}, \href
  {https://ui.adsabs.harvard.edu/abs/2025MNRAS.539.1337S} {539, 1337}
  (\mn@eprint {arXiv} {2410.17109})

\bibitem[\protect\citeauthoryear{{Schaye} et~al.,}{{Schaye}
  et~al.}{2015}]{Schaye+2015}
{Schaye} J.,  et~al., 2015, \mn@doi [\mnras] {10.1093/mnras/stu2058}, \href
  {https://ui.adsabs.harvard.edu/abs/2015MNRAS.446..521S} {446, 521}
  (\mn@eprint {arXiv} {1407.7040})

\bibitem[\protect\citeauthoryear{{Schmidt} \& {Federrath}}{{Schmidt} \&
  {Federrath}}{2011}]{Schmidt+Federrath2011}
{Schmidt} W.,  {Federrath} C.,  2011, \mn@doi [\aap]
  {10.1051/0004-6361/201015630}, \href
  {https://ui.adsabs.harvard.edu/abs/2011A&A...528A.106S} {528, A106}
  (\mn@eprint {arXiv} {1010.4492})

\bibitem[\protect\citeauthoryear{{Schneider} \& {Robertson}}{{Schneider} \&
  {Robertson}}{2015}]{Schneider+2015}
{Schneider} E.~E.,  {Robertson} B.~E.,  2015, \mn@doi [\apjs]
  {10.1088/0067-0049/217/2/24}, \href
  {https://ui.adsabs.harvard.edu/abs/2015ApJS..217...24S} {217, 24} (\mn@eprint
  {arXiv} {1410.4194})

\bibitem[\protect\citeauthoryear{{Scoccimarro}}{{Scoccimarro}}{1998}]{Scoccimarro+1998}
{Scoccimarro} R.,  1998, \mn@doi [\mnras] {10.1046/j.1365-8711.1998.01845.x},
  \href {https://ui.adsabs.harvard.edu/abs/1998MNRAS.299.1097S} {299, 1097}
  (\mn@eprint {arXiv} {astro-ph/9711187})

\bibitem[\protect\citeauthoryear{{Seljak} \& {Zaldarriaga}}{{Seljak} \&
  {Zaldarriaga}}{1996}]{Seljak+1996}
{Seljak} U.,  {Zaldarriaga} M.,  1996, \mn@doi [\apj] {10.1086/177793}, \href
  {https://ui.adsabs.harvard.edu/abs/1996ApJ...469..437S} {469, 437}
  (\mn@eprint {arXiv} {astro-ph/9603033})

\bibitem[\protect\citeauthoryear{{Semenov}}{{Semenov}}{2024}]{Semenov2024}
{Semenov} V.~A.,  2024, \mn@doi [arXiv e-prints] {10.48550/arXiv.2410.23339},
  \href {https://ui.adsabs.harvard.edu/abs/2024arXiv241023339S} {p.
  arXiv:2410.23339} (\mn@eprint {arXiv} {2410.23339})

\bibitem[\protect\citeauthoryear{{Semenov}, {Kravtsov}  \& {Gnedin}}{{Semenov}
  et~al.}{2016}]{Semenov+2016}
{Semenov} V.~A.,  {Kravtsov} A.~V.,   {Gnedin} N.~Y.,  2016, \mn@doi [\apj]
  {10.3847/0004-637X/826/2/200}, \href
  {https://ui.adsabs.harvard.edu/abs/2016ApJ...826..200S} {826, 200}
  (\mn@eprint {arXiv} {1512.03101})

\bibitem[\protect\citeauthoryear{{Shetty} \& {Ostriker}}{{Shetty} \&
  {Ostriker}}{2008}]{Shetty+2008}
{Shetty} R.,  {Ostriker} E.~C.,  2008, \mn@doi [\apj] {10.1086/590383}, \href
  {https://ui.adsabs.harvard.edu/abs/2008ApJ...684..978S} {684, 978}
  (\mn@eprint {arXiv} {0805.3996})

\bibitem[\protect\citeauthoryear{{Sijacki}, {Springel}, {Di Matteo}  \&
  {Hernquist}}{{Sijacki} et~al.}{2007}]{Sijacki+2007}
{Sijacki} D.,  {Springel} V.,  {Di Matteo} T.,   {Hernquist} L.,  2007, \mn@doi
  [\mnras] {10.1111/j.1365-2966.2007.12153.x}, \href
  {https://ui.adsabs.harvard.edu/abs/2007MNRAS.380..877S} {380, 877}
  (\mn@eprint {arXiv} {0705.2238})

\bibitem[\protect\citeauthoryear{{Sijacki}, {Springel}  \&
  {Haehnelt}}{{Sijacki} et~al.}{2009}]{Sijacki+2009}
{Sijacki} D.,  {Springel} V.,   {Haehnelt} M.~G.,  2009, \mn@doi [\mnras]
  {10.1111/j.1365-2966.2009.15452.x}, \href
  {https://ui.adsabs.harvard.edu/abs/2009MNRAS.400..100S} {400, 100}
  (\mn@eprint {arXiv} {0905.1689})

\bibitem[\protect\citeauthoryear{{Sijacki}, {Vogelsberger}, {Genel},
  {Springel}, {Torrey}, {Snyder}, {Nelson}  \& {Hernquist}}{{Sijacki}
  et~al.}{2015}]{Sijacki+2015}
{Sijacki} D.,  {Vogelsberger} M.,  {Genel} S.,  {Springel} V.,  {Torrey} P.,
  {Snyder} G.~F.,  {Nelson} D.,   {Hernquist} L.,  2015, \mn@doi [\mnras]
  {10.1093/mnras/stv1340}, \href
  {https://ui.adsabs.harvard.edu/abs/2015MNRAS.452..575S} {452, 575}
  (\mn@eprint {arXiv} {1408.6842})

\bibitem[\protect\citeauthoryear{{Silk} \& {Rees}}{{Silk} \&
  {Rees}}{1998}]{Silk+Rees1998}
{Silk} J.,  {Rees} M.~J.,  1998, \mn@doi [\aap]
  {10.48550/arXiv.astro-ph/9801013}, \href
  {https://ui.adsabs.harvard.edu/abs/1998A&A...331L...1S} {331, L1} (\mn@eprint
  {arXiv} {astro-ph/9801013})

\bibitem[\protect\citeauthoryear{Smith et~al.,}{Smith
  et~al.}{2017}]{Smith+2016}
Smith B.~D.,  et~al., 2017, \mn@doi [Monthly Notices of the Royal Astronomical
  Society] {10.1093/mnras/stw3291}, 466, 2217 (\mn@eprint {}
  {https://academic.oup.com/mnras/article-pdf/466/2/2217/10868505/stw3291.pdf})

\bibitem[\protect\citeauthoryear{{Smith}, {Sijacki}  \& {Shen}}{{Smith}
  et~al.}{2018}]{Smith+2018}
{Smith} M.~C.,  {Sijacki} D.,   {Shen} S.,  2018, \mn@doi [\mnras]
  {10.1093/mnras/sty994}, \href
  {https://ui.adsabs.harvard.edu/abs/2018MNRAS.478..302S} {478, 302}
  (\mn@eprint {arXiv} {1709.03515})

\bibitem[\protect\citeauthoryear{{Smith}, {Sijacki}  \& {Shen}}{{Smith}
  et~al.}{2019}]{Smith+2019}
{Smith} M.~C.,  {Sijacki} D.,   {Shen} S.,  2019, \mn@doi [\mnras]
  {10.1093/mnras/stz599}, \href
  {https://ui.adsabs.harvard.edu/abs/2019MNRAS.485.3317S} {485, 3317}
  (\mn@eprint {arXiv} {1807.04288})

\bibitem[\protect\citeauthoryear{{Smith} et~al.,}{{Smith}
  et~al.}{2024}]{Smith+2024}
{Smith} M.~C.,  et~al., 2024, \mn@doi [\mnras] {10.1093/mnras/stad3168}, \href
  {https://ui.adsabs.harvard.edu/abs/2024MNRAS.527.1216S} {527, 1216}
  (\mn@eprint {arXiv} {2301.07116})

\bibitem[\protect\citeauthoryear{{Somerville} \& {Dav{\'e}}}{{Somerville} \&
  {Dav{\'e}}}{2015}]{Somerville+2015}
{Somerville} R.~S.,  {Dav{\'e}} R.,  2015, \mn@doi [\araa]
  {10.1146/annurev-astro-082812-140951}, \href
  {https://ui.adsabs.harvard.edu/abs/2015ARA&A..53...51S} {53, 51} (\mn@eprint
  {arXiv} {1412.2712})

\bibitem[\protect\citeauthoryear{{Somerville} \& {Primack}}{{Somerville} \&
  {Primack}}{1999}]{Somerville+1999}
{Somerville} R.~S.,  {Primack} J.~R.,  1999, \mn@doi [\mnras]
  {10.1046/j.1365-8711.1999.03032.x}, \href
  {https://ui.adsabs.harvard.edu/abs/1999MNRAS.310.1087S} {310, 1087}
  (\mn@eprint {arXiv} {astro-ph/9802268})

\bibitem[\protect\citeauthoryear{{Spergel} et~al.,}{{Spergel}
  et~al.}{2003}]{Spergel+2003}
{Spergel} D.~N.,  et~al., 2003, \mn@doi [\apjs] {10.1086/377226}, \href
  {https://ui.adsabs.harvard.edu/abs/2003ApJS..148..175S} {148, 175}
  (\mn@eprint {arXiv} {astro-ph/0302209})

\bibitem[\protect\citeauthoryear{{Springel}}{{Springel}}{2010a}]{Springel2010a}
{Springel} V.,  2010a, \mn@doi [\araa] {10.1146/annurev-astro-081309-130914},
  \href {https://ui.adsabs.harvard.edu/abs/2010ARA&A..48..391S} {48, 391}
  (\mn@eprint {arXiv} {1109.2219})

\bibitem[\protect\citeauthoryear{{Springel}}{{Springel}}{2010b}]{Springel2010}
{Springel} V.,  2010b, \mn@doi [\mnras] {10.1111/j.1365-2966.2009.15715.x},
  \href {https://ui.adsabs.harvard.edu/abs/2010MNRAS.401..791S} {401, 791}
  (\mn@eprint {arXiv} {0901.4107})

\bibitem[\protect\citeauthoryear{{Springel} \& {Hernquist}}{{Springel} \&
  {Hernquist}}{2003}]{Springel+2003}
{Springel} V.,  {Hernquist} L.,  2003, \mn@doi [\mnras]
  {10.1046/j.1365-8711.2003.06206.x}, \href
  {https://ui.adsabs.harvard.edu/abs/2003MNRAS.339..289S} {339, 289}
  (\mn@eprint {arXiv} {astro-ph/0206393})

\bibitem[\protect\citeauthoryear{{Springel}, {White}, {Tormen}  \&
  {Kauffmann}}{{Springel} et~al.}{2001}]{Springel+2001}
{Springel} V.,  {White} S. D.~M.,  {Tormen} G.,   {Kauffmann} G.,  2001,
  \mn@doi [\mnras] {10.1046/j.1365-8711.2001.04912.x}, \href
  {https://ui.adsabs.harvard.edu/abs/2001MNRAS.328..726S} {328, 726}
  (\mn@eprint {arXiv} {astro-ph/0012055})

\bibitem[\protect\citeauthoryear{{Springel} et~al.,}{{Springel}
  et~al.}{2005}]{Springel+2005}
{Springel} V.,  et~al., 2005, \mn@doi [\nat] {10.1038/nature03597}, \href
  {https://ui.adsabs.harvard.edu/abs/2005Natur.435..629S} {435, 629}
  (\mn@eprint {arXiv} {astro-ph/0504097})

\bibitem[\protect\citeauthoryear{{Springel} et~al.,}{{Springel}
  et~al.}{2018}]{Springel+2018}
{Springel} V.,  et~al., 2018, \mn@doi [\mnras] {10.1093/mnras/stx3304}, \href
  {https://ui.adsabs.harvard.edu/abs/2018MNRAS.475..676S} {475, 676}
  (\mn@eprint {arXiv} {1707.03397})

\bibitem[\protect\citeauthoryear{{Springel}, {Pakmor}, {Zier}  \&
  {Reinecke}}{{Springel} et~al.}{2021}]{Gadget4}
{Springel} V.,  {Pakmor} R.,  {Zier} O.,   {Reinecke} M.,  2021, \mn@doi
  [\mnras] {10.1093/mnras/stab1855}, \href
  {https://ui.adsabs.harvard.edu/abs/2021MNRAS.506.2871S} {506, 2871}
  (\mn@eprint {arXiv} {2010.03567})

\bibitem[\protect\citeauthoryear{{Srisawat} et~al.,}{{Srisawat}
  et~al.}{2013}]{Srisawat+2013}
{Srisawat} C.,  et~al., 2013, \mn@doi [\mnras] {10.1093/mnras/stt1545}, \href
  {https://ui.adsabs.harvard.edu/abs/2013MNRAS.436..150S} {436, 150}
  (\mn@eprint {arXiv} {1307.3577})

\bibitem[\protect\citeauthoryear{{Stanway} \& {Eldridge}}{{Stanway} \&
  {Eldridge}}{2018}]{Stanway+2018}
{Stanway} E.~R.,  {Eldridge} J.~J.,  2018, \mn@doi [\mnras]
  {10.1093/mnras/sty1353}, \href
  {https://ui.adsabs.harvard.edu/abs/2018MNRAS.479...75S} {479, 75} (\mn@eprint
  {arXiv} {1805.08784})

\bibitem[\protect\citeauthoryear{{Stinson}, {Seth}, {Katz}, {Wadsley},
  {Governato}  \& {Quinn}}{{Stinson} et~al.}{2006}]{Stinson+2006}
{Stinson} G.,  {Seth} A.,  {Katz} N.,  {Wadsley} J.,  {Governato} F.,   {Quinn}
  T.,  2006, \mn@doi [\mnras] {10.1111/j.1365-2966.2006.11097.x}, \href
  {https://ui.adsabs.harvard.edu/abs/2006MNRAS.373.1074S} {373, 1074}
  (\mn@eprint {arXiv} {astro-ph/0602350})

\bibitem[\protect\citeauthoryear{{Stinson}, {Brook}, {Macci{\`o}}, {Wadsley},
  {Quinn}  \& {Couchman}}{{Stinson} et~al.}{2013}]{Stinson+2013}
{Stinson} G.~S.,  {Brook} C.,  {Macci{\`o}} A.~V.,  {Wadsley} J.,  {Quinn}
  T.~R.,   {Couchman} H.~M.~P.,  2013, \mn@doi [\mnras] {10.1093/mnras/sts028},
  \href {https://ui.adsabs.harvard.edu/abs/2013MNRAS.428..129S} {428, 129}
  (\mn@eprint {arXiv} {1208.0002})

\bibitem[\protect\citeauthoryear{{Su}, {Hopkins}, {Hayward},
  {Faucher-Gigu{\`e}re}, {Kere{\v{s}}}, {Ma}  \& {Robles}}{{Su}
  et~al.}{2017}]{Su+2017}
{Su} K.-Y.,  {Hopkins} P.~F.,  {Hayward} C.~C.,  {Faucher-Gigu{\`e}re} C.-A.,
  {Kere{\v{s}}} D.,  {Ma} X.,   {Robles} V.~H.,  2017, \mn@doi [\mnras]
  {10.1093/mnras/stx1463}, \href
  {https://ui.adsabs.harvard.edu/abs/2017MNRAS.471..144S} {471, 144}
  (\mn@eprint {arXiv} {1607.05274})

\bibitem[\protect\citeauthoryear{{Sutherland} \& {Dopita}}{{Sutherland} \&
  {Dopita}}{1993}]{Sutherland+Dopita1993}
{Sutherland} R.~S.,  {Dopita} M.~A.,  1993, \mn@doi [\apjs] {10.1086/191823},
  \href {https://ui.adsabs.harvard.edu/abs/1993ApJS...88..253S} {88, 253}

\bibitem[\protect\citeauthoryear{{Talbot}, {Pakmor}, {Pfrommer}, {Springel},
  {Werhahn}, {Bieri}  \& {van de Voort}}{{Talbot} et~al.}{2024}]{Talbot+2024}
{Talbot} R.~Y.,  {Pakmor} R.,  {Pfrommer} C.,  {Springel} V.,  {Werhahn} M.,
  {Bieri} R.,   {van de Voort} F.,  2024, \mn@doi [arXiv e-prints]
  {10.48550/arXiv.2410.07316}, \href
  {https://ui.adsabs.harvard.edu/abs/2024arXiv241007316T} {p. arXiv:2410.07316}
  (\mn@eprint {arXiv} {2410.07316})

\bibitem[\protect\citeauthoryear{{Taylor} \& {Kobayashi}}{{Taylor} \&
  {Kobayashi}}{2014}]{Taylor+Kobayashi2014}
{Taylor} P.,  {Kobayashi} C.,  2014, \mn@doi [\mnras] {10.1093/mnras/stu983},
  \href {https://ui.adsabs.harvard.edu/abs/2014MNRAS.442.2751T} {442, 2751}
  (\mn@eprint {arXiv} {1405.4194})

\bibitem[\protect\citeauthoryear{{Teyssier}}{{Teyssier}}{2002}]{Ramses}
{Teyssier} R.,  2002, \mn@doi [\aap] {10.1051/0004-6361:20011817}, \href
  {https://ui.adsabs.harvard.edu/abs/2002A&A...385..337T} {385, 337}
  (\mn@eprint {arXiv} {astro-ph/0111367})

\bibitem[\protect\citeauthoryear{{Teyssier}, {Pontzen}, {Dubois}  \&
  {Read}}{{Teyssier} et~al.}{2013}]{Teyssier+2013}
{Teyssier} R.,  {Pontzen} A.,  {Dubois} Y.,   {Read} J.~I.,  2013, \mn@doi
  [\mnras] {10.1093/mnras/sts563}, \href
  {https://ui.adsabs.harvard.edu/abs/2013MNRAS.429.3068T} {429, 3068}
  (\mn@eprint {arXiv} {1206.4895})

\bibitem[\protect\citeauthoryear{{Thob}, {Sanderson}, {Eden}, {Nikakhtar},
  {Panithanpaisal}, {Garavito-Camargo}  \& {Sharma}}{{Thob}
  et~al.}{2024}]{Thob+2024}
{Thob} A.,  {Sanderson} R.,  {Eden} A.,  {Nikakhtar} F.,  {Panithanpaisal} N.,
  {Garavito-Camargo} N.,   {Sharma} S.,  2024, \mn@doi [The Journal of Open
  Source Software] {10.21105/joss.06234}, \href
  {https://ui.adsabs.harvard.edu/abs/2024JOSS....9.6234T} {9, 6234} (\mn@eprint
  {arXiv} {2312.02268})

\bibitem[\protect\citeauthoryear{{Tollet}, {Cattaneo}, {Macci{\`o}}, {Dutton}
  \& {Kang}}{{Tollet} et~al.}{2019}]{Tollet+2019}
{Tollet} {\'E}.,  {Cattaneo} A.,  {Macci{\`o}} A.~V.,  {Dutton} A.~A.,   {Kang}
  X.,  2019, \mn@doi [\mnras] {10.1093/mnras/stz545}, \href
  {https://ui.adsabs.harvard.edu/abs/2019MNRAS.485.2511T} {485, 2511}
  (\mn@eprint {arXiv} {1902.03888})

\bibitem[\protect\citeauthoryear{{Toomre} \& {Toomre}}{{Toomre} \&
  {Toomre}}{1972}]{Toomre+1972}
{Toomre} A.,  {Toomre} J.,  1972, \mn@doi [\apj] {10.1086/151823}, \href
  {https://ui.adsabs.harvard.edu/abs/1972ApJ...178..623T} {178, 623}

\bibitem[\protect\citeauthoryear{{Trebitsch} et~al.,}{{Trebitsch}
  et~al.}{2021}]{Trebitsch+2021}
{Trebitsch} M.,  et~al., 2021, \mn@doi [\aap] {10.1051/0004-6361/202037698},
  \href {https://ui.adsabs.harvard.edu/abs/2021A&A...653A.154T} {653, A154}
  (\mn@eprint {arXiv} {2002.04045})

\bibitem[\protect\citeauthoryear{{Tremmel}, {Governato}, {Volonteri}  \&
  {Quinn}}{{Tremmel} et~al.}{2015}]{Tremmel+2015}
{Tremmel} M.,  {Governato} F.,  {Volonteri} M.,   {Quinn} T.~R.,  2015, \mn@doi
  [\mnras] {10.1093/mnras/stv1060}, \href
  {https://ui.adsabs.harvard.edu/abs/2015MNRAS.451.1868T} {451, 1868}
  (\mn@eprint {arXiv} {1501.07609})

\bibitem[\protect\citeauthoryear{{Tremmel}, {Karcher}, {Governato},
  {Volonteri}, {Quinn}, {Pontzen}, {Anderson}  \& {Bellovary}}{{Tremmel}
  et~al.}{2017}]{Tremmel+2017}
{Tremmel} M.,  {Karcher} M.,  {Governato} F.,  {Volonteri} M.,  {Quinn} T.~R.,
  {Pontzen} A.,  {Anderson} L.,   {Bellovary} J.,  2017, \mn@doi [\mnras]
  {10.1093/mnras/stx1160}, \href
  {https://ui.adsabs.harvard.edu/abs/2017MNRAS.470.1121T} {470, 1121}
  (\mn@eprint {arXiv} {1607.02151})

\bibitem[\protect\citeauthoryear{{Tr{\"o}ster}, {Ferguson},
  {Harnois-D{\'e}raps}  \& {McCarthy}}{{Tr{\"o}ster}
  et~al.}{2019}]{Troester+2019}
{Tr{\"o}ster} T.,  {Ferguson} C.,  {Harnois-D{\'e}raps} J.,   {McCarthy} I.~G.,
   2019, \mn@doi [\mnras] {10.1093/mnrasl/slz075}, \href
  {https://ui.adsabs.harvard.edu/abs/2019MNRAS.487L..24T} {487, L24}
  (\mn@eprint {arXiv} {1903.12173})

\bibitem[\protect\citeauthoryear{{Tumlinson}, {Peeples}  \& {Werk}}{{Tumlinson}
  et~al.}{2017}]{Tumlinson+2017}
{Tumlinson} J.,  {Peeples} M.~S.,   {Werk} J.~K.,  2017, \mn@doi [\araa]
  {10.1146/annurev-astro-091916-055240}, \href
  {https://ui.adsabs.harvard.edu/abs/2017ARA&A..55..389T} {55, 389} (\mn@eprint
  {arXiv} {1709.09180})

\bibitem[\protect\citeauthoryear{{Vandenbroucke} \& {De
  Rijcke}}{{Vandenbroucke} \& {De Rijcke}}{2016}]{Vandenbroucke+2016}
{Vandenbroucke} B.,  {De Rijcke} S.,  2016, \mn@doi [Astronomy and Computing]
  {10.1016/j.ascom.2016.05.001}, \href
  {https://ui.adsabs.harvard.edu/abs/2016A&C....16..109V} {16, 109} (\mn@eprint
  {arXiv} {1605.03576})

\bibitem[\protect\citeauthoryear{{Villaescusa-Navarro}
  et~al.,}{{Villaescusa-Navarro} et~al.}{2021}]{Villaescusa-Navarro+2021}
{Villaescusa-Navarro} F.,  et~al., 2021, \mn@doi [\apj]
  {10.3847/1538-4357/abf7ba}, \href
  {https://ui.adsabs.harvard.edu/abs/2021ApJ...915...71V} {915, 71} (\mn@eprint
  {arXiv} {2010.00619})

\bibitem[\protect\citeauthoryear{{Vogelsberger} et~al.,}{{Vogelsberger}
  et~al.}{2014}]{Vogelsberger+2014}
{Vogelsberger} M.,  et~al., 2014, \mn@doi [\mnras] {10.1093/mnras/stu1536},
  \href {https://ui.adsabs.harvard.edu/abs/2014MNRAS.444.1518V} {444, 1518}
  (\mn@eprint {arXiv} {1405.2921})

\bibitem[\protect\citeauthoryear{{Vogelsberger}, {Marinacci}, {Torrey}  \&
  {Puchwein}}{{Vogelsberger} et~al.}{2020}]{ReviewVogelsberger+2020}
{Vogelsberger} M.,  {Marinacci} F.,  {Torrey} P.,   {Puchwein} E.,  2020,
  \mn@doi [Nature Reviews Physics] {10.1038/s42254-019-0127-2}, \href
  {https://ui.adsabs.harvard.edu/abs/2020NatRP...2...42V} {2, 42} (\mn@eprint
  {arXiv} {1909.07976})

\bibitem[\protect\citeauthoryear{{Volonteri}, {Dubois}, {Pichon}  \&
  {Devriendt}}{{Volonteri} et~al.}{2016}]{Volonteri+2016}
{Volonteri} M.,  {Dubois} Y.,  {Pichon} C.,   {Devriendt} J.,  2016, \mn@doi
  [\mnras] {10.1093/mnras/stw1123}, \href
  {https://ui.adsabs.harvard.edu/abs/2016MNRAS.460.2979V} {460, 2979}
  (\mn@eprint {arXiv} {1602.01941})

\bibitem[\protect\citeauthoryear{{Volonteri}, {Habouzit}  \&
  {Colpi}}{{Volonteri} et~al.}{2021}]{Volonteri+2021}
{Volonteri} M.,  {Habouzit} M.,   {Colpi} M.,  2021, \mn@doi [Nature Reviews
  Physics] {10.1038/s42254-021-00364-9}, \href
  {https://ui.adsabs.harvard.edu/abs/2021NatRP...3..732V} {3, 732} (\mn@eprint
  {arXiv} {2110.10175})

\bibitem[\protect\citeauthoryear{{Wadekar}, {Villaescusa-Navarro}, {Ho}  \&
  {Perreault-Levasseur}}{{Wadekar} et~al.}{2021}]{Wadekar+2021}
{Wadekar} D.,  {Villaescusa-Navarro} F.,  {Ho} S.,   {Perreault-Levasseur} L.,
  2021, \mn@doi [\apj] {10.3847/1538-4357/ac033a}, \href
  {https://ui.adsabs.harvard.edu/abs/2021ApJ...916...42W} {916, 42} (\mn@eprint
  {arXiv} {2007.10340})

\bibitem[\protect\citeauthoryear{{Wadsley}, {Keller}  \& {Quinn}}{{Wadsley}
  et~al.}{2017}]{Gasoline2}
{Wadsley} J.~W.,  {Keller} B.~W.,   {Quinn} T.~R.,  2017, \mn@doi [\mnras]
  {10.1093/mnras/stx1643}, \href
  {https://ui.adsabs.harvard.edu/abs/2017MNRAS.471.2357W} {471, 2357}
  (\mn@eprint {arXiv} {1707.03824})

\bibitem[\protect\citeauthoryear{{Weinberger} et~al.,}{{Weinberger}
  et~al.}{2017}]{Weinberger+2017}
{Weinberger} R.,  et~al., 2017, \mn@doi [\mnras] {10.1093/mnras/stw2944}, \href
  {https://ui.adsabs.harvard.edu/abs/2017MNRAS.465.3291W} {465, 3291}
  (\mn@eprint {arXiv} {1607.03486})

\bibitem[\protect\citeauthoryear{{Wellons} et~al.,}{{Wellons}
  et~al.}{2023}]{Wellons+2023}
{Wellons} S.,  et~al., 2023, \mn@doi [\mnras] {10.1093/mnras/stad511}, \href
  {https://ui.adsabs.harvard.edu/abs/2023MNRAS.520.5394W} {520, 5394}
  (\mn@eprint {arXiv} {2203.06201})

\bibitem[\protect\citeauthoryear{{Wetzel}, {Hopkins}, {Kim},
  {Faucher-Gigu{\`e}re}, {Kere{\v{s}}}  \& {Quataert}}{{Wetzel}
  et~al.}{2016}]{Wetzel+2016}
{Wetzel} A.~R.,  {Hopkins} P.~F.,  {Kim} J.-h.,  {Faucher-Gigu{\`e}re} C.-A.,
  {Kere{\v{s}}} D.,   {Quataert} E.,  2016, \mn@doi [\apjl]
  {10.3847/2041-8205/827/2/L23}, \href
  {https://ui.adsabs.harvard.edu/abs/2016ApJ...827L..23W} {827, L23}
  (\mn@eprint {arXiv} {1602.05957})

\bibitem[\protect\citeauthoryear{{Wibking} \& {Krumholz}}{{Wibking} \&
  {Krumholz}}{2022}]{Wibkin+2022}
{Wibking} B.~D.,  {Krumholz} M.~R.,  2022, \mn@doi [\mnras]
  {10.1093/mnras/stac439}, \href
  {https://ui.adsabs.harvard.edu/abs/2022MNRAS.512.1430W} {512, 1430}
  (\mn@eprint {arXiv} {2110.01792})

\bibitem[\protect\citeauthoryear{{Wiersma}, {Schaye}  \& {Smith}}{{Wiersma}
  et~al.}{2009}]{Wiersma+2009}
{Wiersma} R. P.~C.,  {Schaye} J.,   {Smith} B.~D.,  2009, \mn@doi [\mnras]
  {10.1111/j.1365-2966.2008.14191.x}, \href
  {https://ui.adsabs.harvard.edu/abs/2009MNRAS.393...99W} {393, 99} (\mn@eprint
  {arXiv} {0807.3748})

\bibitem[\protect\citeauthoryear{{Wright}, {Somerville}, {Lagos}, {Schaller},
  {Dav{\'e}}, {Angl{\'e}s-Alc{\'a}zar}  \& {Genel}}{{Wright}
  et~al.}{2024}]{Wright+2024}
{Wright} R.~J.,  {Somerville} R.~S.,  {Lagos} C. d.~P.,  {Schaller} M.,
  {Dav{\'e}} R.,  {Angl{\'e}s-Alc{\'a}zar} D.,   {Genel} S.,  2024, \mn@doi
  [\mnras] {10.1093/mnras/stae1688}, \href
  {https://ui.adsabs.harvard.edu/abs/2024MNRAS.532.3417W} {532, 3417}
  (\mn@eprint {arXiv} {2402.08408})

\bibitem[\protect\citeauthoryear{{Xu}, {Wise}, {Norman}, {Ahn}  \&
  {O'Shea}}{{Xu} et~al.}{2016}]{Hao+2016}
{Xu} H.,  {Wise} J.~H.,  {Norman} M.~L.,  {Ahn} K.,   {O'Shea} B.~W.,  2016,
  \mn@doi [\apj] {10.3847/1538-4357/833/1/84}, \href
  {https://ui.adsabs.harvard.edu/abs/2016ApJ...833...84X} {833, 84} (\mn@eprint
  {arXiv} {1604.07842})

\bibitem[\protect\citeauthoryear{{Zel'dovich}}{{Zel'dovich}}{1970}]{Zeldovich1970}
{Zel'dovich} Y.~B.,  1970, \aap, \href
  {https://ui.adsabs.harvard.edu/abs/1970A&A.....5...84Z} {5, 84}

\bibitem[\protect\citeauthoryear{{Zhang}, {Zhu}, {Wu}, {Adams}  \&
  {Hu}}{{Zhang} et~al.}{2022}]{Zhang+2022}
{Zhang} C.,  {Zhu} Y.-j.,  {Wu} D.,  {Adams} N.~A.,   {Hu} X.,  2022, \mn@doi
  [Journal of Hydrodynamics] {10.1007/s42241-022-0052-1}, \href
  {https://ui.adsabs.harvard.edu/abs/2022JHyDy..34..767Z} {34, 767} (\mn@eprint
  {arXiv} {2205.03074})

\bibitem[\protect\citeauthoryear{{de Bernardis} et~al.,}{{de Bernardis}
  et~al.}{2000}]{deBernardis+2000}
{de Bernardis} P.,  et~al., 2000, \mn@doi [\nat] {10.1038/35010035}, \href
  {https://ui.adsabs.harvard.edu/abs/2000Natur.404..955D} {404, 955}
  (\mn@eprint {arXiv} {astro-ph/0004404})

\bibitem[\protect\citeauthoryear{{van de Voort}}{{van de
  Voort}}{2017}]{vandeVoort2017}
{van de Voort} F.,  2017, in {Fox} A.,  {Dav{\'e}} R.,  eds,  Astrophysics and
  Space Science Library Vol. 430, Gas Accretion onto Galaxies. p.~301
  (\mn@eprint {arXiv} {1612.00591}), \mn@doi{10.1007/978-3-319-52512-9_13}

\bibitem[\protect\citeauthoryear{{van de Voort}, {Schaye}, {Booth}, {Haas}  \&
  {Dalla Vecchia}}{{van de Voort} et~al.}{2011}]{vandeVoort+2011}
{van de Voort} F.,  {Schaye} J.,  {Booth} C.~M.,  {Haas} M.~R.,   {Dalla
  Vecchia} C.,  2011, \mn@doi [\mnras] {10.1111/j.1365-2966.2011.18565.x},
  \href {https://ui.adsabs.harvard.edu/abs/2011MNRAS.414.2458V} {414, 2458}
  (\mn@eprint {arXiv} {1011.2491})

\bibitem[\protect\citeauthoryear{{van de Voort}, {Springel}, {Mandelker}, {van
  den Bosch}  \& {Pakmor}}{{van de Voort} et~al.}{2019}]{VandeVoort+2019}
{van de Voort} F.,  {Springel} V.,  {Mandelker} N.,  {van den Bosch} F.~C.,
  {Pakmor} R.,  2019, \mn@doi [\mnras] {10.1093/mnrasl/sly190}, \href
  {https://ui.adsabs.harvard.edu/abs/2019MNRAS.482L..85V} {482, L85}
  (\mn@eprint {arXiv} {1808.04369})

\makeatother
\end{thebibliography}
\input{chapter.bbl}

\end{document}